# The Logical Difference for the Lightweight Description Logic $\mathcal{EL}$

**Boris Konev** KONEV@LIVERPOOL.AC.UK
**Michel Ludwig** MICHEL.LUDWIG@LIVERPOOL.AC.UK
*Department of Computer Science*
*University of Liverpool, UK*

**Dirk Walther** DIRK.WALTHER@UPM.ES
*Departamento Inteligencia Artificial, Facultad de Informática*
*Universidad Politécnica de Madrid, Spain*

**Frank Wolter** WOLTER@LIVERPOOL.AC.UK
*Department of Computer Science*
*University of Liverpool, UK*

## Abstract

We study a logic-based approach to versioning of ontologies. Under this view, ontologies provide answers to queries about some vocabulary of interest. The difference between two versions of an ontology is given by the set of queries that receive different answers. We investigate this approach for terminologies given in the description logic $\mathcal{EL}$ extended with role inclusions and domain and range restrictions for three distinct types of queries: subsumption, instance, and conjunctive queries. In all three cases, we present polynomial-time algorithms that decide whether two terminologies give the same answers to queries over a given vocabulary and compute a succinct representation of the difference if it is non-empty. We present an implementation, CEX2, of the developed algorithms for subsumption and instance queries and apply it to distinct versions of SNOMED CT and the NCI ontology.

## 1. Introduction

Terminologies are lightweight ontologies that are used to provide a common vocabulary for a domain of interest together with descriptions of the meaning of terms built from the vocabulary and relationships between them. They are being used in areas such as medical informatics, bio-informatics, and the semantic web to capture domain semantics and promote interoperability. Terminologies are often large and complex. For example, the widely used medical terminology SNOMED CT (Systematized Nomenclature of Medicine Clinical Terms) contains more than 300 000 term definitions (IHTSDO, 2008). Another example is the National Cancer Institute ontology (NCI) consisting of more than 60 000 axioms (Golbeck, Fragaso, Hartel, Hendler, Oberhaler, & Parsia, 2003). Engineering, maintaining, and using such terminologies is a complex and laborious task, which is practically unfeasible without appropriate tool support. In this article, we focus on a principled logic-based approach to support for *terminology versioning*.

Dealing with multiple versions of the same information unit is nothing new in computing, and version control is a well established computer technology. Although modern version control systems provide a range of operations including support for collaborative development, branching, merging, *etc.*, these operations extend and rely on the basic operations of *detecting and representing the differences* between versions. In this paper, we focus on this basic problem of versioning.

The need for versioning support is recognised by the ontology research community and ontology users, and a large number of approaches and tools have been developed. In our review of currently existing support for ontology versioning, we distinguish *three* approaches and describe them according to the difference between ontologies they compute:

1. versioning based on syntactic difference (syntactic diff);
2. versioning based on structural difference (structural diff);
3. versioning based on logical difference (logical diff).

The *syntactic diff* underlies most existing version control systems used in software development (Conradi & Westfechtel, 1998) (such as, for example, RCS, CVS, SCCS). It works with text files and represents the difference between versions as blocks of text present in one version but not another, ignoring any meta-information about the document. As observed already in the work of Noy and Musen (2002), ontology versioning cannot rely on a purely syntactic diff operation since many syntactic differences (e.g., the order of ontology axioms) do not affect the semantics of ontologies. Therefore, ontology versioning based on syntactic difference is essentially limited to comparing rather informal change logs (Oliver, Shahar, Shortliffe, & Musen, 1999).

The *structural diff* extends the syntactic diff by taking into account information about the structure of ontologies. It has been suggested for dealing with structured and hierarchical documents such as UML diagrams, database schemas, or XML documents (see, e.g., Ohst, Welle, & Kelter, 2003, and references within). For ontologies, the main characteristic of the structural diff is that it regards them as structured objects, such as an *is-a* taxonomy (Noy & Musen, 2002), a set of RDF triplets (Klein, Fensel, Kiryakov, & Ognyanov, 2002) or a set of class defining axioms (Redmond, Smith, Drummond, & Tudorache, 2008; Jiménez-Ruiz, Cuenca Grau, Horrocks, & Llavori, 2011). Changes to ontologies are mostly described in terms of structural operations, for example, adding or deleting a class, extending a class, renaming slots, moving a class from one place in the hierarchy to another, adding or deleting an axiom, class renaming, *etc.*; sometimes basic logical properties of ontologies, e.g., the equivalence of different structural forms of concepts, are also taken into account (Palma, Haase, Corcho, & Gómez-Pérez, 2009; Jiménez-Ruiz et al., 2011). Ontology versioning based on structural diff of some form is available in most current ontology editors and ontology management systems either natively or through plugins (Noy & Musen, 2002; Klein et al., 2002; Jiménez-Ruiz et al., 2011).

Though very helpful, the structural diff still has the deficiency of having no unambiguous semantic foundation and being syntax dependent. Moreover, it is tailored towards applications of ontologies which are based on the induced concept hierarchy (or some mild extension of it), but does not capture modern applications such as ontology based data access (OBDA) (Poggi, Lembo, Calvanese, Giacomo, Lenzerini, & Rosati, 2008; Lutz, Toman, & Wolter, 2009) in which ontologies are used to provide a user-oriented view of the data and make it accessible via queries formulated solely in the language of the ontology without any knowledge of the actual structure of the data.[1]

The *logical diff* has only been recently introduced (Konev, Walther, & Wolter, 2008; Kontchakov, Wolter, & Zakharyaschev, 2010) and completely abstracts from the representation of the ontology. Here, an ontology is regarded as a set of axioms formulated in a logical language with a formal and unambiguous semantics. Under this view, ontologies provide answers to queries about some vocabulary of interest. Typical queries include subsumption queries between concepts and, if the ontology is used to access instance data, instance and conjunctive queries. The logical diff is motivated by this view. If two versions of an ontology give the same answers to a class of queries relevant to an application domain, they may be deemed to have no difference regardless of their syntactic or structural form; and queries producing different answers from the versions may be considered as a characterisation of the difference itself. In this way one can, for example, define exactly the differences visible when querying instance data or exactly the differences expressed by subsumptions between concepts.

To make this approach work in practice, at least two problems have to be addressed:

- For most ontology languages and classes of queries the computational complexity of even detecting if two ontology versions differ over a certain vocabulary is at least one exponential harder than ontology classification and is sometimes undecidable; and even if the computational complexity does not increase, searching for differences between ontologies within a certain vocabulary requires techniques that are very different from those used for standard reasoning (Lutz, Walther, & Wolter, 2007; Lutz & Wolter, 2010; Cuenca Grau, Horrocks, Kazakov, & Sattler, 2008).

- If the set of queries producing different answers from the two versions is not empty, it is typically infinite and, therefore, cannot be presented to the user as such. Thus, techniques to succinctly characterise its elements and present them to the user are required.

The aim of this paper is to provide first steps toward solutions to these problems for terminologies (aka classical TBoxes) given in the description logic $\mathcal{ELH}^r$ that extends the description logic $\mathcal{EL}$ underlying the OWL 2 EL profile with role inclusions and domain and range restrictions (Baader, Brandt, & Lutz, 2008). Our main contributions are as follows:

1. It has been argued that syntax-dependence should be regarded as an advantage rather than a deficiency in the context of versioning (Gonçalves, Parsia, & Sattler, 2011; Jiménez-Ruiz et al., 2011). For example, Jiménez-Ruiz et al. argue that logical equivalence between ontologies can be too permissive: "even if $O \equiv O'$ – the strongest assumption from a semantic point of view – conflicts may still exist. This might result from the presence of incompatible annotations (statements that act as comments and do not carry logical meaning), or a mismatch in modelling styles; for example, $O$ may be written in a simple language such as the OWL 2 EL profile and contain $\alpha = (A \sqsubseteq B \sqcap C)$, while $O'$ may contain $\beta = (\neg B \sqcup \neg C \sqsubseteq \neg A)$. Even though $\alpha \equiv \beta$, the explicit use of negation and disjunction means that $O'$ is outside the EL profile." We agree with Jiménez-Ruiz et al. and Gonçalves et al. that there are various applications in which a structural rather than logical difference is appropriate. Even a syntactic diff has applications in ontology versioning. In practice, we see logic-based approaches as *complementary* to structural approaches. An interesting analysis of NCI versions taking into account both structural and logical differences is given in the work of Gonçalves et al.

• for subsumption, instance, and conjunctive queries, we present polynomial-time algorithms that decide whether two $\mathcal{ELH}^r$-terminologies give different answers to some query from the respective class of queries over a given signature of concept and role names (note that we use the terms signature and vocabulary synonymously).

• Besides of a polynomial-time decision procedure detecting differences, we also develop a succinct presentation of the (typically infinite) difference. This presentation can be computed in polynomial time as well.

• We present two different types of polynomial-time algorithms for deciding the existence of logical differences between terminologies and for computing a succinct representation of it: the first type of algorithms is conceptually more transparent as it keeps the two input terminologies separate and reduces (a substantial part of) the difference problem to an instance checking problem for an ABox. Such algorithms are, however, not sufficiently efficient on *very large* inputs. For example, substantial performance problems occur when computing the differences between versions of Snomed CT on their joint signature since the constructed ABox is typically of quadratic size in the input terminologies. The second variant of algorithms, which is based on dynamic programming, is more efficient in practice. It is developed in detail for *acyclic* $\mathcal{ELH}^r$-terminologies.

• We present an implementation, CEX2, that is based on the second type of algorithms and computes a succinct representation of the difference between acyclic $\mathcal{ELH}^r$-terminologies for the concept and instance query case. In addition, a prototype implementation of the ABox-based algorithm is used to estimate its efficiency.

• As an important tool in our investigation, we present description logics, $\mathcal{EL}^{\mathsf{ran}}$ and $\mathcal{EL}^{\mathsf{ran},\sqcap,u}$, that capture as subsumption differences the instance and query difference between $\mathcal{ELH}^r$-terminologies. This result is presented for general $\mathcal{ELH}^r$-TBoxes and can, therefore, be exploited in future work on versioning for general $\mathcal{ELH}^r$-TBoxes.

• We present experiments using CEX2 that illustrate the efficiency of the algorithms and potential applications to terminologies such as Snomed CT and NCI. A plugin for Protégé is discussed. CEX2 extends the functionality of the first version of CEX (Konev, Walther, & Wolter, 2008) and of the OwlDiff plugin (Křemen, Šmíd, & Kouba, 2011), which implements the algorithms developed by Konev, Walther, and Wolter. Based on Snomed CT, we also investigate the performance of the ABox-based algorithms in practice.

This paper is based on, and extends the work of Konev, Walther, and Wolter (2008). To improve readability, a number of proofs have been deferred to an appendix.

## 2. Preliminaries

Let $\mathsf{N_C}$, $\mathsf{N_R}$, and $\mathsf{N_I}$ be countably infinite and mutually disjoint sets of *concept names*, *role names*, and *individual names*. *$\mathcal{EL}$-concepts $C$* are built according to the rule

$$C := \quad A \quad | \quad \top \quad | \quad C \sqcap D \quad | \quad \exists r.C,$$

where $A \in \mathsf{N_C}$, $r \in \mathsf{N_R}$, and $C, D$ range over $\mathcal{EL}$-concepts. The set of *$\mathcal{ELH}^r$-inclusions* consists of

- *concept inclusions* $C \sqsubseteq D$, $\mathsf{ran}(r) \sqsubseteq D$ and $\mathsf{ran}(r) \sqcap C \sqsubseteq D$,

- *concept equations* $C \equiv D$, and
- *role inclusions* $r \sqsubseteq s$,

where $C$ and $D$ are $\mathcal{EL}$-concepts and $r, s \in \mathsf{N_R}$. An $\mathcal{ELH}^r$-TBox $\mathcal{T}$ is a finite set of $\mathcal{ELH}^r$-inclusions. Inclusions of the form $\mathsf{ran}(r) \sqsubseteq D$ and $\mathsf{ran}(r) \sqcap C \sqsubseteq D$ are also referred to as *range restrictions*, and inclusions of the form $\exists r.\top \sqsubseteq D$ are referred to as *domain restrictions.*

An $\mathcal{ELH}^r$-TBox is called an $\mathcal{ELH}^r$*-terminology* if all its concept inclusions and equations are of the form

- $A \sqsubseteq C$ and $A \equiv C$,
- $\mathsf{ran}(r) \sqsubseteq C$, and
- $\exists r.\top \sqsubseteq C$,

where $A \in \mathsf{N_C}$ and $r \in \mathsf{N_R}$, $C$ is an $\mathcal{EL}$-concept such that $C \neq \top$, $C \neq \top \sqcap \top$, etc., and no concept name occurs more than once on the left-hand side. Note that, in concept inclusions of the form $\exists r.\top \sqsubseteq C$, the concept $\exists r.\top$ is often denoted $\mathrm{dom}(r)$. A terminology is *acyclic* (or *unfoldable*) if the process of exhaustively substituting definitions in place of the defined concept names terminates. For example, if a terminology contains a concept inclusion

$$\mathsf{Mother} \sqsubseteq \exists \mathsf{hasMother}.\mathsf{Mother}$$

it is not acyclic. Formally, consider the relation $\prec_{\mathcal{T}}$ between concept names by setting $A \prec_{\mathcal{T}} B$ if there exists an $\mathcal{ELH}^r$-inclusion of the form $A \equiv C$ or $A \sqsubseteq C$ in $\mathcal{T}$ such that $B$ occurs in $C$. A terminology $\mathcal{T}$ is *acyclic* if the transitive closure $\prec_{\mathcal{T}}^+$ of $\prec_{\mathcal{T}}$ is irreflexive.

In description logic, instance data are represented by *ABox assertions* of the form $\top(a)$, $A(a)$ and $r(a, b)$, where $a, b \in \mathsf{N_I}$, $A \in \mathsf{N_C}$, and $r \in \mathsf{N_R}$. An *ABox* $\mathcal{A}$ is a non-empty finite set of ABox-assertions. $\mathcal{A}$ is said to be a *singleton ABox* if it contains exactly one ABox assertion. By $\mathsf{obj}(\mathcal{A})$ we denote the set of individual names in $\mathcal{A}$. A *knowledge base* $\mathcal{K}$ (KB) is a pair $(\mathcal{T}, \mathcal{A})$ consisting of a TBox $\mathcal{T}$ and an ABox $\mathcal{A}$. Assertions of the form $C(a)$ and $r(a, b)$, where $a, b \in \mathsf{N_I}$, $C$ an $\mathcal{EL}$-concept, and $r \in \mathsf{N_R}$, are called *instance assertions.* Note that instance assertions of the form $C(a)$ with $C$ not a concept name nor $C = \top$ do not occur in ABoxes.

The semantics of $\mathcal{ELH}^r$ is given by *interpretations* $\mathcal{I} = (\Delta^{\mathcal{I}}, \cdot^{\mathcal{I}})$, where the *domain* $\Delta^{\mathcal{I}}$ is a non-empty set, and $\cdot^{\mathcal{I}}$ is a function mapping each concept name $A$ to a subset $A^{\mathcal{I}}$ of $\Delta^{\mathcal{I}}$, each role name $r$ to a binary relation $r^{\mathcal{I}} \subseteq \Delta^{\mathcal{I}} \times \Delta^{\mathcal{I}}$, and each individual name $a$ to an element $a^{\mathcal{I}} \in \Delta^{\mathcal{I}}$. The *extension* $C^{\mathcal{I}}$ of a concept $C$ is defined by induction as follows:

$$\begin{array}{rcl}
\top^{\mathcal{I}} & := & \Delta^{\mathcal{I}} \\
(C \sqcap D)^{\mathcal{I}} & := & C^{\mathcal{I}} \cap D^{\mathcal{I}} \\
(\exists r.C)^{\mathcal{I}} & := & \{d \in \Delta^{\mathcal{I}} \mid \exists e \in C^{\mathcal{I}} : (d, e) \in r^{\mathcal{I}}\} \\
\mathsf{ran}(r)^{\mathcal{I}} & := & \{d \in \Delta^{\mathcal{I}} \mid \exists e : (e, d) \in r^{\mathcal{I}}\}
\end{array}$$

$\mathcal{I}$ *satisfies*

- a concept inclusion $C \sqsubseteq D$, in symbols $\mathcal{I} \models C \sqsubseteq D$, if $C^{\mathcal{I}} \subseteq D^{\mathcal{I}}$;

- a concept equation $C \equiv D$, in symbols $\mathcal{I} \models C \equiv D$, if $C^{\mathcal{I}} = D^{\mathcal{I}}$;
- a role inclusion $r \sqsubseteq s$, in symbols $\mathcal{I} \models r \sqsubseteq s$, if $r^{\mathcal{I}} \subseteq s^{\mathcal{I}}$;
- an assertion $C(a)$, in symbols $\mathcal{I} \models C(a)$, if $a^{\mathcal{I}} \in C^{\mathcal{I}}$,
- an assertion $r(a, b)$, in symbols $\mathcal{I} \models r(a, b)$, if $(a^{\mathcal{I}}, b^{\mathcal{I}}) \in r^{\mathcal{I}}$.

We say that an interpretation $\mathcal{I}$ is a *model of a TBox $\mathcal{T}$ (ABox $\mathcal{A}$)* if $\mathcal{I} \models \alpha$ for all $\alpha \in \mathcal{T}$ ($\alpha \in \mathcal{A}$). An $\mathcal{ELH}^r$-inclusion $\alpha$ *follows from* a TBox $\mathcal{T}$ if every model of $\mathcal{T}$ is a model of $\alpha$, in symbols $\mathcal{T} \models \alpha$. $\models \alpha$ is used to denote that $\alpha$ follows from the empty TBox and we sometimes write $r \sqsubseteq_{\mathcal{T}} s$ for $\mathcal{T} \models r \sqsubseteq s$. An instance assertion $\alpha$ follows from a KB $(\mathcal{T}, \mathcal{A})$ if every individual name that occurs in $\alpha$ also occurs in $\mathsf{obj}(\mathcal{A})$ and every model of $(\mathcal{T}, \mathcal{A})$ is a model of $\alpha$, in symbols $(\mathcal{T}, \mathcal{A}) \models \alpha$. The most important ways of querying $\mathcal{ELH}^r$-TBoxes and KBs are

- subsumption: check whether $\mathcal{T} \models \alpha$, for an $\mathcal{ELH}^r$-inclusion $\alpha$ and TBox $\mathcal{T}$,
- instance checking: check whether $(\mathcal{T}, \mathcal{A}) \models \alpha$, for an instance assertion $\alpha$ and KB $(\mathcal{T}, \mathcal{A})$, and
- conjunctive query answering.

To define the latter, call a first-order formula $q(\vec{x})$ a *conjunctive query* if it is of the form $\exists \vec{y} \psi(\vec{x}, \vec{y})$, where $\psi$ is a conjunction of expressions $A(t)$, $A \in \mathsf{N_C}$, and $r(t_1, t_2)$, $r \in \mathsf{N_R}$, with $t, t_1, t_2$ drawn from $\mathsf{N_I}$ and the sequences of variables $\vec{x}$ and $\vec{y}$. Let $\vec{x} = x_1, \dots, x_k$. Let $\mathcal{I}$ be an interpretation and $\pi$ be a mapping from $\vec{x} \cup \vec{y}$ into $\Delta^{\mathcal{I}}$. Set $\pi(a) = a^{\mathcal{I}}$ for all $a \in \mathsf{obj}(\mathcal{A})$. We say that a vector $\vec{a} = a_1, \dots, a_k$ is a *$\pi$-match of* $q(\vec{x})$ and $\mathcal{I}$ if $\pi$ satisfies the following conditions:

- $\pi(t) \in A^{\mathcal{I}}$ for every conjunct $A(t)$ of $\psi$;
- $(\pi(t_1), \pi(t_2)) \in r^{\mathcal{I}}$ for every conjunct $r(t_1, t_2)$ of $\psi$;
- $\pi(x_i) = a_i^{\mathcal{I}}$ for $1 \leq i \leq k$.

We set $\mathcal{I} \models q[\vec{a}]$ if, and only if, there exists a $\pi$ such that $\vec{a}$ is a $\pi$-match of $q(\vec{x})$ and $\mathcal{I}$. Let $(\mathcal{T}, \mathcal{A})$ be a KB. Then a sequence $\vec{a}$ of members of $\mathsf{obj}(\mathcal{A})$ is a *certain answer* to $q(\vec{x})$ of a KB $(\mathcal{T}, \mathcal{A})$, in symbols $(\mathcal{T}, \mathcal{A}) \models q(\vec{a})$, if $\mathcal{I} \models q[\vec{a}]$, for every model $\mathcal{I}$ of $(\mathcal{T}, \mathcal{A})$.

All three types of querying $\mathcal{ELH}^r$-TBoxes have been studied extensively. The complexity of subsumption and instance checking is in PTime (Baader et al., 2008). The combined complexity of answering Boolean conjunctive queries (i.e., deciding whether $(\mathcal{T}, \mathcal{A}) \models q$ for a conjunctive query $q$ without free variables) is coNP-complete (Rosati, 2007) and its data complexity is in PTime (Rosati, 2007). Information on reasoners for subsumption checking for $\mathcal{ELH}^r$ can be found in the work of Delaitre and Kazakov (2009), Kazakov, Krötzsch, and Simancik (2011), and Mendez and Suntisrivaraporn (2009). Lutz et al. (2009) present an approach to efficient conjunctive query answering for $\mathcal{ELH}^r$.

### 2.1 Normal Form

It is often convenient to consider *normalised* $\mathcal{ELH}^r$-terminologies. Let $\mathcal{T}$ be an $\mathcal{ELH}^r$-terminology and $A$ a concept name. Call $A$

- *primitive in* $\mathcal{T}$ if $A \in \mathsf{N_C} \setminus (\{A \in \mathsf{N_C} \mid A \equiv C \in \mathcal{T}\} \cup \{A \in \mathsf{N_C} \mid A \sqsubseteq C \in \mathcal{T}\})$;
- *pseudo-primitive in* $\mathcal{T}$ if $A \in \mathsf{N_C} \setminus \{A \in \mathsf{N_C} \mid A \equiv C \in \mathcal{T}\}$.

Note that concept names that do not occur in $\mathcal{T}$ are primitive and pseudo-primitive in $\mathcal{T}$. Call a concept name $A$ *non-conjunctive in* $\mathcal{T}$ if it is pseudo-primitive in $\mathcal{T}$ or there exists a concept of the form $\exists r.C$ such that $A \equiv \exists r.C \in \mathcal{T}$. Otherwise, $A$ is called *conjunctive in* $\mathcal{T}$. Thus, $A$ is conjunctive in $\mathcal{T}$ if, and only if, there exists a concept name $B$ such that $A \equiv B \in \mathcal{T}$ or there exist $C_1, \ldots, C_n$, $n \geq 2$, such that $A \equiv C_1 \sqcap \cdots \sqcap C_n \in \mathcal{T}$. Let $X$ be a finite set of concepts. We say that a concept $F$ is a *conjunction* of concepts in $X$ if $F$ is of the form $\bigsqcap_{D \in X} D$. Any $D \in X$ is then called a *conjunct* of $F$ and, if $D$ is a concept name, then it is called an *atomic conjunct* of $F$. We sometimes write $D \in F$ instead of $D \in X$.

An $\mathcal{ELH}^r$-terminology $\mathcal{T}$ is *normalised* if it consists of $\mathcal{ELH}^r$-inclusions of the following form:

- $A \equiv \exists r.B$, or $A \equiv F$, where $A, B$ are concept names and $F$ is a non-empty conjunction of concept names such that every conjunct $B'$ of $F$ is non-conjunctive in $\mathcal{T}$;
- $E \sqsubseteq \exists r.B$, $E \sqsubseteq \exists r.\top$, or $E \sqsubseteq F$, where $B$ is a concept name, $E$ is either a concept name, or is of the form $\exists s.\top$, or $\mathsf{ran}(s)$, and $F$ is a non-empty conjunction of concept names such that every conjunct $B'$ of $F$ is non-conjunctive in $\mathcal{T}$.

As the following lemma shows, any $\mathcal{ELH}^r$-terminology can be normalised yielding a model conservative extension of the original terminology.

**Lemma 1.** *For every $\mathcal{ELH}^r$-terminology $\mathcal{T}$, one can construct in polynomial time a normalised terminology $\mathcal{T}'$ of polynomial size in $|\mathcal{T}|$ such that $\mathsf{sig}(\mathcal{T}) \subseteq \mathsf{sig}(\mathcal{T}')$, $\mathcal{T}' \models \mathcal{T}$, and for every model $\mathcal{I}$ of $\mathcal{T}$ there exists a model $\mathcal{J}$ of $\mathcal{T}'$ such that $\Delta^{\mathcal{I}} = \Delta^{\mathcal{J}}$ and $X^{\mathcal{I}} = X^{\mathcal{J}}$ for every $X \in \mathsf{sig}(\mathcal{T})$. Moreover, $\mathcal{T}'$ is acyclic if $\mathcal{T}$ is acyclic.*

Normalised terminologies in the sense defined above are a minor modification of normalised terminologies as defined by Baader (2003). The straightforward extension of the proof given by Baader is provided in the appendix.

### 2.2 Canonical Model

We define a canonical model, $\mathcal{I}_\mathcal{K}$, for $\mathcal{ELH}^r$-knowledge bases $\mathcal{K}$. $\mathcal{I}_\mathcal{K}$ can be constructed in polynomial time and gives the same answers to instance queries as $\mathcal{K}$; i.e., $\mathcal{I}_\mathcal{K} \models \alpha$ if, and only if, $\mathcal{K} \models \alpha$, for any instance assertion $\alpha$. The construction is similar to the canonical model introduced by Lutz et al. (2009).

Let $\mathsf{sub}(\mathcal{T})$ denote the set of all subconcepts of concepts used in $\mathcal{T}$, $\mathsf{rol}(\mathcal{T})$ the set of all role names occurring in $\mathcal{T}$. Take fresh individual names $x_{\mathsf{ran}(r),D}$ for every $r \in \mathsf{rol}(\mathcal{T})$ and $D \in \mathsf{sub}(\mathcal{T})$ and set

$$\mathsf{NI_{aux}} := \{x_{\mathsf{ran}(r),D} \mid r \in \mathsf{rol}(\mathcal{T}) \text{ and } D \in \mathsf{sub}(\mathcal{T})\}.$$

Now define the *generating interpretation* $\mathcal{W}_\mathcal{K}$ of a KB $\mathcal{K} = (\mathcal{T}, \mathcal{A})$ as follows:

$$
\begin{array}{rcl}
\Delta^{\mathcal{W}_\mathcal{K}} & := & \mathsf{obj}(\mathcal{A}) \cup \mathsf{NI}_{\mathsf{aux}}; \\
A^{\mathcal{W}_\mathcal{K}} & := & \{a \in \mathsf{obj}(\mathcal{A}) \mid \mathcal{K} \models A(a)\} \cup \{x_{\mathsf{ran}(r),D} \in \mathsf{NI}_{\mathsf{aux}} \mid \mathcal{T} \models \mathsf{ran}(r) \sqcap D \sqsubseteq A\}; \\
r^{\mathcal{W}_\mathcal{K}} & := & \{(a,b) \in \mathsf{obj}(\mathcal{A}) \times \mathsf{obj}(\mathcal{A}) \mid s(a,b) \in \mathcal{A} \text{ and } \mathcal{T} \models s \sqsubseteq r\} \cup \\
& & \{(a, x_{\mathsf{ran}(s),D}) \in \mathsf{obj}(\mathcal{A}) \times \mathsf{NI}_{\mathsf{aux}} \mid \mathcal{K} \models \exists s.D(a) \text{ and } \mathcal{T} \models s \sqsubseteq r\} \cup \\
& & \{(x_{\mathsf{ran}(s),D}, x_{\mathsf{ran}(s'),D'}) \in \mathsf{NI}_{\mathsf{aux}} \times \mathsf{NI}_{\mathsf{aux}} \mid \mathcal{T} \models \mathsf{ran}(s) \sqcap D \sqsubseteq \exists s'.D',\ \ \mathcal{T} \models s' \sqsubseteq r\}; \\
a^{\mathcal{W}_\mathcal{K}} & := & a, \text{ for all } a \in \mathsf{obj}(\mathcal{A}).
\end{array}
$$

A *path* in $\mathcal{W}_\mathcal{K}$ is a finite sequence $d_0 r_1 d_1 \cdots r_n d_n$, $n \geq 0$, where $d_0 \in \mathsf{obj}(\mathcal{A})$ and, for all $i < n$, $(d_i, d_{i+1}) \in r_{i+1}^{\mathcal{W}_\mathcal{K}}$. We use $\mathsf{paths}(\mathcal{W}_\mathcal{K})$ to denote the set of all paths in $\mathcal{W}_\mathcal{K}$. If $p \in \mathsf{paths}(\mathcal{W}_\mathcal{K})$, then $\mathsf{tail}(p)$ denotes the last element $d_n$ in $p$.

The *canonical model* $\mathcal{I}_\mathcal{K}$ of a knowledge base $\mathcal{K}$ is the restriction of $\mathcal{W}_\mathcal{K}$ to all domain elements $d$ such that there is a path in $\mathcal{W}_\mathcal{K}$ with tail $d$. The following result summarises the main properties of $\mathcal{I}_\mathcal{K}$.

**Theorem 2.** *Let $\mathcal{K} = (\mathcal{T}, \mathcal{A})$ be an $\mathcal{ELH}^r$-KB. Then*

1. *$\mathcal{I}_\mathcal{K}$ is a model of $\mathcal{K}$;*
2. *$\mathcal{I}_\mathcal{K}$ can be computed in polynomial time in the size of $\mathcal{K}$;*
3. *for all $x_{\mathsf{ran}(s),D} \in \Delta^{\mathcal{I}_\mathcal{K}}$ and all $a \in \mathsf{obj}(\mathcal{A})$, if $C$ is an $\mathcal{EL}$-concept or $C = \mathsf{ran}(r)$, then*
   - *$\mathcal{K} \models C(a)$ if, and only if, $a^{\mathcal{I}_\mathcal{K}} \in C^{\mathcal{I}_\mathcal{K}}$.*
   - *$\mathcal{T} \models \mathsf{ran}(s) \sqcap D \sqsubseteq C$ if, and only if, $x_{\mathsf{ran}(s),D} \in C^{\mathcal{I}_\mathcal{K}}$.*

The proof of Theorem 2 is given in the appendix. It follows from Point 3 that $\mathcal{I}_\mathcal{K}$ gives the same answers to instance queries as $\mathcal{K}$ itself.

## 3. Logical Difference

In this section, we introduce three notions of logical difference between TBoxes and the derived notion of $\Sigma$-inseparability. Intuitively, the logical difference between two TBoxes $\mathcal{T}_1$ and $\mathcal{T}_2$ should be the set of all 'relevant formulas' $\varphi$ such that $\mathcal{T}_1 \models \varphi$ and $\mathcal{T}_2 \not\models \varphi$ or *vice versa*. Of course, which formulas $\varphi$ are relevant depends on the application domain. In many applications only subsumptions between concepts are relevant, but if TBoxes are employed to access instance data, then answers to instance or even conjunctive queries can be relevant as well. In addition, in applications of large-scale terminologies such as Snomed CT and NCI typically only a very small subset of the vocabulary of the terminology is relevant. Thus, a meaningful notion of logical difference should take into account only those formulas that are given in a certain *signature* of interest, where a signature $\Sigma$ is a subset of $\mathsf{N_C} \cup \mathsf{N_R}$. Given a concept, role, concept inclusion, TBox, ABox, or query $E$, we denote by $\mathsf{sig}(E)$ the signature of $E$, that is, the set of concept and role names occurring in it. We call $E$ a *$\Sigma$-concept, $\Sigma$-concept inclusion, $\Sigma$-TBox, $\Sigma$-ABox*, or *$\Sigma$-query*, respectively, if $\mathsf{sig}(E) \subseteq \Sigma$. Similarly, an *$\mathcal{EL}_\Sigma$-concept* $C$ is an $\mathcal{EL}$-concept such that $\mathsf{sig}(C) \subseteq \Sigma$ and an *$\mathcal{ELH}^r_\Sigma$-inclusion* $\alpha$ is an $\mathcal{ELH}^r$-inclusion such that $\mathsf{sig}(\alpha) \subseteq \Sigma$.

The first notion of logical difference we introduce corresponds to applications in which only subsumptions are relevant.

**Definition 3** ($\Sigma$-concept difference)**.** *The $\Sigma$-concept difference between $\mathcal{ELH}^r$-TBoxes $\mathcal{T}_1$ and $\mathcal{T}_2$ is the set $\mathsf{cDiff}_\Sigma(\mathcal{T}_1,\mathcal{T}_2)$ of all $\mathcal{ELH}^r_\Sigma$-inclusions $\alpha$ such that $\mathcal{T}_1 \models \alpha$ and $\mathcal{T}_2 \not\models \alpha$. We say that $\mathcal{T}_1$ and $\mathcal{T}_2$ are $\Sigma$-concept inseparable, in symbols $\mathcal{T}_1 \equiv^C_\Sigma \mathcal{T}_2$, if $\mathsf{cDiff}_\Sigma(\mathcal{T}_1,\mathcal{T}_2) = \mathsf{cDiff}_\Sigma(\mathcal{T}_2,\mathcal{T}_1) = \emptyset$.*

$\Sigma$-concept inseparability between $\mathcal{T}_1$ and $\mathcal{T}_2$ means that $\mathcal{T}_1$ can be replaced by $\mathcal{T}_2$ in any application that is only concerned with $\mathcal{ELH}^r_\Sigma$-inclusions.[2] As the following example shows, however, $\Sigma$-concept inseparable terminologies can give different answers for the same instance query and data.

**Example 4.** Let $\mathcal{T}_1 = \{\mathsf{ran}(r) \sqsubseteq A_1, \mathsf{ran}(s) \sqsubseteq A_2, B \equiv A_1 \sqcap A_2\}, \mathcal{T}_2 = \emptyset, \Sigma = \{r,s,B\}$. One can show that $\mathcal{T}_1$ and $\mathcal{T}_2$ are $\Sigma$-concept inseparable. However, for the $\Sigma$-ABox $\mathcal{A} = \{r(a,c), s(b,c)\}$ we have $(\mathcal{T}_1,\mathcal{A}) \models B(c)$ but $(\mathcal{T}_2,\mathcal{A}) \not\models B(c)$.

To take into account the differences between TBoxes that are relevant if TBoxes are used to access instance data, we consider the $\Sigma$-instance difference.

**Definition 5** ($\Sigma$-instance difference)**.** *The $\Sigma$-instance difference between TBoxes $\mathcal{T}_1$ and $\mathcal{T}_2$ is the set $\mathsf{iDiff}_\Sigma(\mathcal{T}_1,\mathcal{T}_2)$ of pairs of the form $(\mathcal{A},\alpha)$, where $\mathcal{A}$ is a $\Sigma$-ABox and $\alpha$ a $\Sigma$-instance assertion such that $(\mathcal{T}_1,\mathcal{A}) \models \alpha$ and $(\mathcal{T}_2,\mathcal{A}) \not\models \alpha$. We say that $\mathcal{T}_1$ and $\mathcal{T}_2$ are* $\Sigma$-instance inseparable*, in symbols $\mathcal{T}_1 \equiv^i_\Sigma \mathcal{T}_2$, if $\mathsf{iDiff}_\Sigma(\mathcal{T}_1,\mathcal{T}_2) = \mathsf{iDiff}_\Sigma(\mathcal{T}_2,\mathcal{T}_1) = \emptyset$.*

In contrast to $\mathcal{ELH}^r$, it has been shown by Lutz and Wolter (2010) that for $\mathcal{EL}$-TBoxes there is no difference between $\Sigma$-concept inseparability and $\Sigma$-instance inseparability. In this paper we extend this result to $\mathcal{ELH}^r$-TBoxes without range restrictions (the proof is given after Corollary 37):

**Theorem 6.** *Let $\mathcal{T}_1$ and $\mathcal{T}_2$ be $\mathcal{ELH}^r$-TBoxes without range restrictions and $\Sigma$ a signature. Then $\mathcal{T}_1 \equiv^C_\Sigma \mathcal{T}_2$ if, and only if, $\mathcal{T}_1 \equiv^i_\Sigma \mathcal{T}_2$.*

Sometimes, instance queries are not sufficiently expressive, and conjunctive queries are employed. In that case, the following notion of difference is appropriate.

**Definition 7** ($\Sigma$-query-difference)**.** *The $\Sigma$-query difference between TBoxes $\mathcal{T}_1$ and $\mathcal{T}_2$ is the set $\mathsf{qDiff}_\Sigma(\mathcal{T}_1,\mathcal{T}_2)$ of pairs of the form $(\mathcal{A},q(\vec{a}))$, where $\mathcal{A}$ is a $\Sigma$-ABox, $q(\vec{x})$ a $\Sigma$-conjunctive query, and $\vec{a}$ a tuple of individual names in $\mathcal{A}$ such that $(\mathcal{T}_1,\mathcal{A}) \models q(\vec{a})$ and $(\mathcal{T}_2,\mathcal{A}) \not\models q(\vec{a})$. We say that $\mathcal{T}_1$ and $\mathcal{T}_2$ are $\Sigma$-query inseparable, in symbols $\mathcal{T}_1 \equiv^q_\Sigma \mathcal{T}$, if $\mathsf{qDiff}_\Sigma(\mathcal{T}_1,\mathcal{T}_2) = \mathsf{qDiff}_\Sigma(\mathcal{T}_2,\mathcal{T}_1) = \emptyset$.*

As observed by Lutz and Wolter (2010) already, even for $\mathcal{EL}$ $\Sigma$-instance inseparability does not imply $\Sigma$-query inseparability. The following is a simple example.

**Example 8.** Let $\mathcal{T}_1 = \{A \sqsubseteq \exists r.B\}, \mathcal{T}_2 = \emptyset, \Sigma = \{A,B\}$. Then $\mathcal{T}_1$ and $\mathcal{T}_2$ are $\Sigma$-instance inseparable, but they are not $\Sigma$-query inseparable. Consider the $\Sigma$-ABox $\mathcal{A} = \{A(a)\}$ and the $\Sigma$-query $q = \exists x.B(x)$. Then $(\mathcal{T}_1,\mathcal{A}) \models q$ but $(\mathcal{T}_2,\mathcal{A}) \not\models q$.

2. We refer the reader to the conclusion of this paper for a brief discussion of this claim.

It is shown by Lutz and Wolter (2010) that Example 8 is essentially the only situation in which there is a difference between $\Sigma$-instance inseparability and $\Sigma$-query inseparability in $\mathcal{EL}$: the two notions become equivalent for $\mathcal{EL}$ if the universal role is admitted in instance queries (e.g., in Example 8, the conjunctive query $\exists x.B(x)$ corresponds to the instance query $\exists u.B(a)$ for the universal role $u$). In contrast, for $\mathcal{ELH}^r$ there are more subtle differences between the instance and the query case.

**Example 9.** Let $\mathcal{T}_1 = \{A \sqsubseteq \exists s.\top, s \sqsubseteq r_1, s \sqsubseteq r_2\}, \mathcal{T}_2 = \{A \sqsubseteq \exists r_1.\top \sqcap \exists r_2.\top\}, \Sigma = \{A, r_1, r_2\}$. Then $\mathcal{T}_1$ and $\mathcal{T}_2$ are $\Sigma$-concept and $\Sigma$-instance inseparable, but they are not $\Sigma$-query inseparable. To show the latter, let $\mathcal{A} = \{A(a)\}$ and let $q = \exists x(r_1(a,x) \wedge r_2(a,x))$. Then $(\mathcal{T}_1, \mathcal{A}) \models q$ but $(\mathcal{T}_2, \mathcal{A}) \not\models q$.

We have seen that $\Sigma$-concept inseparability does not imply $\Sigma$-instance inseparability and that $\Sigma$-instance inseparability does not imply $\Sigma$-query inseparability. The converse implications, however, hold:

**Lemma 10.** *For all $\mathcal{ELH}^r$-TBoxes $\mathcal{T}_1$ and $\mathcal{T}_2$ and all signatures $\Sigma$:*

$$\mathcal{T}_1 \equiv^q_\Sigma \mathcal{T}_2 \quad \Rightarrow \quad \mathcal{T}_1 \equiv^i_\Sigma \mathcal{T}_2 \quad \Rightarrow \quad \mathcal{T}_1 \equiv^C_\Sigma \mathcal{T}_2.$$

*Proof.* The first implication follows from the observation that every instance query can be regarded as a conjunctive query. For the second implication, note first that if $s \sqsubseteq r \in \mathsf{cDiff}_\Sigma(\mathcal{T}_1, \mathcal{T}_2)$, then $(\{s(a,b)\}, r(a,b)) \in \mathsf{iDiff}_\Sigma(\mathcal{T}_1, \mathcal{T}_2)$. Now let $C \sqsubseteq D \in \mathsf{cDiff}_\Sigma(\mathcal{T}_1, \mathcal{T}_2)$. One can construct a $\Sigma$-ABox $\mathcal{A}_C$ with individual $a$ such that for all $\mathcal{EL}$-concepts $D'$: $(\mathcal{T}, \mathcal{A}_C) \models D'(a)$ if, and only if, $\mathcal{T} \models C \sqsubseteq D'$ (cf. Lemma 36). Thus $(\mathcal{A}_C, D(a)) \in \mathsf{iDiff}_\Sigma(\mathcal{T}_1, \mathcal{T}_2)$. □

Having introduced three notions of difference between $\mathcal{ELH}^r$-TBoxes, we now investigate two problems: (i) how to detect whether there is any difference between two $\mathcal{ELH}^r$-terminologies and, if so, (ii) how to represent the differences.

In what follows we assume that the fresh symbols used in the normalised form of terminologies do not occur in the signature $\Sigma$ for which we compute the difference between terminologies. Then we obtain the following lemma as a direct corollary of Lemma 1.

**Lemma 11.** *For any $\mathcal{ELH}^r$-terminologies $\mathcal{T}_1$, $\mathcal{T}_2$ and their normalised forms $\mathcal{T}_1'$, $\mathcal{T}_2'$ as defined in Lemma 1, we have that the following hold:*

- $\mathsf{cDiff}_\Sigma(\mathcal{T}_1, \mathcal{T}_2) = \mathsf{cDiff}_\Sigma(\mathcal{T}_1', \mathcal{T}_2')$;
- $\mathsf{iDiff}_\Sigma(\mathcal{T}_1, \mathcal{T}_2) = \mathsf{iDiff}_\Sigma(\mathcal{T}_1', \mathcal{T}_2')$;
- $\mathsf{qDiff}_\Sigma(\mathcal{T}_1, \mathcal{T}_2) = \mathsf{qDiff}_\Sigma(\mathcal{T}_1', \mathcal{T}_2')$.

From now on, unless stated otherwise, we consider normalised terminologies only.

## 4. The Case of $\mathcal{EL}$-Terminologies

Before investigating the logical difference for $\mathcal{ELH}^r$-terminologies, we illustrate the main ideas behind the proofs by considering the $\Sigma$-concept difference for $\mathcal{EL}$-terminologies. An $\mathcal{EL}$-terminology is an $\mathcal{ELH}^r$-terminology consisting of $\mathcal{EL}$-inclusions only, that is, concept inclusions of the form $A \sqsubseteq C$ and concept equations of the form $A \equiv C$. We start with the observation that even for acyclic $\mathcal{EL}$-terminologies there are $\mathcal{T}_1$ and $\mathcal{T}_2$ in which $\mathsf{cDiff}_\Sigma(\mathcal{T}_1, \mathcal{T}_2)$ contains inclusions of at least exponential size only. Thus, when searching for witness inclusions in $\mathsf{cDiff}_\Sigma(\mathcal{T}_1, \mathcal{T}_2)$, one has to deal with the case in which all witness inclusions have at least exponential size.

**Example 12.** Consider

$$\begin{array}{rcl} \mathcal{T}_1 & = & \{A_0 \sqsubseteq B_0, A_1 \equiv B_n\} \cup \{B_{i+1} \equiv \exists r.B_i \sqcap \exists s.B_i \mid 0 \leq i < n\} \\ \mathcal{T}_2 & = & \{A_1 \sqsubseteq F_0\} \cup \{F_i \sqsubseteq \exists r.F_{i+1} \sqcap \exists s.F_{i+1} \mid 0 \leq i < n\} \end{array}$$

and $\Sigma = \{A_0, A_1, r, s\}$. Then a concept inclusion in $\mathsf{cDiff}_\Sigma(\mathcal{T}_1, \mathcal{T}_2)$ of minimal size is given by $C_n \sqsubseteq A_1$, where

$$C_0 = A_0 \;\text{ and }\; C_{i+1} = \exists r.C_i \sqcap \exists s.C_i, \;\text{ for } i \geq 0.$$

Clearly, $C_n$ is of exponential size. Note, however, that if we use structure sharing and define the size of $C_n$ as the number of its subconcepts, then $C_n$ is only of polynomial size.

We now derive basic properties of $\mathcal{EL}$-terminologies using a sequent calculus.

### 4.1 Proof System for $\mathcal{EL}$

We derive basic properties of $\mathcal{EL}$ from the Gentzen-style sequent calculus presented by Hofmann (2005); see Figure 1. The calculus operates on *sequents* of the form $C \sqsubseteq D$, where $C, D$ are $\mathcal{EL}$-concepts; here the symbol $\sqsubseteq$ is treated as a syntactic separator. A *derivation* (or, equivalently, a *proof*) of a sequent $C \sqsubseteq D$ is a finite rooted tree whose nodes are labelled with sequents, whose root is labelled with $C \sqsubseteq D$, whose leaves are labelled with axioms (instances of Ax or AxTop) and whose internal nodes are labelled with the result of an application of one of the inference rules to the labels of their children. The *length* of a derivation is the number of rule applications in the derivation.

**Example 13.** Let $\mathcal{T} = \{A \equiv B_1 \sqcap B_2, F \sqsubseteq B_1\}$. A derivation $\mathcal{D}$ of the sequent $\exists r.(F \sqcap B_2) \sqsubseteq \exists r.A$ is shown below. The root of the derivation $\mathcal{D}$ is labelled with $\exists r.(F \sqcap B_2) \sqsubseteq \exists r.A$ and the two leaves with $B_1 \sqsubseteq B_1$ and $B_2 \sqsubseteq B_2$, respectively.

$$\dfrac{\dfrac{\dfrac{\dfrac{\dfrac{\dfrac{}{B_1 \sqsubseteq B_1}\,(\textsc{Ax})}{F \sqsubseteq B_1}\,(\textsc{PDefL})}{F \sqcap B_2 \sqsubseteq B_1}\,(\textsc{AndL1}) \quad \dfrac{\dfrac{}{B_2 \sqsubseteq B_2}\,(\textsc{Ax})}{F \sqcap B_2 \sqsubseteq B_2}\,(\textsc{AndL2})}{F \sqcap B_2 \sqsubseteq B_1 \sqcap B_2}\,(\textsc{AndR})}{F \sqcap B_2 \sqsubseteq A}\,(\textsc{DefR})}{\exists r.(F \sqcap B_2) \sqsubseteq \exists r.A}\,(\textsc{Ex})$$

$$\frac{}{C \sqsubseteq C}\ (\textsc{Ax}) \quad \frac{}{C \sqsubseteq \top}\ (\textsc{AxTop}) \quad \frac{C \sqsubseteq E}{C \sqcap D \sqsubseteq E}\ (\textsc{AndL1}) \quad \frac{D \sqsubseteq E}{C \sqcap D \sqsubseteq E}\ (\textsc{AndL2})$$

$$\frac{C \sqsubseteq E \quad C \sqsubseteq D}{C \sqsubseteq D \sqcap E}\ (\textsc{AndR}) \quad \frac{C \sqsubseteq D}{\exists r.C \sqsubseteq \exists r.D}\ (\textsc{Ex})$$

$$\frac{C_A \sqsubseteq D}{A \sqsubseteq D}\ (\textsc{DefL}) \quad \frac{D \sqsubseteq C_A}{D \sqsubseteq A}\ (\textsc{DefR}) \quad \text{where } A \equiv C_A \in \mathcal{T}$$

$$\frac{C_A \sqsubseteq D}{A \sqsubseteq D}\ (\textsc{PDefL}) \quad \text{where } A \sqsubseteq C_A \in \mathcal{T}$$

Figure 1: Gentzen-style proof system for $\mathcal{EL}$-terminologies.

Notice that the basic calculus of Hofmann (2005) considers $\mathcal{EL}$ without the constant $\top$ and for terminologies without concept inclusions. To take care of $\top$, we have added the rule (AxTop), and (PDefL) is the rule representing inclusions of the form $A \sqsubseteq C$. Cut-elimination, completeness, and correctness can now be shown in a straightforward extension of the proof given by Hofmann.

For a terminology $\mathcal{T}$ and concepts $C, D$, we write $\mathcal{T} \vdash C \sqsubseteq D$ if, and only if, there exists a proof of $C \sqsubseteq D$ in the calculus of Figure 1.

**Theorem 14** (Hofmann). *For all $\mathcal{EL}$-terminologies $\mathcal{T}$ and concepts $C, D$, it holds that $\mathcal{T} \models C \sqsubseteq D$ if, and only if, $\mathcal{T} \vdash C \sqsubseteq D$.*

We apply this calculus to derive a description of the syntactic form of concepts $C$ such that $\mathcal{T} \models C \sqsubseteq D$, where $D$ is non-conjunctive in $\mathcal{T}$.

**Lemma 15.** *Let $\mathcal{T}$ be a normalised $\mathcal{EL}$-terminology, $r$ a role name, $A$ a concept name and $D$ an $\mathcal{EL}$-concept.*

1. *Assume*
$$\mathcal{T} \models \bigsqcap_{1 \leq i \leq n} A_i \sqcap \bigsqcap_{1 \leq j \leq m} \exists r_j.C_j \sqsubseteq A,$$
*where $A$ is pseudo-primitive in $\mathcal{T}$, $A_i$ are concept names for $1 \leq i \leq n$, $C_j$ are $\mathcal{EL}$-concepts for $1 \leq j \leq m$, and $m, n \geq 0$. Then there exists $A_i$, $1 \leq i \leq n$, such that $\mathcal{T} \models A_i \sqsubseteq A$.*

2. *Assume now*
$$\mathcal{T} \models \bigsqcap_{1 \leq i \leq n} A_i \sqcap \bigsqcap_{1 \leq j \leq m} \exists r_j.C_j \sqsubseteq \exists r.D,$$
*where $A_i$ are concept names for $1 \leq i \leq n$, $C_j$ are $\mathcal{EL}$-concepts for $1 \leq j \leq m$, and $m, n \geq 0$. Then*
   - *there exists $A_i$, $1 \leq i \leq n$, such that $\mathcal{T} \models A_i \sqsubseteq \exists r.D$ or*
   - *there exists $r_j$, $1 \leq j \leq m$, such that $r_j = r$ and $\mathcal{T} \models C_j \sqsubseteq D$.*

*Proof.* We use Theorem 14. First, we prove Point 1. Let $C = \bigsqcap_{1 \leq i \leq n} A_i \sqcap \bigsqcap_{1 \leq j \leq m} \exists r_j.C_j$ and assume $\mathcal{T} \models C \sqsubseteq A$, where $A$ is pseudo-primitive in $\mathcal{T}$. Let $\mathcal{D}$ be a proof of $C \sqsubseteq A$. Note that, since $A$ is pseudo-primitive in $\mathcal{T}$ (and a concept name), by inspecting the form of the conclusions of the inference rules, one can see that the root of the derivation $\mathcal{D}$ can only have been derived by either Ax, AndL1, AndL2, DefL, or PDefL. We now show that there exists $A_i$, $1 \leq i \leq n$, such that $\mathcal{T} \models A_i \sqsubseteq A$ by induction on $n + m$, i.e. the number of conjuncts in $C$. It is easy to see that $n + m \geq 1$ as $\mathcal{T} \not\models \top \sqsubseteq A$ by definition of terminologies $\mathcal{T}$.

The base case of $n + m = 1$ is trivial: the root of $\mathcal{D}$ can only have been derived by one of Ax, DefL, or PDefL; so, we can conclude that $C = A_1$; i.e. $n = 1$, $m = 0$, and we set $A_i = A_1$.

Assume $n + m > 1$. Then the root of $\mathcal{D}$ can only have been derived by either AndL1 or AndL2. In both cases, the premise used in the application of either inference rule is a sequent $C' \sqsubseteq A$ such that either $C = C' \sqcap D$ or $C = D \sqcap C'$ for an $\mathcal{EL}$-concept $D$. Thus, $C'$ contains less conjuncts than $C$ (but still at least one). We can also conclude that $\mathcal{T} \models C' \sqsubseteq A$ holds by Theorem 14. By applying the induction hypothesis, there hence exists a concept name $A_i$ which is a conjunct of $C'$ such that $\mathcal{T} \models A_i \sqsubseteq A$. Finally, we still note that $A_i$ is also a conjunct of $C$.

We now prove Point 2. Let $C = \bigsqcap_{1 \leq i \leq n} A_i \sqcap \bigsqcap_{1 \leq j \leq m} \exists r_j.C_j$ and assume $\mathcal{T} \models C \sqsubseteq \exists r.D$. Let $\mathcal{D}$ be a proof of $C \sqsubseteq \exists r.D$. Note that due to the form of the right-hand side of the sequent $C \sqsubseteq \exists r.D$, the rule used to derive the root of $\mathcal{D}$ can only have been one of Ax, AndL1, AndL2, DefL, PDefL, or Ex. We now prove that either there exists $A_i$, $1 \leq i \leq n$, such that $\mathcal{T} \models A_i \sqsubseteq \exists r.D$, or there exists $r_j$, $1 \leq j \leq m$, with $r_j = r$ and $\mathcal{T} \models C_j \sqsubseteq D$ by induction on $n + m$ again. Similarly to above, we have $n + m \geq 1$.

If $n+m = 1$, the rule used to derive the root of $\mathcal{D}$ can only have been one of Ax, DefL, PDefL, or Ex. We have two subcases:

- the root of $\mathcal{D}$ was derived with DefL or PDefL: then $n = 1$, $m = 0$ and $C = A_1$; i.e. $\mathcal{T} \models A_i \sqsubseteq \exists r.D$ for $A_i = A_1$.
- the root of $\mathcal{D}$ was derived with Ax or Ex: then $n = 0$, $m = 1$, $C = \exists r_1.C_1$, and $r_1 = r$. If $C_1 = D$, then obviously $\mathcal{T} \models C_1 \sqsubseteq D$ holds. Otherwise, the rule Ex was used to derive the root of $\mathcal{D}$ and $\mathcal{T} \vdash C_1 \sqsubseteq D$ holds, which implies that $\mathcal{T} \models C_1 \sqsubseteq D$. Thus, in any case, $r_j = r$ and $\mathcal{T} \models C_j \sqsubseteq D$ holds for $j = 1$.

The case $n+m > 1$ can be proved by induction analogously to the proof of Point 1 above. □

We apply Lemma 15 to elements of $\mathsf{cDiff}_\Sigma(\mathcal{T}_1, \mathcal{T}_2)$.

**Theorem 16** (Primitive witness for $\mathcal{EL}$)**.** *Let $\mathcal{T}_1$ and $\mathcal{T}_2$ be $\mathcal{EL}$-terminologies and $\Sigma$ a signature. If $\varphi \in \mathsf{cDiff}_\Sigma(\mathcal{T}_1, \mathcal{T}_2)$, then either $C \sqsubseteq A$ or $A \sqsubseteq D$ is a member of $\mathsf{cDiff}_\Sigma(\mathcal{T}_1, \mathcal{T}_2)$, where $A \in \mathsf{sig}(\varphi)$ is a concept name and $C$, $D$ are $\mathcal{EL}$-concepts occurring in $\varphi$.*

*Proof.* Let $\varphi = C \sqsubseteq D \in \mathsf{cDiff}_\Sigma(\mathcal{T}_1, \mathcal{T}_2)$. The proof is by induction on the construction of $D$. We have $D \neq \top$ as $\mathcal{T}_2 \models C \sqsubseteq \top$. If $D = D_1 \sqcap D_2$, then one of $C \sqsubseteq D_i$, $i = 1, 2$, is in $\mathsf{cDiff}_\Sigma(\mathcal{T}_1, \mathcal{T}_2)$ and we can apply the induction hypothesis. If $D = \exists r.D_1$ then, by Lemma 15, either (i) there exists a conjunct $A$ of $C$, $A$ a concept name, such that $\mathcal{T}_1 \models A \sqsubseteq D$, or (ii) there exists a conjunct $\exists r.C_1$ of $C$ with $\mathcal{T}_1 \models C_1 \sqsubseteq D_1$.

In case (i) it follows that $\mathcal{T}_2 \not\models A \sqsubseteq D$ as otherwise $\mathcal{T}_2 \models C \sqsubseteq D$ and $C \sqsubseteq D \notin \mathsf{cDiff}_\Sigma(\mathcal{T}_1, \mathcal{T}_2)$ due to $\models C \sqsubseteq A$. Hence, $A \sqsubseteq D \in \mathsf{cDiff}_\Sigma(\mathcal{T}_1, \mathcal{T}_2)$.

Finally, for case (ii) we obtain $\mathcal{T}_2 \not\models C_1 \sqsubseteq D_1$ as otherwise $\models C \sqsubseteq \exists r.C_1$, $\mathcal{T}_2 \models \exists r.C_1 \sqsubseteq D$ and $C \sqsubseteq D \notin \mathsf{cDiff}_\Sigma(\mathcal{T}_1, \mathcal{T}_2)$ again. Thus, $C_1 \sqsubseteq D_1 \in \mathsf{cDiff}_\Sigma(\mathcal{T}_1, \mathcal{T}_2)$ and we can apply the induction hypothesis. □

By Theorem 16, every inclusion $C \sqsubseteq D$ in the $\Sigma$-concept difference of $\mathcal{T}_1$ and $\mathcal{T}_2$ "contains" a basic witness inclusion that has a concept name either on the right-hand side or the left-hand side. We define

- the set of *left-hand* $\Sigma$-concept difference witnesses, $\mathsf{cWtn}^{\mathsf{lhs}}_\Sigma(\mathcal{T}_1, \mathcal{T}_2)$, as the set of all $A \in \Sigma \cap \mathsf{N_C}$ such that there exists a concept $D$ with $A \sqsubseteq D \in \mathsf{cDiff}_\Sigma(\mathcal{T}_1, \mathcal{T}_2)$ and
- the set of *right-hand* $\Sigma$-concept difference witnesses, $\mathsf{cWtn}^{\mathsf{rhs}}_\Sigma(\mathcal{T}_1, \mathcal{T}_2)$, as the set of all $A \in \Sigma \cap \mathsf{N_C}$ such that there exists a concept $C$ with $C \sqsubseteq A \in \mathsf{cDiff}_\Sigma(\mathcal{T}_1, \mathcal{T}_2)$.

We regard the concept names in $\mathsf{cWtn}^{\mathsf{lhs}}_\Sigma(\mathcal{T}_1, \mathcal{T}_2)$ and $\mathsf{cWtn}^{\mathsf{rhs}}_\Sigma(\mathcal{T}_1, \mathcal{T}_2)$ as a succinct and, in a certain sense, complete representation of the $\Sigma$-concept difference between $\mathcal{T}_1$ and $\mathcal{T}_2$ and define the set of all $\Sigma$-concept difference witnesses as

$$\mathsf{cWtn}_\Sigma(\mathcal{T}_1, \mathcal{T}_2) = (\mathsf{cWtn}^{\mathsf{lhs}}_\Sigma(\mathcal{T}_1, \mathcal{T}_2), \mathsf{cWtn}^{\mathsf{rhs}}_\Sigma(\mathcal{T}_1, \mathcal{T}_2)).$$

In what follows, we first present a polytime algorithm computing $\mathsf{cWtn}^{\mathsf{rhs}}_\Sigma(\mathcal{T}_1, \mathcal{T}_2)$. A polytime algorithm computing $\mathsf{cWtn}^{\mathsf{lhs}}_\Sigma(\mathcal{T}_1, \mathcal{T}_2)$ has already been given by Lutz and Wolter (2010) (for $\mathcal{EL}$-TBoxes). We briefly present it since an extension will be developed when we consider $\mathcal{ELH}^r$-terminologies. Both algorithms together decide $\Sigma$-concept inseparability since, by Theorem 16, $\mathcal{T}_1$ and $\mathcal{T}_2$ are $\Sigma$-concept inseparable if, and only if, $\mathsf{cWtn}_\Sigma(\mathcal{T}_1, \mathcal{T}_2) = \mathsf{cWtn}_\Sigma(\mathcal{T}_2, \mathcal{T}_1) = (\emptyset, \emptyset)$.

### 4.2 Computing $\mathsf{cWtn}^{\mathsf{rhs}}_\Sigma(\mathcal{T}_1, \mathcal{T}_2)$

Let $A \in \Sigma$ and assume we want to decide whether $A \in \mathsf{cWtn}^{\mathsf{rhs}}_\Sigma(\mathcal{T}_1, \mathcal{T}_2)$. Thus, we want to decide whether there exists a $\Sigma$-concept $C$ such that $\mathcal{T}_1 \models C \sqsubseteq A$ and $\mathcal{T}_2 \not\models C \sqsubseteq A$. Our general strategy is as follows. Let

$$\mathsf{noimply}_{\mathcal{T}_2,\Sigma}(A) = \{C \mid \mathcal{T}_2 \not\models C \sqsubseteq A,\ C \text{ an } \mathcal{EL}_\Sigma\text{-concept}\}.$$

We aim at an algorithm that checks whether $\mathsf{noimply}_{\mathcal{T}_2,\Sigma}(A)$ contains some $C$ with $\mathcal{T}_1 \models C \sqsubseteq A$. For two sets $\mathcal{C}$ and $\mathcal{D}$ of concepts we call $\mathcal{C}$ a *cover of* $\mathcal{D}$ if $\mathcal{C} \subseteq \mathcal{D}$ and for all $D \in \mathcal{D}$ there exists a $C \in \mathcal{C}$ such that $\models C \sqsubseteq D$. Thus, $\mathcal{C} \subseteq \mathsf{noimply}_{\mathcal{T}_2,\Sigma}(A)$ is a cover of $\mathsf{noimply}_{\mathcal{T}_2,\Sigma}(A)$ if for all $D \in \mathsf{noimply}_{\mathcal{T}_2,\Sigma}(A)$ there exists a $C \in \mathcal{C}$ such that $\models C \sqsubseteq D$. Note that if $\mathcal{C}$ is a cover of $\mathsf{noimply}_{\mathcal{T}_2,\Sigma}(A)$, then there exists some $\Sigma$-concept $C$ such that $C \sqsubseteq A \in \mathsf{cDiff}_\Sigma(\mathcal{T}_1, \mathcal{T}_2)$ if, and only if, there exists some $C \in \mathcal{C}$ such that $\mathcal{T}_1 \models C \sqsubseteq A$. Thus we have reduced the original problem to the construction of an appropriate cover $\mathcal{C}$ and deciding the subsumption problem $\mathcal{T}_1 \models C \sqsubseteq A$, for $C \in \mathcal{C}$. Unfortunately, in general, no finite cover exists. The following example illustrates the situation.

**Example 17.** (1) Let $\Sigma = \{A, B, r\}$ and $\mathcal{T}_2 = \emptyset$. Then $\mathsf{noimply}_{\mathcal{T}_2,\Sigma}(A)$ contains all $\Sigma$-concepts that do not have $A$ as an atomic conjunct. Clearly, $\mathsf{noimply}_{\mathcal{T}_2,\Sigma}(A)$ contains no finite cover.

(2) Let $\Sigma' = \{A, B, r\}$ and $\mathcal{T}_2' = \{A \equiv \exists r.A\}$. Then $\mathsf{noimply}_{\mathcal{T}_2',\Sigma'}(A)$ contains all $\Sigma' \setminus \{A\}$-concepts and contains no finite cover.

(3) Let $\Sigma'' = \{A, B_1, B_2\}$ and $\mathcal{T}_2'' = \{A \equiv B_1 \sqcap B_2\}$. Then $\{B_1, B_2\}$ is a cover of $\mathsf{noimply}_{\mathcal{T}_2'',\Sigma''}(A)$.

As a consequence, instead of directly constructing a cover of $\mathsf{noimply}_{\mathcal{T}_2,\Sigma}(A)$, we first construct transparent and small covers of

$$\mathsf{noimply}_{\mathcal{T}_2,\Sigma}(A) \cap \{C \mid \mathsf{depth}(C) \leq n\},$$

for all $n \geq 0$, where $\mathsf{depth}(C)$ is the *role-depth* of $C$; i.e., the number of nestings of existential restrictions in $C$.[3] Those covers are denoted $\mathsf{noimply}^n_{\mathcal{T}_2,\Sigma}(A)$, $n \geq 0$, and are singleton sets if $A$ is non-conjunctive in $\mathcal{T}_2$ and finite sets containing at most $k$ concepts if $A \equiv B_1 \sqcap \cdots \sqcap B_k \in \mathcal{T}_2$. Based on this sequence, we present two distinct algorithms for computing $\mathsf{cWtn}^{\mathsf{rhs}}_\Sigma(\mathcal{T}_1, \mathcal{T}_2)$:

1. we encode the infinite sequence $\mathsf{noimply}^n_{\mathcal{T}_2,\Sigma}(A)$, $n \geq 0$, into a polynomial-size ABox $\mathcal{A}_{\mathcal{T}_2,\Sigma}$. In this way we obtain a reduction of the original problem to an instance checking problem for the knowledge base $(\mathcal{T}_1, \mathcal{A}_{\mathcal{T}_2,\Sigma})$. In a certain sense, the ABox $\mathcal{A}_{\mathcal{T}_2,\Sigma}$ encodes a (in general infinite) cover of $\mathsf{noimply}_{\mathcal{T}_2,\Sigma}(A)$.
2. we employ the terminology $\mathcal{T}_1$ in a dynamic programming approach to decide which concepts in $\mathsf{noimply}^n_{\mathcal{T}_2,\Sigma}(A)$ are relevant for deciding whether $A \in \mathsf{cWtn}^{\mathsf{rhs}}_\Sigma(\mathcal{T}_1, \mathcal{T}_2)$. Although less transparent, for large terminologies the latter approach is considerably more efficient. We develop it for acyclic terminologies.

For an $\mathcal{EL}$-terminology $\mathcal{T}$, a concept name $A$ and a signature $\Sigma$, set

$$\mathsf{pre}^\Sigma_\mathcal{T}(A) = \{B \in \Sigma \mid \mathcal{T} \models B \sqsubseteq A\}.$$

The finite covers $\mathsf{noimply}^n_{\mathcal{T}_2,\Sigma}(A)$, $n \geq 0$, are defined in Figure 2. For $n = 0$, the set $\mathsf{noimply}^n_{\mathcal{T}_2,\Sigma}(A)$ consists of concepts without role names. We distinguish between conjunctive and non-conjunctive $A$. Note that if $A$ is non-conjunctive, then $\mathsf{noimply}^n_{\mathcal{T}_2,\Sigma}(A)$ is a singleton set. Example 17 (3) shows that this is not always the case for conjunctive $A$. For $n + 1$, we distinguish between pseudo-primitive concept names, conjunctive concept names, and those that have a definition of the form $A \equiv \exists r.C$. Again, for non-conjunctive $A$, $\mathsf{noimply}^{n+1}_{\mathcal{T}_2,\Sigma}(A)$ is a singleton set. Note that the concepts $\mathsf{all}^n_\Sigma$ are covers of $\{C \mid \mathsf{depth}(C) \leq n,\ C \text{ an } \mathcal{EL}_\Sigma\text{-concept}\}$, for all $n \geq 0$. We illustrate the definitions using the $\mathcal{EL}$-terminologies from Example 17.

**Example 18.** (1) Let $\Sigma = \{A, B, r\}$ and $\mathcal{T}_2 = \emptyset$. Then $A$ and $B$ are non-conjunctive in $\mathcal{T}_2$ and $\mathsf{noimply}^0_{\mathcal{T}_2,\Sigma}(A) = \{B\}$ and $\mathsf{noimply}^0_{\mathcal{T}_2,\Sigma}(B) = \{A\}$. $A$ and $B$ are also pseudo-primitive in $\mathcal{T}_2$, and so $\mathsf{noimply}^1_{\mathcal{T}_2,\Sigma}(A) = \{B \sqcap \exists r.(A \sqcap B)\}$ and $\mathsf{noimply}^1_{\mathcal{T}_2,\Sigma}(B) = \{A \sqcap \exists r.(A \sqcap B)\}$.

3. More precisely $\mathsf{depth}(A) = 0$, $\mathsf{depth}(C_1 \sqcap C_2) = \max\{\mathsf{depth}(C_1), \mathsf{depth}(C_2)\}$, and $\mathsf{depth}(\exists r.D) = \mathsf{depth}(D) + 1$.

Set, inductively,

$$\mathsf{all}^0_\Sigma = \bigsqcap_{A' \in \Sigma} A' \quad \text{and} \quad \mathsf{all}^{n+1}_\Sigma = \bigsqcap_{A' \in \Sigma} A' \sqcap \bigsqcap_{s \in \Sigma} \exists s.\mathsf{all}^n_\Sigma.$$

Define $\mathsf{noimply}^0_{\mathcal{T}_2,\Sigma}(A)$ as follows:

- if $A$ is non-conjunctive in $\mathcal{T}_2$, then

$$\mathsf{noimply}^0_{\mathcal{T}_2,\Sigma}(A) = \{ \bigsqcap_{A' \in \Sigma \setminus \mathsf{pre}^\Sigma_{\mathcal{T}_2}(A)} A' \};$$

- if $A$ is conjunctive and $A \equiv F \in \mathcal{T}_2$, then

$$\mathsf{noimply}^0_{\mathcal{T}_2,\Sigma}(A) = \bigcup_{B \in F} \mathsf{noimply}^0_{\mathcal{T}_2,\Sigma}(B);$$

and define, inductively, $\mathsf{noimply}^{n+1}_{\mathcal{T}_2,\Sigma}(A)$ by

- if $A$ is pseudo-primitive in $\mathcal{T}_2$, then

$$\mathsf{noimply}^{n+1}_{\mathcal{T}_2,\Sigma}(A) = \{ \bigsqcap_{A' \in (\Sigma \setminus \mathsf{pre}^\Sigma_{\mathcal{T}_2}(A))} A' \sqcap \bigsqcap_{s \in \Sigma} \exists s.\mathsf{all}^n_\Sigma \}.$$

- If $A$ is conjunctive and $A \equiv F \in \mathcal{T}_2$, then

$$\mathsf{noimply}^{n+1}_{\mathcal{T}_2,\Sigma}(A) = \bigcup_{B \in F} \mathsf{noimply}^{n+1}_{\mathcal{T}_2,\Sigma}(B).$$

- If $A \equiv \exists r.B \in \mathcal{T}_2$, then

$$\mathsf{noimply}^{n+1}_{\mathcal{T}_2,\Sigma}(A) = \{C^{n+1}_{\Sigma,\mathcal{T}_2}\}, \text{ where}$$

$$C^{n+1}_{\Sigma,\mathcal{T}_2} = ( \bigsqcap_{A' \in (\Sigma \setminus \mathsf{pre}^\Sigma_{\mathcal{T}_2}(A))} A' \sqcap \bigsqcap_{r \neq s \in \Sigma} \exists s.\mathsf{all}^n_\Sigma \sqcap \bigsqcap_{\substack{r \in \Sigma \\ E \in \mathsf{noimply}^n_{\mathcal{T}_2,\Sigma}(B)}} \exists r.E).$$

Figure 2: Definition of $\mathsf{noimply}^n_{\mathcal{T}_2,\Sigma}(A)$

(2) Let $\Sigma' = \{A, B, r\}$ and $\mathcal{T}_2' = \{A \equiv \exists r.A\}$. Then $A$ and $B$ are non-conjunctive in $\mathcal{T}_2'$ and $\mathsf{noimply}^0_{\mathcal{T}_2',\Sigma'}(A) = \{B\}$ and $\mathsf{noimply}^0_{\mathcal{T}_2',\Sigma'}(B) = \{A\}$. $B$ is pseudo-primitive in $\mathcal{T}_2'$ and so $\mathsf{noimply}^1_{\mathcal{T}_2,\Sigma}(B) = \{A \sqcap \exists r.(A \sqcap B)\}$. $A \equiv \exists r.A \in \mathcal{T}_2'$ and so $\mathsf{noimply}_{\mathcal{T}_2',\Sigma'}(A) = \{B \sqcap \exists r.B\}$.

(3) Let $\Sigma'' = \{A, B_1, B_2\}$ and $\mathcal{T}_2'' = \{A \equiv B_1 \sqcap B_2\}$. $B_1$ and $B_2$ are non-conjunctive in $\mathcal{T}_2''$ and so $\mathsf{noimply}^0_{\mathcal{T}_2'',\Sigma''}(B_1) = \{B_2\}$ and $\mathsf{noimply}^0_{\mathcal{T}_2'',\Sigma''}(B_2) = \{B_1\}$. $A$ is conjunctive in $\mathcal{T}_2''$

and, by definition, $\mathsf{noimply}^0_{\mathcal{T}_2'',\Sigma''}(A) = \{B_1, B_2\}$. Since $\Sigma$ does not contain any role names, we have $\mathsf{noimply}^0_{\mathcal{T}_2'',\Sigma''}(X) = \mathsf{noimply}^n_{\mathcal{T}_2'',\Sigma''}(X)$, for all $X \in \{A, B_1, B_2\}$ and $n > 0$.

The following lemma shows the correctness of the definition of $\mathsf{noimply}^n_{\mathcal{T}_2,\Sigma}(A)$.

**Lemma 19.** *Let $\mathcal{T}_2$ be a normalised $\mathcal{EL}$-terminology, $\Sigma$ be a signature, and $A \in \mathsf{N_C}$. Then* $\mathsf{noimply}^n_{\mathcal{T}_2,\Sigma}(A)$ *is a cover of* $\mathsf{noimply}_{\mathcal{T}_2,\Sigma}(A) \cap \{C \mid \mathsf{depth}(C) \leq n\}$. *Namely, for all $n \geq 0$,*

**C1.** *$\mathcal{T}_2 \not\models C \sqsubseteq A$, for all $C \in \mathsf{noimply}^n_{\mathcal{T}_2,\Sigma}(A)$.*

**C2.** *For all $\mathcal{EL}_\Sigma$-concepts $D$ with $n = \mathsf{depth}(D)$, if $\mathcal{T}_2 \not\models D \sqsubseteq A$, then $\models C \sqsubseteq D$ for some $C \in \mathsf{noimply}^n_{\mathcal{T}_2,\Sigma}(A)$.*

*In particular, $\bigcup_{n\geq 0} \mathsf{noimply}^n_{\mathcal{T}_2,\Sigma}(A)$ is a cover of $\mathsf{noimply}_{\mathcal{T}_2,\Sigma}(A)$.*

*Proof.* **C1**. Assume first that $A$ is pseudo-primitive in $\mathcal{T}_2$. Then $\mathsf{noimply}^n_{\mathcal{T}_2,\Sigma}(A)$ consists of $C = \bigsqcap_{A' \in (\Sigma \setminus \mathsf{pre}^\Sigma_{\mathcal{T}_2}(A))} A' \sqcap F$, where $F$ is a (possibly empty) conjunction of concepts of the form $\exists s.F_i$. By Lemma 15, $\mathcal{T}_2 \not\models C \sqsubseteq A$ because the only atomic conjuncts of $C$ are in $\Sigma \setminus \mathsf{pre}^\Sigma_{\mathcal{T}_2}(A)$.

We now prove **C1** for concept names $A$ which are not pseudo-primitive in $\mathcal{T}_2$. The proof is by induction on $n$. For $n = 0$ and $A \equiv \exists r.B \in \mathcal{T}_2$, assume $\mathcal{T}_2 \models \bigsqcap_{A' \in (\Sigma \setminus \mathsf{pre}^\Sigma_{\mathcal{T}_2}(A))} A' \sqsubseteq A$. As $A \equiv \exists r.B \in \mathcal{T}_2$, we have by Lemma 15 that there must exist $A' \in \Sigma \setminus \mathsf{pre}^\Sigma_{\mathcal{T}_2}(A)$ with $\mathcal{T}_2 \models A' \sqsubseteq A$. But this contradicts the definition of the set $\mathsf{pre}^\Sigma_{\mathcal{T}_2}(A)$). For $n = 0$ and $A$ conjunctive with $A \equiv F \in \mathcal{T}_2$, let $C \in \mathsf{noimply}^n_{\mathcal{T}_2,\Sigma}(A) = \bigcup_{B \in F} \mathsf{noimply}^n_{\mathcal{T}_2,\Sigma}(B)$. There hence exists an atomic conjunct $B$ of $F$ such that $C \in \mathsf{noimply}^n_{\mathcal{T}_2,\Sigma}(B)$. As $\mathcal{T}_2$ is normalised, $B$ is non-conjunctive, i.e. property **C1** has already been proved above for $B$. Thus, $\mathcal{T}_2 \not\models C \sqsubseteq B$, which implies that $\mathcal{T}_2 \not\models C \sqsubseteq A$ as otherwise $\mathcal{T}_2 \models C \sqsubseteq B$ would hold.

For the induction step, assume **C1** has been proved for $n \geq 0$.

Let $A \equiv \exists r.B \in \mathcal{T}_2$ and let $C^{n+1}_{\mathcal{T}_2,\Sigma}$ be the only element of $\mathsf{noimply}^{n+1}_{\mathcal{T}_2,\Sigma}(A)$. Assume $\mathcal{T}_2 \models C^{n+1}_{\mathcal{T}_2,\Sigma} \sqsubseteq A$. By Lemma 15 there are two possibilities:

- $\mathcal{T}_2 \models \bigsqcap_{A' \in (\Sigma \setminus \mathsf{pre}^\Sigma_{\mathcal{T}_2}(A))} A' \sqsubseteq \exists r.B$. Similarly to above, the claim follows from Lemma 15 and the fact that $A \equiv \exists r.B \in \mathcal{T}_2$.
- $r \in \Sigma$ and there exists $E \in \mathsf{noimply}^n_{\mathcal{T}_2,\Sigma}(B)$ such that $\mathcal{T}_2 \models E \sqsubseteq B$. This is excluded by the induction hypothesis.

We have derived a contradiction. The case $A \equiv F \in \mathcal{T}_2$, $A$ conjunctive in $\mathcal{T}_2$, is considered analogously to the case $n = 0$.

**C2**. Let $n = 0$ and assume first that $A$ is non-conjunctive. Let $D$ be a $\Sigma$-concept with $\mathsf{depth}(D) = 0$ and $\mathcal{T}_2 \not\models D \sqsubseteq A$. Then all conjuncts of $D$ are in $\Sigma \setminus \mathsf{pre}^\Sigma_{\mathcal{T}_2}(A)$ and we obtain $\models \bigsqcap_{A' \in \Sigma \setminus \mathsf{pre}^\Sigma_{\mathcal{T}_2}(A)} A' \sqsubseteq D$. Now assume $A$ is conjunctive in $\mathcal{T}_2$ and $A \equiv F \in \mathcal{T}_2$. Let $D$ be a $\Sigma$-concept with $\mathsf{depth}(D) = 0$ and $\mathcal{T}_2 \not\models D \sqsubseteq A$. Then $\mathcal{T}_2 \not\models D \sqsubseteq B$, for some conjunct $B$ of $F$. By induction, $\models C \sqsubseteq D$ for the (unique as $B$ must be non-conjunctive) $C \in \mathsf{noimply}^0_{\mathcal{T}_2,\Sigma}(B)$, and therefore $\models C \sqsubseteq D$ for some $C \in \mathsf{noimply}^0_{\mathcal{T}_2,\Sigma}(A)$.

For the induction step, assume that **C2** has been shown for $n$. Let $D$ be a $\Sigma$-concept with $\mathcal{T}_2 \not\models D \sqsubseteq A$ and $\mathsf{depth}(D) = n + 1$.

(a) Let $A$ be pseudo-primitive in $\mathcal{T}_2$. Then the atomic conjuncts of $D$ are included in $\Sigma \setminus \mathsf{pre}^{\Sigma}_{\mathcal{T}_2}(A)$. Now $\models C \sqsubseteq D$ follows immediately for $C = \bigsqcap_{A' \in \Sigma \setminus \mathsf{pre}^{\Sigma}_{\mathcal{T}_2}(A)} A' \sqcap \bigsqcap_{s \in \Sigma} \exists s.\mathsf{all}^n_{\Sigma}$.

(b) Let $A \equiv \exists r.B \in \mathcal{T}_2$. Let $C^{n+1}_{\mathcal{T}_2,\Sigma}$ be the only element of $\mathsf{noimply}^{n+1}_{\mathcal{T}_2,\Sigma}(A)$ and assume

$$D = \bigsqcap_{E \in Q_0} E \sqcap \bigsqcap_{(s,D') \in Q_1} \exists s.D'.$$

Then $Q_0 \subseteq \Sigma \setminus \mathsf{pre}^{\Sigma}_{\mathcal{T}_2}(A)$. Hence, $\models C^{n+1}_{\mathcal{T}_2,\Sigma} \sqsubseteq \bigsqcap_{E \in Q_0} E$. Now consider a conjunct $\exists s.D'$ of $D$. We distinguish two cases:

- if $s \neq r$, then $\models C^{n+1}_{\mathcal{T}_2,\Sigma} \sqsubseteq \exists s.D'$, as required.
- if $s = r$, then $s \in \Sigma$ and it is sufficient to show that there exists $E \in \mathsf{noimply}^n_{\mathcal{T}_2,\Sigma}(B)$ such that $\models E \sqsubseteq D'$. Suppose there does not exist such an $E$. Then, by (the contraposition of) the induction hypothesis, $\mathcal{T}_2 \models D' \sqsubseteq B$. But this contradicts $\mathcal{T}_2 \not\models D \sqsubseteq A$ (as $A \equiv \exists r.B \in \mathcal{T}_2$).

(c) $A$ is conjunctive in $\mathcal{T}_2$ and $A \equiv F \in \mathcal{T}_2$. This case is analogous to the case in which $A$ is conjunctive in $\mathcal{T}_2$ and $n = 0$. ☐

**Corollary 20.** *For all normalised $\mathcal{EL}$-terminologies $\mathcal{T}_1$ and $\mathcal{T}_2$ and all $A \in \mathsf{N_C}$ the following conditions are equivalent:*

- *there exists an $\mathcal{EL}_\Sigma$-concept $C$ such that $\mathcal{T}_1 \models C \sqsubseteq A$ and $\mathcal{T}_2 \not\models C \sqsubseteq A$;*
- *there exists $n \geq 0$ and $C \in \mathsf{noimply}^n_{\mathcal{T}_2,\Sigma}(A)$ such that $\mathcal{T}_1 \models C \sqsubseteq A$.*

Observe that a direct application of Corollary 20 does not yield a procedure for computing $\mathsf{cWtn}^{\mathsf{rhs}}_{\Sigma}(\mathcal{T}_1, \mathcal{T}_2)$ as it gives no bound on $n$ for the set $\mathsf{noimply}^n_{\mathcal{T}_2,\Sigma}(A)$. At this point we present two ways of avoiding this problem (as well as the problem that concepts in $\mathsf{noimply}^n_{\mathcal{T}_2,\Sigma}(A)$ can be of exponential size). Firstly, instead of working with covers we construct an ABox encoding covers. In contrast to concepts, ABoxes admit the encoding of structure sharing and cycles and so, intuitively, admit the polynomial reconstruction of the 'infinite concept' $\bigsqcap_{n \geq 0, C \in \mathsf{noimply}^n_{\mathcal{T}_2,\Sigma}(A)} C$.

The ABox $\mathcal{A}_{\mathcal{T}_2,\Sigma}$ is constructed in Figure 3, where for a normalised $\mathcal{EL}$-terminology $\mathcal{T}$ and a concept name $A \in \mathsf{sig}(\mathcal{T})$, we set

$$\mathsf{non\text{-}conj}_{\mathcal{T}}(A) = \begin{cases} \{A\}, A \text{ is non-conjunctive in } \mathcal{T} \\ \{B_1, \ldots, B_n\}, A \equiv B_1 \sqcap \cdots \sqcap B_n \in \mathcal{T} \end{cases}$$

Note that the construction of $\mathcal{A}_{\mathcal{T}_2,\Sigma}$ is very similar to the construction of $\mathsf{noimply}^n_{\mathcal{T}_2,\Sigma}(A)$. The assertions for the individual $\xi_\Sigma$ play the role of the concepts $\mathsf{all}^n_\Sigma$, $n \geq 0$, and the assertions for the individuals $\xi_A$ play the role of the sets $\mathsf{noimply}^n_{\mathcal{T}_2,\Sigma}(A)$, $n \geq 0$. In fact, one can readily show that $\mathcal{A}_{\mathcal{T}_2,\Sigma} \models C(\xi_A)$ for any $C \in \mathsf{noimply}^n_{\mathcal{T}_2,\Sigma}(A)$ and $A$ non-conjunctive in $\mathcal{T}_2$ and, conversely, (a more involved proof) shows that whenever $\mathcal{A}_{\mathcal{T}_2,\Sigma} \models D(\xi_A)$ for some $\mathcal{EL}$-concept $D$, then there exist $n \geq 0$ and $C \in \mathsf{noimply}^n_{\mathcal{T}_2,\Sigma}(A)$ such that $\models C \sqsubseteq D$. We illustrate the construction of $\mathcal{A}_{\mathcal{T}_2,\Sigma}$ using the $\mathcal{EL}$-terminologies from Example 17.

Let

$$\{\xi_A \mid A \in \mathsf{sig}(\mathcal{T}_2) \cup \Sigma \text{ and non-conjunctive in } \mathcal{T}_2\} \cup \{\xi_\Sigma\} \subseteq \mathsf{N_I}.$$

be a set of individual names. For $A$ non-conjunctive in $\mathcal{T}_2$, define sets $\mathcal{A}_{\mathcal{T}_2,\Sigma}(A)$ of assertions as follows

- if $A$ is pseudo-primitive in $\mathcal{T}_2$, then

$$\mathcal{A}_{\mathcal{T}_2,\Sigma}(A) = \{A'(\xi_A) \mid A' \in \Sigma \setminus \mathsf{pre}^{\Sigma}_{\mathcal{T}_2}(A)\} \cup \{r(\xi_A, \xi_\Sigma) \mid r \in \Sigma\},$$

- if $A \equiv \exists r.B \in \mathcal{T}_2$, then

$$\begin{aligned}\mathcal{A}_{\mathcal{T}_2,\Sigma}(A) =& \{A'(\xi_A) \mid A' \in \Sigma \setminus \mathsf{pre}^{\Sigma}_{\mathcal{T}_2}(A)\} \\ & \cup \{s(\xi_A, \xi_\Sigma) \mid r \neq s \in \Sigma\} \\ & \cup \{r(\xi_A, \xi_{B'}) \mid B' \in \mathsf{non\text{-}conj}_{\mathcal{T}_2}(B), \text{ if } r \in \Sigma\}\end{aligned}$$

Let

$$\mathcal{A}_{\mathcal{T}_2,\Sigma} = \{\, A'(\xi_\Sigma) \mid A' \in \Sigma \,\} \cup \{\, r(\xi_\Sigma, \xi_\Sigma) \mid r \in \Sigma \,\} \cup \bigcup_{\substack{A \in \Sigma \cup \mathsf{sig}(\mathcal{T}_2) \\ A \text{ is non-conjunctive in } \mathcal{T}_2}} \mathcal{A}_{\mathcal{T}_2,\Sigma}(A)$$

Figure 3: Construction of $\mathcal{A}_{\mathcal{T}_2,\Sigma}$.

**Example 21.** (1) Let $\Sigma = \{A, B, r\}$ and $\mathcal{T}_2 = \emptyset$. Then

$$\mathcal{A}_{\mathcal{T}_2,\Sigma} = \{A(\xi_B), B(\xi_A), r(\xi_A, \xi_\Sigma), r(\xi_B, \xi_\Sigma)\} \cup \mathcal{A}_\Sigma,$$

where $\mathcal{A}_\Sigma = \{A(\xi_\Sigma), B(\xi_\Sigma), r(\xi_\Sigma, \xi_\Sigma)\}$.

(2) Let $\Sigma' = \{A, B, r\}$ and $\mathcal{T}_2' = \{A \equiv \exists r.A\}$. Then

$$\mathcal{A}_{\mathcal{T}_2',\Sigma'} = \{A(\xi_B), B(\xi_A), r(\xi_A, \xi_A), r(\xi_B, \xi_{\Sigma'})\} \cup \mathcal{A}_{\Sigma'},$$

where $\mathcal{A}_{\Sigma'} = \{A(\xi_{\Sigma'}), B(\xi_{\Sigma'}), r(\xi_{\Sigma'}, \xi_{\Sigma'})\}$.

(3) Let $\Sigma'' = \{A, B_1, B_2\}$ and $\mathcal{T}_2'' = \{A \equiv B_1 \sqcap B_2\}$. Then

$$\mathcal{A}_{\mathcal{T}_2'',\Sigma''} = \{B_1(\xi_{B_2}), B_2(\xi_{B_1})\} \cup \mathcal{A}_{\Sigma''},$$

where $\mathcal{A}_{\Sigma''} = \{A(\xi_{\Sigma''}), B_1(\xi_{\Sigma''}), B_2(\xi_{\Sigma''})\}$.

We now obtain the following characterisation of $\mathsf{cWtn}^{\mathsf{rhs}}_\Sigma(\mathcal{T}_1, \mathcal{T}_2)$.

**Theorem 22.** *Let $\mathcal{T}_1$ and $\mathcal{T}_2$ be normalised $\mathcal{EL}$-terminologies and $\Sigma$ a signature. Then the following conditions are equivalent for any $A \in \Sigma$:*

- *$A \in \mathsf{cWtn}^{\mathsf{rhs}}_\Sigma(\mathcal{T}_1, \mathcal{T}_2)$;*
- *there exist $n \geq 0$ and $C \in \mathsf{noimply}^n_{\mathcal{T}_2,\Sigma}(A)$ such that $\mathcal{T}_1 \models C \sqsubseteq A$;*

- $(\mathcal{T}_1, \mathcal{A}_{\mathcal{T}_2,\Sigma}) \models A(\xi_B)$ *for some* $B \in \mathsf{non\text{-}conj}_{\mathcal{T}_2}(A)$.

The equivalence of Points 1 and 2 follows from Corollary 20. We do not give a detailed proof of the equivalence of Points 2 and 3 as this follows from the more general results for $\mathcal{ELH}^r$-terminologies we present below.

**Example 23.** For a normalised form of the terminologies from Example 12,

$$\begin{array}{rcl} \mathcal{T}_1 & = & \{A_0 \sqsubseteq B_0, A_1 \equiv B_n\} \cup \{B_{i+1} \equiv B'_{i+1} \sqcap B''_{i+1} \mid 0 \leq i < n\} \\ & & \cup \{B'_{i+1} \equiv \exists r.B_i \mid 0 \leq i < n\} \cup \{B''_{i+1} \equiv \exists s.B_i \mid 0 \leq i < n\} \\ \mathcal{T}_2 & = & \{A_1 \sqsubseteq F_0\} \cup \{F_i \equiv F'_i \sqcap F''_i \mid 0 \leq i < n\} \\ & & \cup \{F'_i \sqsubseteq \exists r.F_{i+1} \mid 0 \leq i < n\} \cup \{F''_i \sqsubseteq \exists s.F_{i+1} \mid 0 \leq i < n\}, \end{array}$$

and $\Sigma = \{A_0, A_1, r, s\}$, the ABox $\mathcal{A}_{\mathcal{T}_2,\Sigma}$ can be graphically represented as

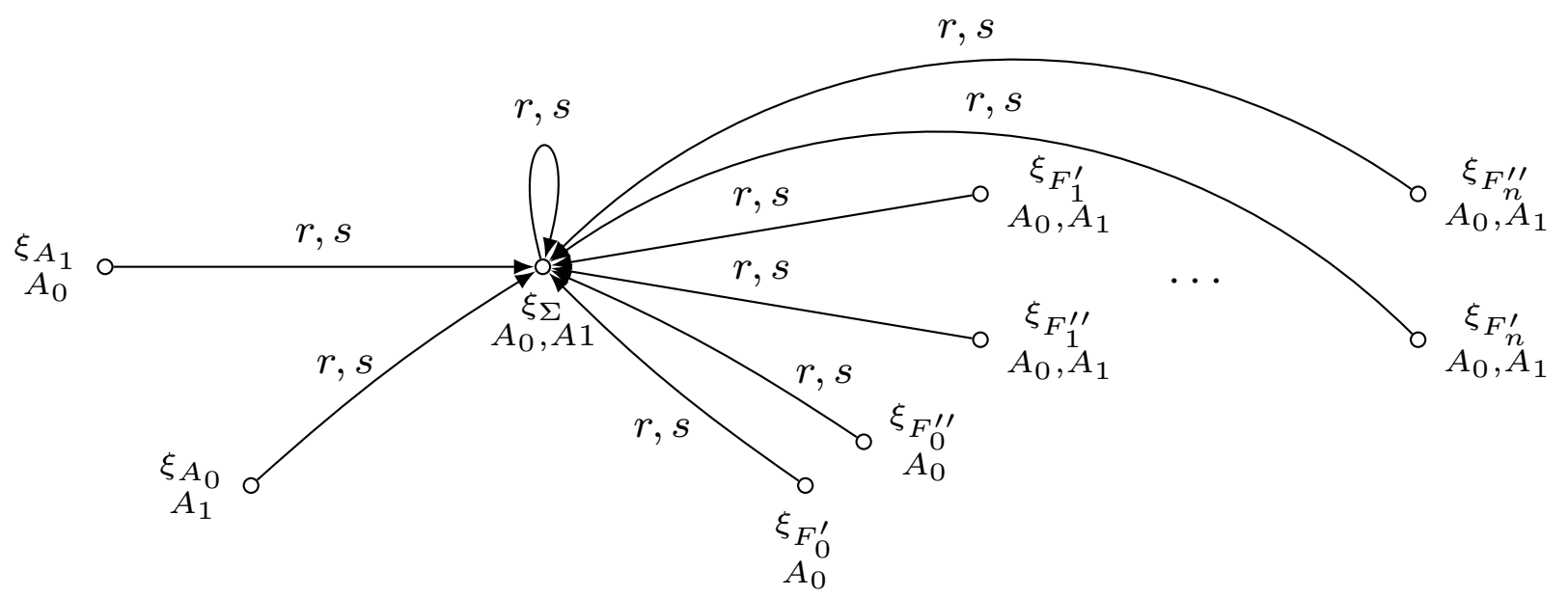

It should be clear that $(\mathcal{T}_1, \mathcal{A}_{\mathcal{T}_2,\Sigma}) \models A_1(\xi_{A_1})$. In fact, $(\mathcal{T}_1, \mathcal{A}) \models A_1(\xi_{A_1})$ holds already for the restriction $\mathcal{A}$ of $\mathcal{A}_{\mathcal{T}_2,\Sigma}$ to the individuals $\{\xi_{A_1}, \xi_\Sigma\}$.

**Theorem 24.** *For $\mathcal{EL}$-terminologies $\mathcal{T}_1$ and $\mathcal{T}_2$ and a signature $\Sigma$, the set $\mathsf{cWtn}^{\mathsf{rhs}}_\Sigma(\mathcal{T}_1, \mathcal{T}_2)$ can be computed in polynomial time.*

*Proof.* It suffices to give a polynomial time algorithm that decides for every $A \in \Sigma$ whether $A \in \mathsf{cWtn}^{\mathsf{rhs}}_\Sigma(\mathcal{T}_1, \mathcal{T}_2)$. First, the ABox $\mathcal{A}_{\mathcal{T}_2,\Sigma}$ can be computed in polynomial time and is of quadratic size in $\mathcal{T}_2$. By Theorem 22, $A \in \mathsf{cWtn}^{\mathsf{rhs}}_\Sigma(\mathcal{T}_1, \mathcal{T}_2)$ iff $(\mathcal{T}_1, \mathcal{A}_{\mathcal{T}_2,\Sigma}) \models A(\xi_B)$ for some $B \in \mathsf{non\text{-}conj}_{\mathcal{T}_2}(A)$, and the latter condition can be checked in polynomial time since instance checking is in polynomial time for $\mathcal{EL}$-TBoxes. □

Regarding the efficiency of this approach, observe that for typical terminologies and large $\Sigma$, the ABox $\mathcal{A}_{\mathcal{T}_2,\Sigma}$ is indeed of quadratic size in $\mathcal{T}_2$ since $\Sigma \setminus \mathsf{pre}^\Sigma_{\mathcal{T}_2}(A)$ will typically contain most of the concept names in $\Sigma$. Thus, for *very large* terminologies and $\Sigma$ a straightforward implementation of this rather elegant algorithm does not work efficiently as one would have to store an ABox of quadratic size and do instance checking for it. We refer the reader to Table 3 and its discussion where a prototype implementation of this approach is applied to modules of Snomed CT.

We now describe our second approach for computing $\mathsf{cWtn}^{\mathsf{rhs}}_\Sigma(\mathcal{T}_1, \mathcal{T}_2)$, which only works for *acyclic* $\mathcal{EL}$-terminologies. Recall that $A \in \mathsf{cWtn}^{\mathsf{rhs}}_\Sigma(\mathcal{T}_1, \mathcal{T}_2)$ if, and only if, there exists an $\mathcal{EL}_\Sigma$-concept $C$ such that $\mathcal{T}_2 \not\models C \sqsubseteq A$ and $\mathcal{T}_1 \models C \sqsubseteq A$. Thus, we have $A \notin \mathsf{cWtn}^{\mathsf{rhs}}_\Sigma(\mathcal{T}_1, \mathcal{T}_2)$ if, and only if, for every $\mathcal{EL}_\Sigma$-concept $C$ with $C \in \mathsf{noimply}_{\mathcal{T}_2,\Sigma}(A)$

```
procedure NotWitness(E)
    if E is pseudo-primitive in T_1 then
        NotWitness(E) := { A ∈ Ξ | pre^Σ_{T_1}(E) ⊆ pre^Σ_{T_2}(A) }
    end if

    if (E ≡ E_1 ⊓ ··· ⊓ E_k ∈ T_1) then
        NotWitness(E) := ⋃_{i=1}^{k} NotWitness(E_i)
    end if

    if E ≡ ∃r.E' ∈ T_1 then
        if r ∉ Σ or All ∈ NotWitness(E') then
            NotWitness(E) := { A ∈ Ξ | pre^Σ_{T_1}(E) ⊆ pre^Σ_{T_2}(A) }
        else
            NotWitness(E) := { A ∈ Ξ | A ≡ ∃r.A' ∈ T_2
                                       non-conj_{T_2}(A') ⊆ NotWitness(E')
                                       pre^Σ_{T_1}(E) ⊆ pre^Σ_{T_2}(A) }
        end if
    end if
end procedure
```

Figure 4: Computation of $\mathsf{NotWitness}(E)$.

it holds that $C \in \mathsf{noimply}_{\mathcal{T}_1,\Sigma}(A)$. Our approach is now based on computing a 'not witness' relation $\mathsf{NW} \subseteq ((\mathsf{sig}(\mathcal{T}_1) \cup \Sigma) \cap \mathsf{N_C}) \times ((\mathsf{sig}(\mathcal{T}_2) \cup \Sigma) \cap \mathsf{N_C})$, which is defined as follows:

$$(E, A) \in \mathsf{NW} \quad \text{if, and only if,} \quad (\dagger)\ \mathsf{noimply}_{\mathcal{T}_2,\Sigma}(A) \subseteq \mathsf{noimply}_{\mathcal{T}_1,\Sigma}(E)$$

Observe that $A \in \mathsf{cWtn}^{\mathsf{rhs}}_{\Sigma}(\mathcal{T}_1, \mathcal{T}_2)$ if, and only if, $(A, A) \notin \mathsf{NW}$; hence, to compute the set $\mathsf{cWtn}^{\mathsf{rhs}}_{\Sigma}(\mathcal{T}_1, \mathcal{T}_2)$ it is sufficient to compute the relation $\mathsf{NW}$. In practice, it is crucial to compute the relation $\mathsf{NW}$ rather than its complement: in typical terminologies most concept names are unrelated in the sense that they do not subsume each other. Thus, the relation $\mathsf{NW}$ is much smaller than its complement (which contains, among others, all pairs $(E, A)$ that do not subsume each other in $\mathcal{T}_1$ and $\mathcal{T}_2$).

To determine the pairs $(E, A) \in \mathsf{NW}$, we aim at computing for every concept name $E \in \mathsf{sig}(\mathcal{T}_1) \cup \Sigma$ the set of concept names $A \in \mathsf{sig}(\mathcal{T}_2) \cup \Sigma$ for which the property $(\dagger)$ holds. This set will be called $\mathsf{NotWitness}(E)$ and is computed in Figure 4, with the following modifications: (1) we only consider those $A \in \mathsf{sig}(\mathcal{T}_2) \cup \Sigma$ which are *non-conjunctive* in $\mathcal{T}_2$ and take conjunctive concept names into account later. (2) We consider a fresh concept name $\mathsf{All}$ not occurring in $\Sigma \cup \mathsf{sig}(\mathcal{T}_1) \cup \mathsf{sig}(\mathcal{T}_2)$ – informally standing for 'all possible $\Sigma$-concepts'.

Thus, the procedure, $\mathsf{NotWitness}(E)$ given in Figure 4 recursively associates with every $E \in \mathsf{sig}(\mathcal{T}_1) \cup \Sigma$ a subset of the set

$$\Xi = \{\mathsf{All}\} \cup \{\, A \mid A \in (\mathsf{sig}(\mathcal{T}_2) \cup \Sigma), A \text{ is non-conjunctive in } \mathcal{T}_2 \,\}$$

and $\mathsf{NW}$ is a relation over

$$((\mathsf{sig}(\mathcal{T}_1) \cup \Sigma) \cap \mathsf{N_C}) \times (((\mathsf{sig}(\mathcal{T}_2) \cup \Sigma) \cap \mathsf{N_C}) \cup \{\mathsf{All}\}).$$

Note that unlike in the approach for computing $\mathsf{cWtn}^{\mathsf{rhs}}_{\Sigma}(\mathcal{T}_1, \mathcal{T}_2)$ that was presented previously, the approach described here does *not* handle the two terminologies separately. In the previous approach the ABox $\mathcal{A}_{\mathcal{T}_2,\Sigma}$ could be precomputed for $\mathcal{T}_2$ and then be re-used to compare $\mathcal{T}_2$ against any other terminology $\mathcal{T}_1$, whereas here both terminologies are analysed simultaneously. We now prove the correctness of the procedure $\mathsf{NotWitness}(E)$.

**Lemma 25.** *For any normalised acyclic $\mathcal{EL}$-terminologies $\mathcal{T}_1$ and $\mathcal{T}_2$, any signature $\Sigma$, any $E \in \mathsf{sig}(\mathcal{T}_1) \cup \Sigma$ and any $A \in \Xi$ the following holds: $A \in \mathsf{NotWitness}(E)$ if, and only if, $(E, A) \in \mathit{NW}$.*

*Proof.* We prove that for any $E \in \mathsf{sig}(\mathcal{T}_1) \cup \Sigma$ and any $A \in \Xi$ the following two conditions are equivalent:

- $A \in \mathsf{NotWitness}(E)$;
- for all $n \geq 0$ and all $C \in \mathsf{noimply}^n_{\mathcal{T}_2,\Sigma}(A)$: $\mathcal{T}_1 \not\models C \sqsubseteq E$.

This is sufficient since $\bigcup_{n\geq 0} \mathsf{noimply}^n_{\mathcal{T}_2,\Sigma}(A)$ is a cover of $\mathsf{noimply}_{\mathcal{T}_2,\Sigma}(A)$ (Lemma 19).

For $E \notin \mathsf{sig}(\mathcal{T}_1)$ the claim is trivial. For $E \in \mathsf{sig}(\mathcal{T}_1)$ the proof is by induction relative to the relation $\succ_{\mathcal{T}_1} \subseteq \mathsf{sig}(\mathcal{T}_1) \times \mathsf{sig}(\mathcal{T}_1)$ (whose definition can be found on page 637). Note that since the considered terminologies are acyclic and $\mathsf{sig}(\mathcal{T}_1)$ is finite, the relation $\succ_{\mathcal{T}_1}$ is well-founded.

We distinguish between the possible definitions of $E$ in $\mathcal{T}_1$. Suppose $E$ is pseudo-primitive in $\mathcal{T}_1$. For $A \in \Xi$, it follows from the definition of $\mathsf{noimply}^n_{\mathcal{T}_2,\Sigma}(A)$ and from Lemma 15 that there exist $n \geq 0$ and $C \in \mathsf{noimply}^n_{\mathcal{T}_2,\Sigma}(A)$ such that $\mathcal{T}_1 \models C \sqsubseteq E$ if, and only if, $\mathcal{T}_1 \models B \sqsubseteq E$ for some $B \in (\Sigma \setminus \mathsf{pre}^{\Sigma}_{\mathcal{T}_2}(A))$. Note that for all $B \in (\Sigma \setminus \mathsf{pre}^{\Sigma}_{\mathcal{T}_2}(A))$, $\mathcal{T}_1 \not\models B \sqsubseteq E$ holds if, and only if, for every $B \in \Sigma$, $\mathcal{T}_1 \models B \sqsubseteq E$ implies that $B \in \mathsf{pre}^{\Sigma}_{\mathcal{T}_2}(A)$. Thus, for every $n$ and $C \in \mathsf{noimply}^n_{\mathcal{T}_2,\Sigma}(A)$, $\mathcal{T}_1 \not\models C \sqsubseteq E$ if, and only if, $\mathsf{pre}^{\Sigma}_{\mathcal{T}_1}(E) \subseteq \mathsf{pre}^{\Sigma}_{\mathcal{T}_2}(A)$ if, and only if, $A \in \mathsf{NotWitness}(E)$.

Assume that $E \equiv E_1 \sqcap \dots \sqcap E_k \in \mathcal{T}_1$. Then, for any concept $C$, $\mathcal{T}_1 \not\models C \sqsubseteq E$ if, and only if, $\mathcal{T}_1 \not\models C \sqsubseteq E_i$ for some $1 \leq i \leq k$. Hence, by applying the induction hypothesis we obtain for every $n$ and $C \in \mathsf{noimply}^n_{\mathcal{T}_2,\Sigma}(A)$, $\mathcal{T}_1 \not\models C \sqsubseteq E$ if, and only if, $A \in \mathsf{NotWitness}(E_i)$ for some $1 \leq i \leq k$, if, and only if, $A \in \mathsf{NotWitness}(E)$.

Finally, assume that $E \equiv \exists r.E' \in \mathcal{T}_1$. Notice that, since $\mathsf{All} \notin (\Sigma \cup \mathsf{sig}(\mathcal{T}_1) \cup \mathsf{sig}(\mathcal{T}_2))$ (in particular, $\mathsf{All}$ is pseudo-primitive in $\mathcal{T}_2$), we have $\mathsf{pre}^{\Sigma}_{\mathcal{T}_2}(\mathsf{All}) = \emptyset$. Thus, by definition for every $n \geq 0$, $\mathsf{noimply}^n_{\mathcal{T}_2,\Sigma}(\mathsf{All}) = \{\mathsf{all}^n_{\Sigma}\}$. By applying the induction hypothesis we can assume that the lemma holds for $E'$, which implies that $\mathsf{All} \notin \mathsf{NotWitness}(E')$ if, and only if, for some $n \geq 0$, $\mathcal{T}_1 \models \mathsf{all}^n_{\Sigma} \sqsubseteq E'$. We now distinguish between the following cases, analogously to the case distinction in procedure $\mathsf{NotWitness}(E)$ (see Figure 4).

If $r \notin \Sigma$, for any $\Sigma$-concept of the form $\exists s.G$, where $s \in \mathsf{N_R} \cap \Sigma$, we have $r \neq s$ and $\mathcal{T}_1 \not\models \exists s.G \sqsubseteq \exists r.E'$. Similarly, if $\mathsf{All} \in \mathsf{NotWitness}(E')$, it holds for every $n \geq 0$ that $\mathcal{T}_1 \not\models \mathsf{all}^n_{\Sigma} \sqsubseteq E'$. Hence, for any $\Sigma$-concept of the form $\exists s.G$, we obtain $\mathcal{T}_1 \not\models \exists s.G \sqsubseteq \exists r.E'$ as otherwise $\mathcal{T}_1 \models \mathsf{all}^n_{\Sigma} \sqsubseteq E'$ would hold for $n = \mathsf{depth}(\exists s.G)$ (where $\mathsf{depth}(\exists s.G)$ is the role-depth of $\exists s.G$). So, by Lemma 15, these two cases are analogous to the case of $E$ being pseudo-primitive considered above.

Assume now that $r \in \Sigma$ and $\mathsf{All} \notin \mathsf{NotWitness}(E')$, that is, for some $n_0 \geq 0$ we have $\mathcal{T}_1 \models \mathsf{all}^{n_0}_{\Sigma} \sqsubseteq E'$.

First, we observe that if $A$ does not have a definition of the form $A \equiv \exists r.A'$ in $\mathcal{T}_2$, then for the unique $C \in \mathsf{noimply}^{n_0+1}_{\mathcal{T}_2,\Sigma}(A)$ we have $\mathcal{T}_1 \models C \sqsubseteq E$ as $\exists r.\mathsf{all}^{n_0}_{\Sigma}$ is a conjunct of $C$ (and as $A$ is non-conjunctive in $\mathcal{T}_2$ by definition of the set $\Xi$). If $A$ has a definition of the form $A \equiv \exists r.A'$ in $\mathcal{T}_2$, for any $n \geq 0$ and $C \in \mathsf{noimply}^{n}_{\mathcal{T}_2,\Sigma}(A)$, we have by Lemma 15 that $\mathcal{T}_1 \not\models C \sqsubseteq E$ if, and only if, $\mathsf{pre}^{\Sigma}_{\mathcal{T}_1}(E) \subseteq \mathsf{pre}^{\Sigma}_{\mathcal{T}_2}(A)$, and, if $n > 0$, for every $C' \in \mathsf{noimply}^{n-1}_{\mathcal{T}_2,\Sigma}(A')$ we have $\mathcal{T}_1 \not\models C' \sqsubseteq A'$.

We can conclude that in case $r \in \Sigma$ and $\mathsf{All} \notin \mathsf{NotWitness}(E')$, for any $A \in \Xi$, any $n \geq 0$, and any $C \in \mathsf{noimply}^{n}_{\mathcal{T}_2,\Sigma}(A)$, we have $\mathcal{T}_1 \not\models C \sqsubseteq E$, if, and only if, $A \equiv \exists r.A' \in \mathcal{T}_2$, $\mathsf{pre}^{\Sigma}_{\mathcal{T}_1}(E) \subseteq \mathsf{pre}^{\Sigma}_{\mathcal{T}_2}(A)$ and for any $m \geq 0$ and any $C' \in \mathsf{noimply}^{m}_{\mathcal{T}_2,\Sigma}(A')$ we have $\mathcal{T}_1 \not\models C' \sqsubseteq E'$. Notice further that, by definition for any $m \geq 0$, $\mathsf{noimply}^{m}_{\mathcal{T}_2,\Sigma}(A') = \bigcup_{B \in \mathsf{non\text{-}conj}_{\mathcal{T}_2}(A')} \mathsf{noimply}^{m}_{\mathcal{T}_2,\Sigma}(B)$. Thus, for any $m \geq 0$ and any $C' \in \mathsf{noimply}^{m}_{\mathcal{T}_2,\Sigma}(A')$, $\mathcal{T}_1 \not\models C' \sqsubseteq E'$ holds if, and only if, for any $m \geq 0$, any $B \in \mathsf{non\text{-}conj}_{\mathcal{T}_2}(A')$ and any $C' \in \mathsf{noimply}^{m}_{\mathcal{T}_2,\Sigma}(B)$, $\mathcal{T}_1 \not\models C' \sqsubseteq E'$, if, and only if, for any $B \in \mathsf{non\text{-}conj}_{\mathcal{T}_2}(A')$, $B \in \mathsf{NotWitness}(E')$ holds by applying the induction hypothesis.

Thus, $\mathcal{T}_1 \not\models C \sqsubseteq E$, for any $n \geq 0$ and $C \in \mathsf{noimply}^{n}_{\mathcal{T}_2,\Sigma}(A)$, if, and only if, $A \in \mathsf{NotWitness}(E)$. □

**Corollary 26.** *Let $\mathcal{T}_1$ and $\mathcal{T}_2$ be normalised acyclic $\mathcal{EL}$-terminologies and $\Sigma$ a signature. Then* $\mathsf{cWtn}^{\mathsf{rhs}}_{\Sigma}(\mathcal{T}_1,\mathcal{T}_2) = \{ A \in \mathsf{sig}(\mathcal{T}_1) \cap \Sigma \mid \exists B \in \mathsf{non\text{-}conj}_{\mathcal{T}_2}(A) \text{ with } B \notin \mathsf{NotWitness}(A) \}$.

*Proof.* First, we observe that if $A \in \mathsf{cWtn}^{\mathsf{rhs}}_{\Sigma}(\mathcal{T}_1,\mathcal{T}_2)$, $A \in \mathsf{sig}(\mathcal{T}_1)$ must hold as otherwise for any $\Sigma$-concept $C$ we have $\mathcal{T}_1 \models C \sqsubseteq A$ if, and only if, $\models C \sqsubseteq A$, and thus $A \notin \mathsf{cWtn}^{\mathsf{rhs}}_{\Sigma}(\mathcal{T}_1,\mathcal{T}_2)$. Now, for all $A \in \mathsf{N_C}$ we have:

| | | |
|---|---|---|
| $A \in \mathsf{cWtn}^{\mathsf{rhs}}_{\Sigma}(\mathcal{T}_1,\mathcal{T}_2)$ | iff | $A \in \mathsf{sig}(\mathcal{T}_1) \cap \Sigma$ (by our observation) and, by definition, there exists a $\Sigma$-concept $C$ with $\mathcal{T}_2 \not\models C \sqsubseteq A$ and $\mathcal{T}_1 \models C \sqsubseteq A$ |
| | iff | $A \in \mathsf{sig}(\mathcal{T}_1) \cap \Sigma$ and there exists $B \in \mathsf{non\text{-}conj}_{\mathcal{T}_2}(A)$ and a $\Sigma$-concept $C$ with $\mathcal{T}_2 \not\models C \sqsubseteq B$ and $\mathcal{T}_1 \models C \sqsubseteq A$ (as otherwise $\mathcal{T}_2 \models C \sqsubseteq A$ would hold) |
| | iff | $A \in \mathsf{sig}(\mathcal{T}_1) \cap \Sigma$ and there exists $B \in \mathsf{non\text{-}conj}_{\mathcal{T}_2}(A)$ with $(A,B) \notin \mathsf{NW}$ (by definition of the relation $\mathsf{NW}$) |
| | iff | $A \in \mathsf{sig}(\mathcal{T}_1) \cap \Sigma$ and there exists $B \in \mathsf{non\text{-}conj}_{\mathcal{T}_2}(A)$ with $B \notin \mathsf{NotWitness}(A)$, by Lemma 25. □ |

For acyclic terminologies, we now obtain an alternative proof of Theorem 24.

**Theorem 27.** *For acyclic $\mathcal{EL}$-terminologies $\mathcal{T}_1$ and $\mathcal{T}_2$ and a signature $\Sigma$,* $\mathsf{cWtn}^{\mathsf{rhs}}_{\Sigma}(\mathcal{T}_1,\mathcal{T}_2)$ *can be computed in polynomial time using the procedure* $\mathsf{NotWitness}(E)$.

*Proof.* To compute the set $\mathsf{cWtn}^{\mathsf{rhs}}_{\Sigma}(\mathcal{T}_1,\mathcal{T}_2)$, it is sufficient by Corollary 26 to compute the sets $\mathsf{NotWitness}(E)$ for every $E \in \mathsf{sig}(\mathcal{T}_1)$. Assuming that $\mathcal{T}_1$ and $\mathcal{T}_2$ are classified and the result of classification is cached, $\mathsf{NotWitness}(E)$ can be computed for all $E \in \mathsf{sig}(\mathcal{T}_1)$, in the worst case, in time $O((|\mathcal{T}_1| + |\mathcal{T}_2|)^3)$. □

**Example 28.** For the acyclic terminologies $\mathcal{T}_1, \mathcal{T}_2$ and the signature $\Sigma$ from Example 23,

$$\mathsf{NotWitness}(A_0) = \{A_0\}, \quad \mathsf{NotWitness}(B_0) = \{A_0\}$$

and for all other concept names $X \in \mathsf{sig}(\mathcal{T}_1)$, $\mathsf{NotWitness}(X) = \emptyset$. In particular $A_1 \notin \mathsf{NotWitness}(A_1)$, so we conclude that $A_1$ is a concept difference witness.

### 4.3 Computing $\mathsf{cWtn}^{\mathsf{lhs}}_{\Sigma}(\mathcal{T}_1, \mathcal{T}_2)$

Recall that the set of left-hand $\Sigma$-concept difference witnesses, $\mathsf{cWtn}^{\mathsf{lhs}}_{\Sigma}(\mathcal{T}_1, \mathcal{T}_2)$, is the set of all $A \in \Sigma \cap \mathsf{N_C}$ such that there exists a concept $C$ with $A \sqsubseteq C \in \mathsf{cDiff}_{\Sigma}(\mathcal{T}_1, \mathcal{T}_2)$. The tractability of computing $\mathsf{cWtn}^{\mathsf{lhs}}_{\Sigma}(\mathcal{T}_1, \mathcal{T}_2)$ for $\mathcal{EL}$ has been proved by Lutz and Wolter (2010) for *arbitrary $\mathcal{EL}$-TBoxes* by reduction to simulation checking. Here we formulate the main steps again because we employ the same technique when dealing with the logical difference for $\mathcal{ELH}^r$-terminologies.

For any two interpretations $\mathcal{I}_1$ and $\mathcal{I}_2$ we say that a relation $S$ between $\mathcal{I}_1$ and $\mathcal{I}_2$ is a $\Sigma$-*simulation* if, and only if, the following conditions hold:

- if $(d, e) \in S$ and $d \in A^{\mathcal{I}_1}$ with $A \in \Sigma$, then $e \in A^{\mathcal{I}_2}$;
- if $(d, e) \in S$ and $(d, d') \in r^{\mathcal{I}_1}$ with $r \in \Sigma$, then there exists $e'$ with $(d', e') \in S$ and $(e, e') \in r^{\mathcal{I}_2}$.

For $d \in \Delta^{\mathcal{I}_1}$ and $e \in \Delta^{\mathcal{I}_2}$ we write $(\mathcal{I}_1, d) \leq_{\Sigma} (\mathcal{I}_2, e)$ if there exists a $\Sigma$-simulation relation $S$ between $\mathcal{I}_1$ and $\mathcal{I}_2$ such that $(d, e) \in S$. It can be checked in polynomial time whether $(\mathcal{I}_1, d) \leq_{\Sigma} (\mathcal{I}_2, e)$ and various polynomial-time algorithms checking the existence of simulations have been developed (Clarke & Schlingloff, 2001; Crafa, Ranzato, & Tapparo, 2011; van Glabbeek & Ploeger, 2008). Simulations characterise the expressive power of $\mathcal{EL}$-concepts in the following sense.

**Lemma 29** (Lutz & Wolter, 2010). *Let $\mathcal{I}_1$ and $\mathcal{I}_2$ be interpretations, $\Sigma$ a signature, $d \in \Delta^{\mathcal{I}_1}$, and $e \in \Delta^{\mathcal{I}_2}$. Then*

$$(\mathcal{I}_1, d) \leq_{\Sigma} (\mathcal{I}_2, e) \quad \Leftrightarrow \quad \text{for all } \mathcal{EL}_{\Sigma}\text{-concepts } C\text{: } d \in C^{\mathcal{I}_1} \Rightarrow e \in C^{\mathcal{I}_2}.$$

It follows that for any $A \in \Sigma$, we have

$$A \in \mathsf{cWtn}^{\mathsf{lhs}}_{\Sigma}(\mathcal{T}_1, \mathcal{T}_2) \quad \Leftrightarrow \quad (\mathcal{I}_{\mathcal{K}_1}, a) \not\leq_{\Sigma} (\mathcal{I}_{\mathcal{K}_2}, a)$$

where $\mathcal{K}_i = (\mathcal{T}_i, \mathcal{A})$ for $\mathcal{A} = \{A(a)\}$ and $\mathcal{I}_{\mathcal{K}_i}$ is the canonical model for $\mathcal{K}_i$, $i = 1, 2$. To see this, recall that by Theorem 2 for every $\mathcal{EL}$-concept $C$, $a \in C^{\mathcal{I}_{\mathcal{K}_i}}$ if, and only if, $(\mathcal{T}_i, \mathcal{A}) \models C(a)$. The latter condition is equivalent to $\mathcal{T}_i \models A \sqsubseteq C$. We have, therefore, proved:

**Theorem 30** (Lutz & Wolter, 2010). *For $\mathcal{EL}$-TBoxes $\mathcal{T}_1$ and $\mathcal{T}_2$ and signatures $\Sigma$, the set $\mathsf{cWtn}^{\mathsf{lhs}}_{\Sigma}(\mathcal{T}_1, \mathcal{T}_2)$ can be computed in polynomial time.*

The following example illustrates the use of simulations between canonical models to determine $\mathsf{cWtn}^{\mathsf{lhs}}_{\Sigma}(\mathcal{T}_1, \mathcal{T}_2)$.

**Example 31.** Let $\Sigma = \{A, r, B_1, B_2\}$ and

$$\begin{aligned}
\mathcal{T}_1 &= \{A \sqsubseteq \exists r.F_0,\ F_0 \sqsubseteq F_1 \sqcap F_2,\ F_1 \sqsubseteq \exists r.B_1,\ F_2 \sqsubseteq \exists r.B_2\},\\
\mathcal{T}_2 &= \{A \sqsubseteq G_1 \sqcap G_2,\ G_1 \sqsubseteq \exists r.G_1',\ G_2 \sqsubseteq \exists r.G_2',\ G_1' \sqsubseteq \exists r.B_1,\ G_2' \sqsubseteq \exists r.B_2\}
\end{aligned}$$

To check whether $A \in \mathsf{cWtn}^{\mathsf{lhs}}_\Sigma(\mathcal{T}_1, \mathcal{T}_2)$ consider the KBs $\mathcal{K}_1 = (\mathcal{T}_1, \{A(a)\})$ and $\mathcal{K}_2 = (\mathcal{T}_2, \{A(a)\})$. Then $A \in \mathsf{cWtn}^{\mathsf{lhs}}_\Sigma(\mathcal{T}_1, \mathcal{T}_2)$ iff $(\mathcal{I}_{\mathcal{K}_1}, a) \not\leq_\Sigma (\mathcal{I}_{\mathcal{K}_2}, a)$, for the canonical models $\mathcal{I}_{\mathcal{K}_1}$ and $\mathcal{I}_{\mathcal{K}_2}$ of $\mathcal{K}_1$ and $\mathcal{K}_2$, respectively. Illustrations of the canonical models $\mathcal{I}_{\mathcal{K}_1}$ and $\mathcal{I}_{\mathcal{K}_2}$ are shown below.

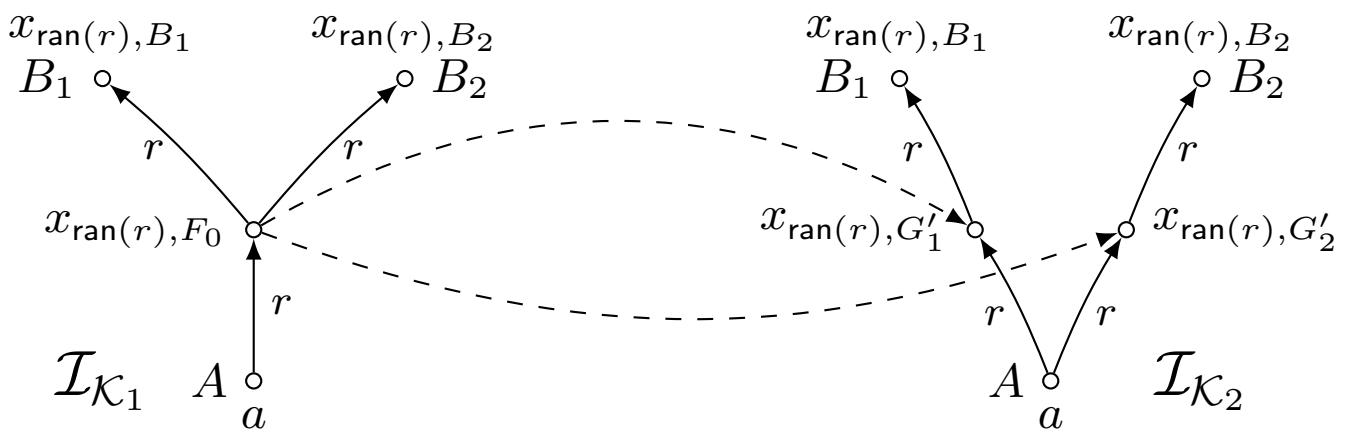

But $(\mathcal{I}_{\mathcal{K}_1}, a) \not\leq_\Sigma (\mathcal{I}_{\mathcal{K}_2}, a)$ because the point $x_{\mathsf{ran}(r),F_0}$ is neither $\Sigma$-simulated by $x_{\mathsf{ran}(r),G'_1}$ nor $\Sigma$-simulated by $x_{\mathsf{ran}(r),G'_2}$. A concept inclusion in $\mathsf{cDiff}_\Sigma(\mathcal{T}_1, \mathcal{T}_2)$ with $A$ on the left-hand side is given by $A \sqsubseteq \exists r.((\exists r.B_1) \sqcap (\exists r.B_2))$.

## 5. $\mathcal{ELH}^r$-Instance Difference

Our polynomial-time algorithms for inseparability and logical difference in $\mathcal{ELH}^r$ are based on extensions of the ideas used in Section 4 for $\mathcal{EL}$. There is, however, one important difference: we introduce new logics, $\mathcal{EL}^{\mathsf{ran}}$ and $\mathcal{EL}^{\mathsf{ran},\sqcap,u}$, for which the concept difference captures exactly the instance and, respectively, query difference in $\mathcal{ELH}^r$. To prove an analogue of Theorem 16 for those languages and, thereby, for the instance and query difference for $\mathcal{ELH}^r$, we introduce a sequent calculus which characterises all $\mathcal{EL}^{\mathsf{ran}}$-consequences of $\mathcal{ELH}^r$-terminologies. We start our investigation with the instance difference case since it is more transparent than the concept difference case (recall that for $\mathcal{EL}$ there is no difference between the instance and the concept difference).

### 5.1 $\mathcal{EL}^{\mathsf{ran}}$-Concept Difference

Recall Example 4 showing that $\mathcal{ELH}^r$ $\Sigma$-concept inseparability does not imply $\Sigma$-instance inseparability:

$$\mathcal{T}_1 = \{\mathsf{ran}(r) \sqsubseteq A_1, \mathsf{ran}(s) \sqsubseteq A_2, B \equiv A_1 \sqcap A_2\}, \quad \mathcal{T}_2 = \emptyset, \quad \Sigma = \{r, s, B\}.$$

Notice that for the ABox $\mathcal{A} = \{r(a,c), s(b,c)\}$, exhibiting the instance difference between $\mathcal{T}_1$ and $\mathcal{T}_2$, $c$ is in the range of both $r$ and $s$. This example suggests that if $\mathsf{ran}(r)$ and $\mathsf{ran}(s)$ could be used in complex concepts, this kind of difference can be made visible for a concept language.

**Definition 32** ($\mathcal{EL}^{\mathsf{ran}}$). *$\mathcal{C}^{\mathsf{ran}}$-concepts are constructed using the following syntax rule*

$$C := \quad A \quad | \quad \mathsf{ran}(r) \quad | \quad C \sqcap D \quad | \quad \exists r.C,$$

*where $A \in \mathsf{N_C}$, $C, D$ range over $\mathcal{C}^{\mathsf{ran}}$-concepts and $r \in \mathsf{N_R}$. The set of $\mathcal{EL}^{\mathsf{ran}}$-inclusions consists of all concept inclusions $C \sqsubseteq D$ and role inclusions $r \sqsubseteq s$, where $C$ is a $\mathcal{C}^{\mathsf{ran}}$-concept, $D$ an $\mathcal{EL}$-concept, and $r, s \in \mathsf{N_R}$.*

Clearly, every $\mathcal{ELH}^r$-inclusion is an $\mathcal{EL}^{\mathsf{ran}}$-inclusion. Additionally, in $\mathcal{EL}^{\mathsf{ran}}$-inclusions the concept $\mathsf{ran}(r)$ can occur everywhere in concepts on the left-hand side of inclusions. This gives us additional concept inclusions for the $\Sigma$-concept difference.

**Example 33.** For $\mathcal{T}_1$ and $\mathcal{T}_2$ from Example 4, we have $\mathcal{T}_1 \models \mathsf{ran}(r) \sqcap \mathsf{ran}(s) \sqsubseteq B$, but $\mathcal{T}_2 \not\models \mathsf{ran}(r) \sqcap \mathsf{ran}(s) \sqsubseteq B$. Thus, using the $\mathcal{C}^{\mathsf{ran}}$-concept $\mathsf{ran}(r) \sqcap \mathsf{ran}(s)$ we can simulate the ABox $\{r(a,c), s(b,c)\}$ from Example 4 and make the $\Sigma$-difference that could not be observed in $\mathcal{ELH}^r$ visible in $\mathcal{EL}^{\mathsf{ran}}$.

We now show that Example 33 can be generalised to arbitrary TBoxes. To this end, we consider the following straightforward generalisation of the $\Sigma$-concept difference to differences over $\mathcal{EL}^{\mathsf{ran}}$.

**Definition 34** ($\mathcal{EL}^{\mathsf{ran}}_\Sigma$-difference)**.** *The $\mathcal{EL}^{\mathsf{ran}}_\Sigma$-difference between $\mathcal{ELH}^r$-TBoxes $\mathcal{T}_1$ and $\mathcal{T}_2$ is the set $\mathsf{cDiff}^{\mathsf{ran}}_\Sigma(\mathcal{T}_1,\mathcal{T}_2)$ of all $\mathcal{EL}^{\mathsf{ran}}_\Sigma$-inclusions $\alpha$ such that $\mathcal{T}_1 \models \alpha$ and $\mathcal{T}_2 \not\models \alpha$.*

To prove the equivalence between $\Sigma$-instance difference in $\mathcal{ELH}^r$ and $\Sigma$-concept difference in $\mathcal{EL}^{\mathsf{ran}}$, we first associate with every ABox $\mathcal{A}$ and individual $a$ in $\mathcal{A}$ a set $\mathcal{C}^{\mathsf{ran}}_{\mathcal{A},a}$ of $\mathcal{C}^{\mathsf{ran}}$-concepts. Assume $\mathcal{A}$ is given. Let, inductively, for $a \in \mathsf{obj}(\mathcal{A})$:

$$C^{0,\mathsf{ran}}_{\mathcal{A},a} = (\bigsqcap_{A(a)\in\mathcal{A}} A) \sqcap (\bigsqcap_{r(b,a)\in\mathcal{A}} \mathsf{ran}(r));$$

and

$$C^{n+1,\mathsf{ran}}_{\mathcal{A},a} = (\bigsqcap_{A(a)\in\mathcal{A}} A) \sqcap (\bigsqcap_{r(b,a)\in\mathcal{A}} \mathsf{ran}(r)) \sqcap (\bigsqcap_{r(a,b)\in\mathcal{A}} \exists r.C^{n,\mathsf{ran}}_{\mathcal{A},b}),$$

and set

$$\mathcal{C}^{\mathsf{ran}}_{\mathcal{A},a} = \{C^{n,\mathsf{ran}}_{\mathcal{A},a} \mid n \geq 0\}$$

Observe that $\mathcal{A} \models C^{n,\mathsf{ran}}_{\mathcal{A},a}(a)$ for all $n > 0$. Moreover, the lemma below shows that, intuitively, the infinite conjunction $\bigsqcap \mathcal{C}^{\mathsf{ran}}_{\mathcal{A},a}$ is the most specific "concept" with $\mathcal{A} \models \bigsqcap \mathcal{C}^{\mathsf{ran}}_{\mathcal{A},a}(a)$.

Conversely, we associate an ABox with a $\mathcal{C}^{\mathsf{ran}}$-concept. The construction is straightforward; however, some care has to be taken since we do not introduce structure sharing but associate distinct individual names with distinct occurrences of subconcepts. Given a $\mathcal{C}^{\mathsf{ran}}$-concept $C$, we first define a *path in $C$* as a finite sequence $C_0 \cdot r_1 \cdot C_1 \cdot \cdots \cdot r_n \cdot C_n$, where $C_0 = C$, $n \geq 0$, and $\exists r_{i+1}.C_{i+1}$ is a conjunct of $C_i$, for $0 \leq i < n$. We use $\mathsf{paths}(C)$ to denote the set of all paths in $C$. If $p \in \mathsf{paths}(C)$, then $\mathsf{tail}(p)$ denotes the last element $C_n$ in $p$.

Now, let $a_{\mathsf{ran}}$ and $a_p$ for $p \in \mathsf{paths}(C)$ be individual names and set inductively:

$$\begin{aligned} \mathcal{A}_C = \;& \{\, s(a_p, a_q) \mid p, q \in \mathsf{paths}(C); q = p \cdot s \cdot C', \text{ for some } C' \,\} \\ & \cup \{\, A(a_p) \mid A \text{ is a conjunct of } \mathsf{tail}(p), p \in \mathsf{paths}(C) \,\} \\ & \cup \{\, \top(a_p) \mid \top \text{ is a conjunct of } \mathsf{tail}(p), p \in \mathsf{paths}(C) \,\} \\ & \cup \{\, r(a_{\mathsf{ran}}, a_p) \mid \mathsf{ran}(r) \text{ is a conjunct of } \mathsf{tail}(p), p \in \mathsf{paths}(C) \,\} \end{aligned}$$

**Example 35.** Let $C = (\exists r.(A \sqcap \mathsf{ran}(v))) \sqcap (\exists s.((\exists t.(A \sqcap \mathsf{ran}(v))) \sqcap (\exists t.(B \sqcap \mathsf{ran}(s)))))$ be a $\mathcal{C}^{\mathsf{ran}}$-concept. Then $\mathcal{A}_C$ can be represented graphically as follows.

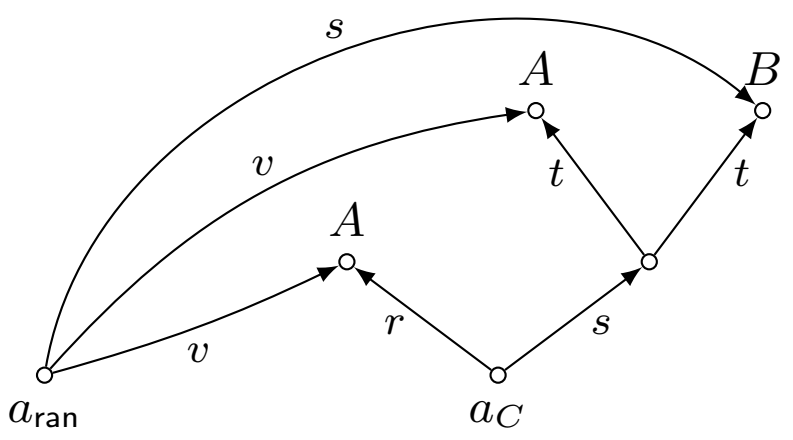

We only indicate $a_C$ and $a_{\mathsf{ran}}$; other individuals are identified by paths in $C$. Note that different occurrences of $A \sqcap \mathsf{ran}(v)$ in $C$ correspond to different individuals in $\mathcal{A}_C$.

**Lemma 36.** *Let $\mathcal{T}$ be an $\mathcal{ELH}^r$-TBox, $\mathcal{A}$ be an ABox, $C_0$ and $D_0$ be $\mathcal{C}^{\mathsf{ran}}$-concepts, and let $a_0 \in \mathsf{obj}(\mathcal{A})$. Then*

- *$(\mathcal{T}, \mathcal{A}) \models D_0(a_0)$ if, and only if, there exists $n \geq 0$ such that $\mathcal{T} \models C^{n,\mathsf{ran}}_{\mathcal{A},a_0} \sqsubseteq D_0$;*
- *$\mathcal{T} \models C_0 \sqsubseteq D_0$ if, and only if, $(\mathcal{T}, \mathcal{A}_{C_0}) \models D_0(a_{C_0})$.*

Below, we will employ this lemma to transfer an analogue of Theorem 16 for $\mathcal{EL}^{\mathsf{ran}}$ to $\mathcal{ELH}^r$-instance differences. For now, we only note the following consequence:

**Corollary 37.** *For any two $\mathcal{ELH}^r$-TBoxes $\mathcal{T}_1$ and $\mathcal{T}_2$, $\mathsf{cDiff}^{\mathsf{ran}}_\Sigma(\mathcal{T}_1, \mathcal{T}_2) = \emptyset$ if, and only if, $\mathsf{iDiff}_\Sigma(\mathcal{T}_1, \mathcal{T}_2) = \emptyset$.*

*Proof.* If $(\mathcal{A}, D_0(a_0)) \in \mathsf{iDiff}_\Sigma(\mathcal{T}_1, \mathcal{T}_2)$, then there exists an $n \geq 0$ such that $C^{n,\mathsf{ran}}_{\mathcal{A},a_0} \sqsubseteq D_0 \in \mathsf{cDiff}^{\mathsf{ran}}_\Sigma(\mathcal{T}_1, \mathcal{T}_2)$. Conversely, if $C_0 \sqsubseteq D_0 \in \mathsf{cDiff}^{\mathsf{ran}}_\Sigma(\mathcal{T}_1, \mathcal{T}_2)$, then $(\mathcal{A}_{C_0}, D_0(a_{C_0})) \in \mathsf{iDiff}_\Sigma(\mathcal{T}_1, \mathcal{T}_2)$. □

Note that Theorem 6 follows from Corollary 37 since for any $\mathcal{ELH}^r$-TBox $\mathcal{T}$ without range restrictions $\mathcal{T} \models C \sqsubseteq D$ if, and only if, $\mathcal{T} \models C' \sqsubseteq D$, where $C'$ is obtained from $C$ by replacing any concept of the form $\mathsf{ran}(r)$ in $C$ by $\top$.

### 5.2 Proof System for $\mathcal{ELH}^r$

The Gentzen-style proof system for $\mathcal{ELH}^r$ consists of the rules given in Figures 1 and 5. Cut elimination, correctness, and completeness of the proof system can be shown similarly to the corresponding proofs given by Hofmann (2005).

**Theorem 38.** *For all $\mathcal{ELH}^r$-terminologies $\mathcal{T}$ and $\mathcal{C}^{\mathsf{ran}}$-concepts $C$ and $D$, it holds that $\mathcal{T} \models C \sqsubseteq D$ if, and only if, $\mathcal{T} \vdash C \sqsubseteq D$.*

We now generalise Lemma 15 to $\mathcal{ELH}^r$-terminologies.

**Lemma 39.** *Let $\mathcal{T}$ be an $\mathcal{ELH}^r$-terminology, $A$ a concept name and $\exists r.D$ an $\mathcal{EL}$-concept. Assume*

$$\mathcal{T} \models \bigsqcap_{1 \leq i \leq l} \mathsf{ran}(s_i) \sqcap \bigsqcap_{1 \leq j \leq n} A_j \sqcap \bigsqcap_{1 \leq k \leq m} \exists r_k.C_k \sqsubseteq \exists r.D,$$

*where $C_k$, $1 \leq k \leq m$, are $\mathcal{C}^{\mathsf{ran}}$-concepts and $l, m, n \geq 0$. Then at least one of the following conditions holds:*

$$\frac{\exists r.(C \sqcap \mathsf{ran}(r)) \sqsubseteq D}{\exists r.C \sqsubseteq D}\ (\textsc{ExRan})$$

$$\frac{B \sqsubseteq D}{\exists r.C \sqsubseteq D}\ (\textsc{Dom}) \quad \text{where } \exists r.\top \sqsubseteq B \in \mathcal{T}$$

$$\frac{A \sqsubseteq D}{\mathsf{ran}(r) \sqsubseteq D}\ (\textsc{Ran}) \quad \text{where } \mathsf{ran}(r) \sqsubseteq A \in \mathcal{T}$$

$$\frac{\exists s.C \sqsubseteq D}{\exists r.C \sqsubseteq D}\ (\textsc{Sub}) \qquad \frac{\mathsf{ran}(s) \sqsubseteq D}{\mathsf{ran}(r) \sqsubseteq D}\ (\textsc{RanSub}) \quad \text{where } r \sqsubseteq s \in \mathcal{T}$$

Figure 5: Additional rules for $\mathcal{ELH}^r$-terminologies.

**(e1)** *there exists* $r_k$, $1 \leq k \leq m$, *such that* $\mathcal{T} \models r_k \sqsubseteq r$ *and* $\mathcal{T} \models C_k \sqcap \mathsf{ran}(r_k) \sqsubseteq D$*;*

**(e2)** *there exists* $A_j$, $1 \leq j \leq n$, *such that* $\mathcal{T} \models A_j \sqsubseteq \exists r.D$*;*

**(e3)** *there exists* $r_k$, $1 \leq k \leq m$, *such that* $\mathcal{T} \models \exists r_k.\top \sqsubseteq \exists r.D$*;*

**(e4)** *there exists* $s_i$, $1 \leq i \leq l$, *such that* $\mathcal{T} \models \mathsf{ran}(s_i) \sqsubseteq \exists r.D$.

*Now assume that* $A$ *is pseudo-primitive and*

$$\mathcal{T} \models \bigsqcap_{1 \leq i \leq l} \mathsf{ran}(s_i) \sqcap \bigsqcap_{1 \leq j \leq n} A_j \sqcap \bigsqcap_{1 \leq k \leq m} \exists r_k.C_k \sqsubseteq A,$$

*where* $C_k$, $1 \leq k \leq m$, *are* $\mathcal{C}^{\mathsf{ran}}$*-concepts and* $l, m, n \geq 0$. *Then at least one of the following conditions holds:*

**(a1)** *there exists* $A_j$, $1 \leq j \leq n$ *such that* $\mathcal{T} \models A_j \sqsubseteq A$*;*

**(a2)** *there exists* $r_k$, $1 \leq k \leq m$ *such that* $\mathcal{T} \models \exists r_k.\top \sqsubseteq A$*;*

**(a3)** *there exists* $s_i$, $1 \leq i \leq l$ *such that* $\mathcal{T} \models \mathsf{ran}(s_i) \sqsubseteq A$.

*Proof.* We prove the first part of the lemma, the second part can then be proved analogously.

Let $C = \bigsqcap_{1 \leq i \leq l} \mathsf{ran}(s_i) \sqcap \bigsqcap_{1 \leq j \leq n} A_j \sqcap \bigsqcap_{1 \leq k \leq m} \exists r_k.C_k$ and assume that $\mathcal{T} \models C \sqsubseteq \exists r.D$ holds. Then, we have $\mathcal{T} \vdash C \sqsubseteq \exists r.D$ by Theorem 38, which implies that there exists a derivation $\mathcal{D}$ of the sequent $C \sqsubseteq \exists r.D$. The proof now proceeds by induction on the depth of $\mathcal{D}$, i.e. the maximal length of any path from the root to one of the leaves of $\mathcal{D}$.

Notice that if $l + n + m \geq 2$, the root of $\mathcal{D}$ can only have been derived by AndL1 or AndL2. The lemma follows then from the induction hypothesis.

Otherwise, we have $l + n + m = 1$. Note that $l + m + n = 0$ is not possible since $\mathcal{T} \not\models \top \sqsubseteq \exists r.D$ by definition of the terminology $\mathcal{T}$. If $C = A_1$ or $C = \mathsf{ran}(s_1)$, then **(e2)** or **(e4)**, respectively, hold already. It remains to consider the case where $C = \exists r_1.C_1$. Then, the rule used to derive the root of $\mathcal{D}$ can only have been one of Ax, Ex, ExRan, Dom or Sub. We consider those cases one by one:

- the root of $\mathcal{D}$ was derived with Ax: then by considering the form of the inference rule, $r_1 = r$ and $C_1 = D$. Hence $\mathcal{T} \models r_1 \sqsubseteq r$ and $\mathcal{T} \models C_1 \sqcap \mathsf{ran}(r_1) \sqsubseteq D$, which implies that **(e1)** holds.

- the root of $\mathcal{D}$ was derived with Ex: then $r_1 = r$ and $\mathcal{T} \vdash C_1 \sqsubseteq D$. Hence, $\mathcal{T} \models r_1 \sqsubseteq r$ and $\mathcal{T} \models C_1 \sqsubseteq D$ holds by Theorem 38. Thus, $\mathcal{T} \models C_1 \sqcap \mathsf{ran}(r_1) \sqsubseteq D$ and we can infer that **(e1)** holds again.

- the root of $\mathcal{D}$ was derived with Dom: we have $\mathcal{T} \vdash B \sqsubseteq \exists r.D$ and $\exists r_1.\top \sqsubseteq B \in \mathcal{T}$. Then by Theorem 38, $\mathcal{T} \models B \sqsubseteq \exists r.D$ and hence, $\mathcal{T} \models \exists r_1.\top \sqsubseteq \exists r.D$, that is, **(e3)** holds.

- the root of $\mathcal{D}$ was derived with ExRan: we obtain $\mathcal{T} \vdash \exists r_1.(C_1 \sqcap \mathsf{ran}(r_1)) \sqsubseteq \exists r.D$. Since the sequent $\exists r_1.(C_1 \sqcap \mathsf{ran}(r_1)) \sqsubseteq \exists r.D$ has a derivation that is of shorter length than $\mathcal{D}$, we can apply the induction hypothesis. Hence, either $\mathcal{T} \models \exists r_1.\top \sqsubseteq \exists r.D$, that is, **(e3)** holds, or $\mathcal{T} \models r_1 \sqsubseteq r$ and $\mathcal{T} \models (C_1 \sqcap \mathsf{ran}(r_1)) \sqcap \mathsf{ran}(r_1) \sqsubseteq D$. Hence **(e1)** holds as $\models C_1 \sqcap \mathsf{ran}(r_1) \sqsubseteq (C_1 \sqcap \mathsf{ran}(r_1)) \sqcap \mathsf{ran}(r_1)$.

- the root of $\mathcal{D}$ was derived with Sub: we obtain $\mathcal{T} \vdash \exists s.C_1 \sqsubseteq \exists r.D$ and $r_1 \sqsubseteq s \in \mathcal{T}$. By the induction hypothesis, either $\mathcal{T} \models \exists s.\top \sqsubseteq \exists r.D$, or $\mathcal{T} \models s \sqsubseteq r$ and $\mathcal{T} \models C_1 \sqcap \mathsf{ran}(s) \sqsubseteq D$. It can be seen that $\mathcal{T} \models \exists r_1.\top \sqsubseteq \exists r.D$, or $\mathcal{T} \models r_1 \sqsubseteq r$ and $\mathcal{T} \models C_1 \sqcap \mathsf{ran}(r_1) \sqsubseteq D$, respectively. Hence **(e3)** or **(e1)** holds.

□

We now prove an extension of Theorem 16 to $\mathcal{EL}^{\mathsf{ran}}$-consequences of $\mathcal{ELH}^r$-terminologies. We give a rather detailed description of the simple witness inclusions contained in members of $\mathsf{cDiff}^{\mathsf{ran}}_\Sigma(\mathcal{T}_1, \mathcal{T}_2)$ since we are going to use this result again when analysing the concept difference in $\mathcal{ELH}^r$.

**Theorem 40** (Primitive witness for $\mathcal{EL}^{\mathsf{ran}}$-differences)**.** *Let $\mathcal{T}_1$ and $\mathcal{T}_2$ be $\mathcal{ELH}^r$-terminologies and $\Sigma$ a signature. If $\varphi \in \mathsf{cDiff}^{\mathsf{ran}}_\Sigma(\mathcal{T}_1, \mathcal{T}_2)$, then either there exist $\{r, s\} \subseteq \mathsf{sig}(\varphi)$ with $r \sqsubseteq s \in \mathsf{cDiff}^{\mathsf{ran}}_\Sigma(\mathcal{T}_1, \mathcal{T}_2)$ or $\varphi$ is of the form $C \sqsubseteq D$, and one of*

1. *$C' \sqsubseteq A$ or $\mathsf{ran}(r) \sqcap C' \sqsubseteq A$,*
2. *$A \sqsubseteq D'$, $\exists r.\top \sqsubseteq D'$ or $\mathsf{ran}(r) \sqsubseteq D'$*

*is a member of $\mathsf{cDiff}^{\mathsf{ran}}_\Sigma(\mathcal{T}_1, \mathcal{T}_2)$, where $r \in \mathsf{sig}(\varphi)$, $A \in \mathsf{sig}(\varphi)$ is a concept name, $C'$ is a subconcept of $C$ and $D'$ is a subconcept of $D$.*

*Proof.* Let $C \sqsubseteq D \in \mathsf{cDiff}^{\mathsf{ran}}_\Sigma(\mathcal{T}_1, \mathcal{T}_2)$, where $C$ is a $\mathcal{C}^{\mathsf{ran}}$-concept and $D$ an $\mathcal{EL}$-concept. We prove the theorem by induction on the structure of $D$.

Notice that $D \neq \top$ as $\mathcal{T}_2 \models C \sqsubseteq \top$. If $D$ is a concept name, then an inclusion from Point 1 exists. If $D = D_1 \sqcap D_2$, then one of $C \sqsubseteq D_i$, $i = 1, 2$, is in $\mathsf{cDiff}^{\mathsf{ran}}_\Sigma(\mathcal{T}_1, \mathcal{T}_2)$. We can apply the induction hypothesis and we can infer that an inclusion from Point 1 or Point 2 exists. If $D = \exists r.D_1$, let $C = \bigsqcap_{1 \leq i \leq l} \mathsf{ran}(s_i) \sqcap \bigsqcap_{1 \leq j \leq n} A_j \sqcap \bigsqcap_{1 \leq k \leq m} \exists r_k.C_k$. Then, by Lemma 39, one of **(e1)**–**(e4)** holds. Cases **(e2)**–**(e4)** directly entail that an inclusion from Point 1 or Point 2 exists. In case of **(e1)**, either $r_k \sqsubseteq r \in \mathsf{cDiff}^{\mathsf{ran}}_\Sigma(\mathcal{T}_1, \mathcal{T}_2)$ or $\mathcal{T}_1 \models C_k \sqcap \mathsf{ran}(r_k) \sqsubseteq D_1$ but $\mathcal{T}_2 \not\models C_k \sqcap \mathsf{ran}(r_k) \sqsubseteq D_1$ (as otherwise $\mathcal{T}_2 \models C \sqsubseteq D$ would hold). Now we can apply the induction hypothesis to $D_1$ and conclude that an inclusion from Point 1 or Point 2 exists. □

### 5.3 Instance Difference Witnesses

Similarly to Theorem 16 for the concept difference between $\mathcal{EL}$-terminologies and derived from its extension, Theorem 40, for $\mathcal{EL}^{\mathsf{ran}}$, we show that every member $(\mathcal{A}, \alpha)$ of $\mathsf{iDiff}_\Sigma(\mathcal{T}_1, \mathcal{T}_2)$ gives rise to a basic witness in which either the ABox or the instance query are atomic. To keep the formulation succinct we give an abstract description of the relationship between $(\mathcal{A}, \alpha) \in \mathsf{iDiff}_\Sigma(\mathcal{T}_1, \mathcal{T}_2)$ and its witness using only the signature of $(\mathcal{A}, \alpha)$. The interested reader will have no problem to derive a stronger relationship between $(\mathcal{A}, \alpha)$ and its witness from the proof.

**Theorem 41** (Primitive witness for $\mathcal{ELH}^r$-instance differences)**.** *Let $\mathcal{T}_1$ and $\mathcal{T}_2$ be $\mathcal{ELH}^r$-terminologies and $\Sigma$ a signature. If $\varphi \in \mathsf{iDiff}_\Sigma(\mathcal{T}_1, \mathcal{T}_2)$, then at least one of the following conditions holds:*

1. $(\{r(a,b)\}, s(a,b)) \in \mathsf{iDiff}_\Sigma(\mathcal{T}_1, \mathcal{T}_2)$*, for some* $r, s \in \mathsf{sig}(\varphi)$*;*
2. $(\mathcal{A}, A(b)) \in \mathsf{iDiff}_\Sigma(\mathcal{T}_1, \mathcal{T}_2)$*, for some concept name* $A \in \mathsf{sig}(\varphi)$*, individual* $b$*, and ABox* $\mathcal{A}$ *with* $\mathsf{sig}(\mathcal{A}) \subseteq \mathsf{sig}(\varphi)$*.*
3. $(\mathcal{A}, D(b)) \in \mathsf{iDiff}_\Sigma(\mathcal{T}_1, \mathcal{T}_2)$*, for some* singleton *ABox* $\mathcal{A}$*, individual* $b$ *in* $\mathcal{A}$*, and* $\mathcal{EL}$*-concept* $D$ *such that* $\mathsf{sig}(\mathcal{A}), \mathsf{sig}(D) \subseteq \mathsf{sig}(\varphi)$*;*

*Proof.* Let $(\mathcal{A}, \alpha) \in \mathsf{iDiff}_\Sigma(\mathcal{T}_1, \mathcal{T}_2)$. We distinguish the following cases.

(a) If $\alpha = s(a,b)$, then $(\mathcal{T}_1, \mathcal{A}) \models s(a,b)$ if, and only if, for some $r(a,b) \in \mathcal{A}$ we have $\mathcal{T}_1 \models r \sqsubseteq s$. As $(\mathcal{T}_2, \mathcal{A}) \not\models s(a,b)$ we obtain $\mathcal{T}_2 \not\models r \sqsubseteq s$. Thus, $(\{r(a,b)\}, s(a,b)) \in \mathsf{iDiff}_\Sigma(\mathcal{T}_1, \mathcal{T}_2)$ and Point 1 holds.

(b) Assume $\alpha = D(b)$ for some $\mathcal{EL}$-concept $D$. By Lemma 36, for some $n \geq 0$ we have $\mathcal{T}_1 \models C^{n,\mathsf{ran}}_{\mathcal{A},b} \sqsubseteq D$ and $\mathcal{T}_2 \not\models C^{n,\mathsf{ran}}_{\mathcal{A},b} \sqsubseteq D$. By Theorem 40, one of (i) $r \sqsubseteq s$, (ii) $A \sqsubseteq D'$, (iii) $\exists r.\top \sqsubseteq D'$, (iv) $\mathsf{ran}(r) \sqsubseteq D'$, (v) $C \sqsubseteq A$, or (vi) $\mathsf{ran}(r) \sqcap C \sqsubseteq A$ is a member of $\mathsf{cDiff}^{\mathsf{ran}}_\Sigma(\mathcal{T}_1, \mathcal{T}_2)$, where $r \in \mathsf{sig}(\varphi)$, $A \in \mathsf{sig}(\varphi)$ is a concept name, $C$ is a subconcept of $C^{n,\mathsf{ran}}_{\mathcal{A},b}$ and $D'$ is a subconcept of $D$. If (i) $r \sqsubseteq s \in \mathsf{cDiff}^{\mathsf{ran}}_\Sigma(\mathcal{T}_1, \mathcal{T}_2)$, then $(\{r(a,b)\}, s(a,b)) \in \mathsf{iDiff}_\Sigma(\mathcal{T}_1, \mathcal{T}_2)$ and Point 1 holds.

Now let $F \sqsubseteq G$ denote the member of $\mathsf{cDiff}^{\mathsf{ran}}_\Sigma(\mathcal{T}_1, \mathcal{T}_2)$ in the cases (ii)-(vi) above. Consider the ABox $\mathcal{A}_F$ associated with the $\mathcal{C}^{\mathsf{ran}}$-concept $F$ in Point 2 of Lemma 36. Then $\mathsf{sig}(\mathcal{A}_F) \subseteq \mathsf{sig}(\varphi)$ and $(\mathcal{A}_F, G(a_F)) \in \mathsf{iDiff}_\Sigma(\mathcal{T}_1, \mathcal{T}_2)$.

In case (ii), we obtain that $F = A$ is a concept name. Hence $\mathcal{A}_F = \{A(a_F)\}$ and Point 3 holds. For case (iii), we obtain $\mathcal{A}_F = \{r(a_F, a_\top), \top(a_\top)\}$ and Point 3 of the lemma applies again (after removing $\top(a_\top)$ from $\mathcal{A}_F$). Similarly, if (iv), then $\mathcal{A}_F = \{r(a_{\mathsf{ran}}, a_F)\}$, and Point 3 of the lemma holds. Finally, for the cases (v) and (vi), $G \in \mathsf{sig}(\varphi)$ is a concept name. Hence Point 2 of the lemma applies. □

Theorem 41 justifies the following finite representation of the $\Sigma$-instance difference between $\mathcal{ELH}^r$-terminologies. It corresponds exactly to the three distinct points of the theorem. Assume $\mathcal{T}_1$ and $\mathcal{T}_2$ are given. Let

- the set of *role* $\Sigma$-instance difference witnesses, $\mathsf{iWtn}^{\mathsf{R}}_{\Sigma}(\mathcal{T}_1,\mathcal{T}_2)$, consist of all $r \sqsubseteq s$ such that $\mathcal{T}_1 \models r \sqsubseteq s$ and $\mathcal{T}_2 \not\models r \sqsubseteq s$;
- the set of *right-hand* $\Sigma$-instance difference witnesses, $\mathsf{iWtn}^{\mathsf{rhs}}_{\Sigma}(\mathcal{T}_1,\mathcal{T}_2)$, consist of all $A \in \Sigma$ such that there exists $\mathcal{A}$ with $(\mathcal{A}, A(a)) \in \mathsf{iDiff}_{\Sigma}(\mathcal{T}_1,\mathcal{T}_2)$;
- the set of *left-hand* $\Sigma$-instance difference witnesses, $\mathsf{iWtn}^{\mathsf{lhs}}_{\Sigma}(\mathcal{T}_1,\mathcal{T}_2)$, consist of all $A \in \Sigma$ such that there exists $C(a)$ with $(\{A(a)\}, C(a)) \in \mathsf{iDiff}_{\Sigma}(\mathcal{T}_1,\mathcal{T}_2)$ and all $r \in \Sigma$ such that there exists $C(c)$ with $c = a$ or $c = b$ such that $(\{r(a,b)\}, C(c)) \in \mathsf{iDiff}_{\Sigma}(\mathcal{T}_1,\mathcal{T}_2)$.

The set of all $\Sigma$-instance difference witnesses is defined as

$$\mathsf{iWtn}_{\Sigma}(\mathcal{T}_1,\mathcal{T}_2) = (\mathsf{iWtn}^{\mathsf{R}}_{\Sigma}(\mathcal{T}_1,\mathcal{T}_2), \mathsf{iWtn}^{\mathsf{rhs}}_{\Sigma}(\mathcal{T}_1,\mathcal{T}_2), \mathsf{iWtn}^{\mathsf{lhs}}_{\Sigma}(\mathcal{T}_1,\mathcal{T}_2)).$$

By Theorem 41, observe that $\mathsf{iWtn}_{\Sigma}(\mathcal{T}_1,\mathcal{T}_2) = (\emptyset,\emptyset,\emptyset)$ if, and only if, $\mathsf{iDiff}_{\Sigma}(\mathcal{T}_1,\mathcal{T}_2) = \emptyset$. The set $\mathsf{iWtn}^{\mathsf{R}}_{\Sigma}(\mathcal{T}_1,\mathcal{T}_2)$ can be easily computed in polynomial time and will not be analysed further in this paper. Thus, our aim now is to present polynomial-time algorithms computing $\mathsf{iWtn}^{\mathsf{rhs}}_{\Sigma}(\mathcal{T}_1,\mathcal{T}_2)$ and $\mathsf{iWtn}^{\mathsf{lhs}}_{\Sigma}(\mathcal{T}_1,\mathcal{T}_2)$.

### 5.4 Computing $\mathsf{iWtn}^{\mathsf{rhs}}_{\Sigma}(\mathcal{T}_1,\mathcal{T}_2)$

We compute $\mathsf{iWtn}^{\mathsf{rhs}}_{\Sigma}(\mathcal{T}_1,\mathcal{T}_2)$ in two different ways: first, we present the more transparent ABox approach that works for arbitrary $\mathcal{ELH}^r$-terminologies, and second we present the more efficient dynamic programming approach that works for acyclic $\mathcal{ELH}^r$-terminologies only. Both approaches have been introduced in Section 4.2 for $\mathcal{EL}$-terminologies. We start with the ABox approach and exhibit a $\Sigma$-ABox $\mathcal{A}_{\mathcal{T}_2,\Sigma}$ depending on $\mathcal{T}_2$ and $\Sigma$ only such that for non-conjunctive $A \in \Sigma$ there exists an ABox $\mathcal{A}$ such that $(\mathcal{A}, A(d)) \in \mathsf{iDiff}_{\Sigma}(\mathcal{T}_1,\mathcal{T}_2)$ if, and only if, $(\mathcal{T}_1, \mathcal{A}_{\mathcal{T}_2,\Sigma}) \models A(\xi_A)$ for a certain individual name $\xi_A$. The case of conjunctive $A$ is reduced to this condition for its defining concept names.

To deal with $\mathcal{ELH}^r$-terminologies rather than with $\mathcal{EL}$-terminologies we have to extend the structure of $\mathcal{A}_{\mathcal{T}_2,\Sigma}$ significantly. To describe the model-theoretic properties of $\mathcal{A}_{\mathcal{T}_2,\Sigma}$, we require the notion of a $\Sigma$-*range simulation*. They capture model-theoretically the expressive power of $\mathcal{C}^{\mathsf{ran}}$-concepts (the concepts that have been used to describe the $\Sigma$-instance difference in terms of subsumption, cf. Lemma 36). For any two ABoxes $\mathcal{A}_1, \mathcal{A}_2$ with designated individual names $a_1$ and $a_2$, we say that a relation $S$ between $\mathsf{obj}(\mathcal{A}_1)$ and $\mathsf{obj}(\mathcal{A}_2)$ is a $\Sigma$-*simulation* if, and only if,

**(S1)** $(a_1, a_2) \in S$;

**(S2)** for all $A \in \Sigma$: if $(a,b) \in S$ and $A(a) \in \mathcal{A}_1$, then $A(b) \in \mathcal{A}_2$;

**(S3)** for all $r \in \Sigma$: if $(a,b) \in S$ and $r(a,a') \in \mathcal{A}_1$, then there exists $b'$ with $(a',b') \in S$ and $r(b,b') \in \mathcal{A}_2$.

We say that $S$ is a $\Sigma$-*range simulation* if, in addition,

**(RS)** for all $r \in \Sigma$: if $(a,b) \in S$ and there exists $c$ such that $r(c,a) \in \mathcal{A}_1$, then there exists $c'$ with $r(c',b) \in \mathcal{A}_2$.

In what follows we write

- $(\mathcal{A}_1, a_1) \leq_\Sigma (\mathcal{A}_2, a_2)$ if there exists a $\Sigma$-simulation between $(\mathcal{A}_1, a_1)$ and $(\mathcal{A}_2, a_2)$; and
- $(\mathcal{A}_1, a_1) \leq_\Sigma^{\mathsf{ran}} (\mathcal{A}_2, a_2)$ if there exists a $\Sigma$-range simulation between $(\mathcal{A}_1, a_1)$ and $(\mathcal{A}_2, a_2)$.

The following lemma shows that range simulations characterise $\mathcal{C}^{\mathsf{ran}}$-concepts.

**Lemma 42.** *Let $\mathcal{A}_1$ and $\mathcal{A}_2$ be $\Sigma$-ABoxes with designated individual names $a_1$ and $a_2$. If $(\mathcal{A}_1, a_1) \leq_\Sigma^{\mathsf{ran}} (\mathcal{A}_2, a_2)$, then $(\mathcal{T}, \mathcal{A}_1) \models C(a_1)$ implies $(\mathcal{T}, \mathcal{A}_2) \models C(a_2)$ for all $\mathcal{C}^{\mathsf{ran}}$-concepts $C$.*

*Proof.* We apply Lemma 36. Let $S$ be a $\Sigma$-range simulation between $\mathcal{A}_1$ and $\mathcal{A}_2$ with $(a_1, a_2) \in S$. One can prove by induction on $n$ that for all $n \geq 0$, for all $a \in \mathsf{obj}(\mathcal{A}_1)$ and for all $b \in \mathsf{obj}(\mathcal{A}_2)$,

$$(*) \text{ If } (a, b) \in S, \text{ then } \mathcal{A}_2 \models C^{n,\mathsf{ran}}_{\mathcal{A}_1,a}(b).$$

Now assume that $(\mathcal{A}_1, a_1) \leq_\Sigma^{\mathsf{ran}} (\mathcal{A}_2, a_2)$ and that $(\mathcal{T}, \mathcal{A}_1) \models C(a_1)$ holds for a $\mathcal{C}^{\mathsf{ran}}$-concept $C$. Then, by Lemma 36, there exists $n \geq 0$ such that $\mathcal{T} \models C^{n,\mathsf{ran}}_{\mathcal{A}_1,a_1} \sqsubseteq C$. Moreover, as $(\mathcal{A}_1, a_1) \leq_\Sigma^{\mathsf{ran}} (\mathcal{A}_2, a_2)$, we have by $(*)$ that $\mathcal{A}_2 \models C^{n,\mathsf{ran}}_{\mathcal{A}_1,a_1}(a_2)$, which then implies that $(\mathcal{T}, \mathcal{A}_2) \models C(a_2)$, as required. □

The construction of $\mathcal{A}_{\mathcal{T},\Sigma}$ is now given in Figure 6, where $\mathcal{T}$ is a normalised $\mathcal{ELH}^r$-terminology and $\Sigma$ a signature. We advise the reader to recall the definition of $\mathcal{A}_{\mathcal{T},\Sigma}$ given in Figure 3 for $\mathcal{EL}$-terminologies $\mathcal{T}$ and then consider the additional ingredients required for $\mathcal{ELH}^r$-terminologies. We remind the reader of the definition of $\mathsf{non\text{-}conj}_\mathcal{T}(A)$ from Section 4.2:

$$\mathsf{non\text{-}conj}_\mathcal{T}(A) = \begin{cases} \{A\}, A \text{ is non-conjunctive in } \mathcal{T} \\ \{B_1, \ldots, B_n\}, A \equiv B_1 \sqcap \cdots \sqcap B_n \in \mathcal{T} \end{cases}$$

In Figure 6, we also use the following sets, for $A \in \mathsf{N_C}$ and $r \in \mathsf{N_R}$:

- $\mathsf{preC}^\Sigma_\mathcal{T}(A) = \{ B \in \Sigma \cap \mathsf{N_C} \mid \mathcal{T} \models B \sqsubseteq A \}$,
- $\mathsf{preDom}^\Sigma_\mathcal{T}(A) = \{ r \in \Sigma \cap \mathsf{N_R} \mid \mathcal{T} \models \exists r.\top \sqsubseteq A \}$,
- $\mathsf{preRan}^\Sigma_\mathcal{T}(A) = \{ r \in \Sigma \cap \mathsf{N_R} \mid \mathcal{T} \models \mathsf{ran}(r) \sqsubseteq A \}$,
- $\mathsf{preRole}^\Sigma_\mathcal{T}(r) = \{ s \in \Sigma \cap \mathsf{N_R} \mid \mathcal{T} \models s \sqsubseteq r \}$.

The following example illustrates the definition of $\mathcal{A}_{\mathcal{T},\Sigma}$.

**Example 43.** For $\mathcal{T}_1 = \{\mathsf{ran}(r) \sqsubseteq A_1, \mathsf{ran}(s) \sqsubseteq A_2, B \equiv A_1 \sqcap A_2\}$, $\mathcal{T}_2 = \emptyset$, and $\Sigma = \{r, s, B\}$ defined as in Example 4, we have

$$\mathcal{A}_{\mathcal{T}_2,\Sigma} = \{B(\xi_\Sigma),\, r(\xi_\Sigma, \xi_\Sigma), s(\xi_\Sigma, \xi_\Sigma),\, r(\xi_B, \xi_\Sigma),\, s(\xi_B, \xi_\Sigma),\, r(\xi_\Sigma, \xi_B),\, s(\xi_\Sigma, \xi_B)\}.$$

It holds that $(\mathcal{T}_1, \mathcal{A}_{\mathcal{T}_2,\Sigma}) \models B(\xi_B)$ and $(\mathcal{T}_2, \mathcal{A}_{\mathcal{T}_2,\Sigma}) \not\models B(\xi_B)$.

Let

$$\{\xi_A \mid A \in \mathsf{sig}(\mathcal{T}) \cup \Sigma \text{ and non-conjunctive in } \mathcal{T}\} \cup \{\xi_\Sigma\} \subseteq \mathsf{N_I}.$$

be a set of individual names. For $A$ non-conjunctive in $\mathcal{T}$, define sets $\mathcal{A}_{\mathcal{T},\Sigma}(A)$ of assertions as follows

- if $A$ is pseudo-primitive in $\mathcal{T}$, then

$$\begin{aligned}\mathcal{A}_{\mathcal{T},\Sigma}(A) = \; & \{\, A'(\xi_A) \mid A' \in \Sigma \setminus \mathsf{preC}^{\Sigma}_{\mathcal{T}}(A) \,\} \\ & \cup \{\, r(\xi_A, \xi_\Sigma) \mid r \in \Sigma \setminus \mathsf{preDom}^{\Sigma}_{\mathcal{T}}(A) \,\} \\ & \cup \{\, r(\xi_\Sigma, \xi_A) \mid r \in \Sigma \setminus \mathsf{preRan}^{\Sigma}_{\mathcal{T}}(A) \,\},\end{aligned}$$

- if $A \equiv \exists r.B \in \mathcal{T}$, then for $s \in \mathsf{N_R}$ let

  $$\Xi_A = \{\, (s, \xi_{B'}) \mid B' \in \mathsf{non\text{-}conj}_{\mathcal{T}}(B), s \in \mathsf{preRole}^{\Sigma}_{\mathcal{T}}(r) \setminus (\mathsf{preDom}^{\Sigma}_{\mathcal{T}}(A) \cup \mathsf{preRan}^{\Sigma}_{\mathcal{T}}(B')) \,\}$$

  and set

$$\begin{aligned}\mathcal{A}_{\mathcal{T},\Sigma}(A) = \; & \{\, A'(\xi_A) \mid A' \in \Sigma \setminus \mathsf{preC}^{\Sigma}_{\mathcal{T}}(A) \,\} \\ & \cup \{\, s(\xi_\Sigma, \xi_A) \mid s \in \Sigma \setminus \mathsf{preRan}^{\Sigma}_{\mathcal{T}}(A) \,\} \\ & \cup \{\, s(\xi_A, \xi_\Sigma) \mid s \in \Sigma \setminus (\mathsf{preRole}^{\Sigma}_{\mathcal{T}}(r) \cup \mathsf{preDom}^{\Sigma}_{\mathcal{T}}(A)) \,\} \\ & \cup \{\, s(\xi_A, \xi) \mid (s, \xi) \in \Xi_A \,\}.\end{aligned}$$

Let

$$\mathcal{A}_{\mathcal{T},\Sigma} = \{\, A'(\xi_\Sigma) \mid A' \in \Sigma \,\} \cup \{\, r(\xi_\Sigma, \xi_\Sigma) \mid r \in \Sigma \,\} \cup \bigcup_{\substack{A \in \Sigma \cup \mathsf{sig}(\mathcal{T}) \\ A \text{ is non-conjunctive in } \mathcal{T}}} \mathcal{A}_{\mathcal{T},\Sigma}(A)$$

Figure 6: Construction of $\mathcal{A}_{\mathcal{T},\Sigma}$ for $\mathcal{ELH}^r$-terminologies.

**Lemma 44.** *For every normalised $\mathcal{ELH}^r$-terminology $\mathcal{T}$ and signature $\Sigma$ the following conditions are equivalent for all $\Sigma$-ABoxes $\mathcal{A}$, $A \in \mathsf{sig}(\mathcal{T}) \cup \Sigma$ non-conjunctive in $\mathcal{T}$, and all $a \in \mathsf{obj}(\mathcal{A})$:*

1. $(\mathcal{T}, \mathcal{A}) \not\models A(a)$;
2. $\xi_A \in \mathsf{obj}(\mathcal{A}_{\mathcal{T},\Sigma})$ *and* $(\mathcal{A}, a) \leq^{\mathsf{ran}}_{\Sigma} (\mathcal{A}_{\mathcal{T},\Sigma}, \xi_A)$.

Lemma 44 is proved in the appendix.

**Lemma 45.** *Let $\mathcal{T}_1$ and $\mathcal{T}_2$ be normalised $\mathcal{ELH}^r$-terminologies, $\Sigma$ a signature and $A \in \Sigma$. Let $\mathcal{A}_{\mathcal{T}_2,\Sigma}$ be the ABox constructed in Figure 6. Then the following conditions are equivalent:*

- *there exists a $\Sigma$-ABox $\mathcal{A}$ such that $(\mathcal{T}_1, \mathcal{A}) \models A(a)$ and $(\mathcal{T}_2, \mathcal{A}) \not\models A(a)$;*
- $(\mathcal{T}_1, \mathcal{A}_{\mathcal{T}_2,\Sigma}) \models A(\xi_B)$ *for some* $B \in \mathsf{non\text{-}conj}_{\mathcal{T}_2}(A)$.

*Proof.* Assume there exists a $\Sigma$-ABox $\mathcal{A}$ and $a \in \mathsf{obj}(\mathcal{A})$ with $(\mathcal{T}_1, \mathcal{A}) \models A(a)$ and $(\mathcal{T}_2, \mathcal{A}) \not\models A(a)$. Then, as $(\mathcal{T}_2, \mathcal{A}) \not\models A(a)$, for some $B \in \mathsf{non\text{-}conj}_{\mathcal{T}_2}(A)$, $(\mathcal{T}_2, \mathcal{A}) \not\models B(a)$. Hence, by Lemma 44, $(\mathcal{A}, a) \leq_\Sigma^{\mathsf{ran}} (\mathcal{A}_{\mathcal{T}_2,\Sigma}, \xi_B)$. But then, by Lemma 42, $(\mathcal{T}_1, \mathcal{A}_{\mathcal{T}_2,\Sigma}) \models A(\xi_B)$, as required.

Conversely, suppose that $(\mathcal{T}_1, \mathcal{A}_{\mathcal{T}_2,\Sigma}) \models A(\xi_B)$ for some $B \in \mathsf{non\text{-}conj}_{\mathcal{T}_2}(A)$ with $\xi_B \in \mathsf{obj}(\mathcal{A}_{\mathcal{T}_2,\Sigma})$. Notice that, by Lemma 44, $(\mathcal{T}_2, \mathcal{A}_{\mathcal{T}_2,\Sigma}) \not\models B(\xi_B)$. Hence $(\mathcal{T}_2, \mathcal{A}_{\mathcal{T}_2,\Sigma}) \not\models A(\xi_B)$ and so $\mathcal{A}_{\mathcal{T}_2,\Sigma}$ and $\xi_B$ witness Point 1. □

**Theorem 46.** *Let $\mathcal{T}_1$ and $\mathcal{T}_2$ be normalised $\mathcal{ELH}^r$-terminologies and $\Sigma$ a signature. Then $\mathsf{iWtn}_\Sigma^{\mathsf{rhs}}(\mathcal{T}_1, \mathcal{T}_2)$ can be computed in polynomial time.*

*Proof.* By Lemma 45, $A \in \mathsf{iWtn}_\Sigma^{\mathsf{rhs}}(\mathcal{T}_1, \mathcal{T}_2)$ if, and only if, for some $B \in \mathsf{non\text{-}conj}_{\mathcal{T}_2}(A)$ we have $(\mathcal{T}_1, \mathcal{A}_{\mathcal{T}_2,\Sigma}) \models A(\xi_B)$. It remains to observe that $\mathcal{A}_{\mathcal{T}_2,\Sigma}$ can be constructed in polynomial time and checking whether $(\mathcal{T}_1, \mathcal{A}_{\mathcal{T}_2,\Sigma}) \models A(\xi_B)$ is in polynomial time. □

We now briefly describe how the dynamic programming approach for computing the set $\mathsf{iWtn}_\Sigma^{\mathsf{rhs}}(\mathcal{T}_1, \mathcal{T}_2)$ for acyclic terminologies is extended from $\mathcal{EL}$ to $\mathcal{ELH}^r$. The extension of the $\mathsf{NotWitness}(E)$ algorithm from Figure 4 to $\mathcal{ELH}^r$ is given in Figure 7. As in Figure 4, the procedure $\mathsf{NotWitness}(E)$ recursively associates with every $E \in \mathsf{sig}(\mathcal{T}_1) \cup \Sigma$ a subset of

$$\Xi = \{\mathsf{All}\} \cup \{ A \mid A \in (\mathsf{sig}(\mathcal{T}_2) \cup \Sigma), A \text{ is non-conjunctive in } \mathcal{T}_2 \}.$$

The conditions for $A \in \mathsf{NotWitness}(E)$ become more complex since now one has to take into account the sets $\mathsf{preRan}_\mathcal{T}^\Sigma(E)$ and $\mathsf{preDom}_\mathcal{T}^\Sigma(E)$. To prove the correctness of the $\mathsf{NotWitness}$ algorithm, we observe the following consequence of Lemma 36.

**Corollary 47.** *Let $\mathcal{T}_1$ and $\mathcal{T}_2$ be normalised acyclic $\mathcal{ELH}^r$-terminologies and $\Sigma$ a signature. Then $\mathsf{iWtn}_\Sigma^{\mathsf{rhs}}(\mathcal{T}_1, \mathcal{T}_2) = \{ A \in \Sigma \mid \exists \mathcal{C}^{\mathsf{ran}}$-concept $C$ such that $C \sqsubseteq A \in \mathsf{cDiff}_\Sigma^{\mathsf{ran}}(\mathcal{T}_1, \mathcal{T}_2) \}$.*

*Proof.* First, let $A \in \mathsf{iWtn}_\Sigma^{\mathsf{rhs}}(\mathcal{T}_1, \mathcal{T}_2)$. Then there exists a $\Sigma$-ABox $\mathcal{A}$ such that $(\mathcal{T}_1, \mathcal{A}) \models A(a)$ and $(\mathcal{T}_2, \mathcal{A}) \not\models A(a)$. Hence, by Point 1 of Lemma 36 there exists $n \geq 0$ such that $C_{\mathcal{A},a}^{n,\mathsf{ran}} \sqsubseteq A \in \mathsf{cDiff}_\Sigma^{\mathsf{ran}}(\mathcal{T}_1, \mathcal{T}_2)$. Note that $C_{\mathcal{A},a}^{n,\mathsf{ran}}$ is a $\mathcal{C}^{\mathsf{ran}}$-concept. Conversely, assume $A \in \Sigma$ such that there exists a $\mathcal{C}^{\mathsf{ran}}$-concept $C$ with $C \sqsubseteq A \in \mathsf{cDiff}_\Sigma^{\mathsf{ran}}(\mathcal{T}_1, \mathcal{T}_2)$. Then by Point 2 of Lemma 36, $(\mathcal{A}_C, A(a_C)) \in \mathsf{iDiff}_\Sigma(\mathcal{T}_1, \mathcal{T}_2)$, i.e. $A \in \mathsf{iWtn}_\Sigma^{\mathsf{rhs}}(\mathcal{T}_1, \mathcal{T}_2)$. □

We now formulate the correctness of the $\mathsf{NotWitness}$ algorithm in the same way as in Corollary 26.

**Theorem 48.** *Let $\mathcal{T}_1$ and $\mathcal{T}_2$ be normalised acyclic $\mathcal{ELH}^r$-terminologies and $\Sigma$ a signature. Then $\mathsf{iWtn}_\Sigma^{\mathsf{rhs}}(\mathcal{T}_1, \mathcal{T}_2) = \{ A \in \mathsf{sig}(\mathcal{T}_1) \cap \Sigma \mid \exists B \in \mathsf{non\text{-}conj}_{\mathcal{T}_2}(A)$ with $B \notin \mathsf{NotWitness}(A) \}$.*

The proof is an extension of the proofs of Lemma 25 and Corollary 26. Namely, one can show that for all $A \in \mathsf{sig}(\mathcal{T}_1) \cup \Sigma$ and all $B \in \mathsf{sig}(\mathcal{T}_2) \cup \Sigma$ such that $B$ is non-conjunctive in $\mathcal{T}_2$ the following conditions are equivalent:

- $B \in \mathsf{NotWitness}(A)$;
- for all $\mathcal{C}_\Sigma^{\mathsf{ran}}$-concepts $C$: $\mathcal{T}_2 \not\models C \sqsubseteq B$ implies $\mathcal{T}_1 \not\models C \sqsubseteq A$.

Using Corollary 47, we thus obtain for every $A \in \Sigma$: $A \in \mathsf{iWtn}_\Sigma^{\mathsf{rhs}}(\mathcal{T}_1, \mathcal{T}_2)$ if, and only if, there exists a $\mathcal{C}_\Sigma^{\mathsf{ran}}$-concept $C$ with $\mathcal{T}_2 \not\models C \sqsubseteq A$ and $\mathcal{T}_1 \models C \sqsubseteq A$ if, and only if, there exists $B \in \mathsf{non\text{-}conj}_{\mathcal{T}_2}(A)$ with $B \notin \mathsf{NotWitness}(A)$.

**procedure** $\mathsf{Aux_{PP}}(E)$
  **if** $\mathsf{preC}^{\Sigma}_{\mathcal{T}_1}(E) = \emptyset$ **and** $\mathsf{preRan}^{\Sigma}_{\mathcal{T}_1}(E) = \emptyset$ **and** $\mathsf{preDom}^{\Sigma}_{\mathcal{T}_1}(E) = \emptyset$ **then**
    **return** $\{\mathsf{All}\}$
  **else**
    $\mathrm{Aux}_{\mathrm{concept}} := \{ A \in \Xi \mid \mathsf{preC}^{\Sigma}_{\mathcal{T}_1}(E) \subseteq \mathsf{preC}^{\Sigma}_{\mathcal{T}_2}(A) \}$
    $\mathrm{Aux}_{\mathrm{ran}} := \{ A \in \Xi \mid \mathsf{preRan}^{\Sigma}_{\mathcal{T}_1}(E) \subseteq \mathsf{preRan}^{\Sigma}_{\mathcal{T}_2}(A) \}$
    $\mathrm{Aux}_{\mathrm{dom}} := \{ A \in \Xi \mid \mathsf{preDom}^{\Sigma}_{\mathcal{T}_1}(E) \subseteq \mathsf{preDom}^{\Sigma}_{\mathcal{T}_2}(A) \}$
    **return** $\mathrm{Aux}_{\mathrm{concept}} \cap \mathrm{Aux}_{\mathrm{ran}} \cap \mathrm{Aux}_{\mathrm{dom}}$
  **end if**
**end procedure**

**procedure** $\mathsf{NotWitness}(E)$
  **if** $E$ is pseudo-primitive in $\mathcal{T}_1$ **then**
    $\mathsf{NotWitness}(E) := \mathsf{Aux_{PP}}(E)$

  **else if** $E \equiv E_1 \sqcap \cdots \sqcap E_k \in \mathcal{T}_1$ **then**
    $\mathsf{NotWitness}(E) := \bigcup_{i=1}^{k} \mathsf{NotWitness}(E_i)$

  **else if** $E \equiv \exists r.E' \in \mathcal{T}_1$ **then**
    **if** $\mathsf{preRole}^{\Sigma}_{\mathcal{T}_1}(r) = \emptyset$ **or** $\mathsf{All} \in \mathsf{NotWitness}(E')$ **then**
      $\mathsf{NotWitness}(E) := \mathsf{Aux_{PP}}(E)$
    **else**

$$\mathrm{Aux}_{\mathrm{role,prim}} := \left\{ A \in \Xi \;\middle|\; \begin{array}{l} A \text{ is pseudo-primitive in } \mathcal{T}, \text{ and} \\ \mathsf{preRole}^{\Sigma}_{\mathcal{T}_1}(r) \subseteq \mathsf{preDom}^{\Sigma}_{\mathcal{T}_2}(A) \end{array} \right\}$$

$$\mathrm{Aux}_{\mathrm{role,exist}} := \left\{ A \in \Xi \;\middle|\; \begin{array}{l} A \equiv \exists t.B \in \mathcal{T}_2, \\ \mathsf{preRole}^{\Sigma}_{\mathcal{T}_1}(r) \subseteq \mathsf{preRole}^{\Sigma}_{\mathcal{T}_2}(t) \cup \mathsf{preDom}^{\Sigma}_{\mathcal{T}_2}(A), \text{ and} \\ \text{for all } s \in \mathsf{preRole}^{\Sigma}_{\mathcal{T}_1}(r) \cap \mathsf{preRole}^{\Sigma}_{\mathcal{T}_2}(t) \text{ with} \\ \quad s \notin \mathsf{preDom}^{\Sigma}_{\mathcal{T}_2}(A) \text{ and } B' \in \mathsf{non\text{-}conj}_{\mathcal{T}_2}(B) \text{ with} \\ \quad s \notin \mathsf{preRan}^{\Sigma}_{\mathcal{T}_2}(B'), \text{ there exists } E'' \in \mathsf{non\text{-}conj}_{\mathcal{T}_1}(E') \\ \quad \text{with } B' \in \mathsf{NotWitness}(E'') \text{ and } s \notin \mathsf{preRan}^{\Sigma}_{\mathcal{T}_1}(E'') \end{array} \right\}$$

      $\mathsf{NotWitness}(E) := (\mathrm{Aux}_{\mathrm{role,prim}} \cup \mathrm{Aux}_{\mathrm{role,exist}}) \cap \mathsf{Aux_{PP}}(E)$
    **end if**
  **end if**
**end procedure**

Figure 7: Computation of $\mathsf{NotWitness}(E)$ for $\mathcal{ELH}^r$.

### 5.5 Tractability of $\mathsf{iWtn}^{\mathsf{lhs}}_{\Sigma}(\mathcal{T}_1, \mathcal{T}_2)$

We prove the tractability of $\mathsf{iWtn}^{\mathsf{lhs}}_{\Sigma}(\mathcal{T}_1, \mathcal{T}_2)$ by the same reduction to simulation checking as in the case of $\mathcal{EL}$-terminologies (Theorem 30).

**Theorem 49.** *Let $\mathcal{T}_1$ and $\mathcal{T}_2$ be $\mathcal{ELH}^r$-TBoxes and let $\Sigma$ be a signature. Then the set $\mathsf{iWtn}^{\mathsf{lhs}}_{\Sigma}(\mathcal{T}_1, \mathcal{T}_2)$ can be computed in polynomial time.*

*Proof.* For any concept name $A \in \Sigma$ we have

$$A \in \mathsf{iWtn}^{\mathsf{lhs}}_{\Sigma}(\mathcal{T}_1, \mathcal{T}_2) \quad \Leftrightarrow \quad (\mathcal{I}_{\mathcal{K}_1}, a) \not\leq_{\Sigma} (\mathcal{I}_{\mathcal{K}_2}, a)$$

where $\mathcal{K}_i = (\mathcal{T}_i, \mathcal{A})$ for $\mathcal{A} = \{A(a)\}$ and $\mathcal{I}_{\mathcal{K}_i}$ is the canonical model for $\mathcal{K}_i$, $i = 1, 2$. Indeed, $(\{A(a)\}, C(a)) \in \mathsf{iDiff}_\Sigma(\mathcal{T}_1, \mathcal{T}_2)$, for some $\mathcal{EL}_\Sigma$-concept $C$, if, and only if, by Theorem 2, $a \in C^{\mathcal{I}_{\mathcal{K}_1}}$ and $a \notin C^{\mathcal{I}_{\mathcal{K}_2}}$. But this condition is, by Lemma 29, equivalent to $(\mathcal{I}_{\mathcal{K}_1}, a) \not\leq_\Sigma (\mathcal{I}_{\mathcal{K}_2}, a)$. The latter condition can be checked in polynomial time.

Similarly, for any role name $r \in \Sigma$ we have

$$r \in \mathsf{iWtn}^{\mathsf{lhs}}_\Sigma(\mathcal{T}_1, \mathcal{T}_2) \quad \Leftrightarrow \quad (\mathcal{I}_{\mathcal{K}_1}, a) \not\leq_\Sigma (\mathcal{I}_{\mathcal{K}_2}, a) \text{ or } (\mathcal{I}_{\mathcal{K}_1}, b) \not\leq_\Sigma (\mathcal{I}_{\mathcal{K}_2}, b)$$

where $\mathcal{K}_i = (\mathcal{T}_i, \mathcal{A})$, $\mathcal{A} = \{r(a, b)\}$, and $\mathcal{I}_{\mathcal{K}_i}$ is the canonical model for $\mathcal{K}_i$, $i = 1, 2$. Again, the latter condition can be checked in polynomial time. □

## 6. $\mathcal{ELH}^r$-Concept Difference

In this section we present polynomial-time algorithms deciding Σ-concept inseparability and computing a succinct representation of the concept difference between $\mathcal{ELH}^r$-terminologies. The algorithms are essentially by reduction to the instance difference case.

We start by introducing the succinct representation of the Σ-concept difference. Let $\mathcal{T}_1$ and $\mathcal{T}_2$ be $\mathcal{ELH}^r$-terminologies. Since $\mathsf{cDiff}_\Sigma(\mathcal{T}_1, \mathcal{T}_2) \subseteq \mathsf{cDiff}^{\mathsf{ran}}_\Sigma(\mathcal{T}_1, \mathcal{T}_2)$, it follows from Theorem 40 for $C \sqsubseteq D \in \mathsf{cDiff}_\Sigma(\mathcal{T}_1, \mathcal{T}_2)$ that there exists an inclusion of at least one of the following forms

(i) $C' \sqsubseteq A$,

(ii) $\mathsf{ran}(r) \sqcap C' \sqsubseteq A$,

(iii) $A \sqsubseteq D'$,

(iv) $\exists r.\top \sqsubseteq D'$, or

(v) $\mathsf{ran}(r) \sqsubseteq D'$

in $\mathsf{cDiff}_\Sigma(\mathcal{T}_1, \mathcal{T}_2)$, where $r \in \mathsf{sig}(\varphi)$, $A \in \mathsf{sig}(\varphi)$ is a concept name, $C'$ is a subconcept of $C$ and $D'$ is a subconcept of $D$. Notice in particular for case (ii) that $C'$ is an $\mathcal{EL}$-concept. Hence, just as in the case of the Σ-instance difference, we obtain the following representation of the Σ-concept difference. Assume $\mathcal{T}_1$ and $\mathcal{T}_2$ are given. Let

- the set of *role inclusion* Σ-concept difference witnesses, $\mathsf{cWtn}^{\mathsf{R}}_\Sigma(\mathcal{T}_1, \mathcal{T}_2)$, consist of all $r \sqsubseteq s$ such that $\mathcal{T}_1 \models r \sqsubseteq s$ and $\mathcal{T}_2 \not\models r \sqsubseteq s$;

- the set of *right-hand* Σ-concept difference witnesses, $\mathsf{cWtn}^{\mathsf{rhs}}_\Sigma(\mathcal{T}_1, \mathcal{T}_2)$, consist of all $A \in \Sigma$ such that there exists an $\mathcal{EL}$-concept $C$ with either $C \sqsubseteq A \in \mathsf{cDiff}_\Sigma(\mathcal{T}_1, \mathcal{T}_2)$ or there additionally exists a role name $r \in \Sigma$ such that $\mathsf{ran}(r) \sqcap C \sqsubseteq A \in \mathsf{cDiff}_\Sigma(\mathcal{T}_1, \mathcal{T}_2)$.

- the set of *left-hand* Σ-concept difference witnesses, $\mathsf{cWtn}^{\mathsf{lhs}}_\Sigma(\mathcal{T}_1, \mathcal{T}_2)$, consist of all $A \in \Sigma$ such that there exists an $\mathcal{EL}$-concept $C$ with $A \sqsubseteq C \in \mathsf{cDiff}_\Sigma(\mathcal{T}_1, \mathcal{T}_2)$, and of all role names $r \in \Sigma$ such that there exists an $\mathcal{EL}$-concept $C$ with either $\exists r.\top \sqsubseteq C \in \mathsf{cDiff}_\Sigma(\mathcal{T}_1, \mathcal{T}_2)$ or $\mathsf{ran}(r) \sqsubseteq C \in \mathsf{cDiff}_\Sigma(\mathcal{T}_1, \mathcal{T}_2)$.

The set of all Σ-concept difference witnesses is defined as

$$\mathsf{cWtn}_\Sigma(\mathcal{T}_1, \mathcal{T}_2) = (\mathsf{cWtn}^{\mathsf{R}}_\Sigma(\mathcal{T}_1, \mathcal{T}_2), \mathsf{cWtn}^{\mathsf{rhs}}_\Sigma(\mathcal{T}_1, \mathcal{T}_2), \mathsf{cWtn}^{\mathsf{lhs}}_\Sigma(\mathcal{T}_1, \mathcal{T}_2)).$$

Observe that $\mathsf{cWtn}_\Sigma(\mathcal{T}_1, \mathcal{T}_2) = (\emptyset, \emptyset, \emptyset)$ if, and only if, $\mathsf{cDiff}_\Sigma(\mathcal{T}_1, \mathcal{T}_2) = \emptyset$. We also obtain that the sets $\mathsf{cWtn}^{\mathsf{R}}_\Sigma(\mathcal{T}_1, \mathcal{T}_2)$ and $\mathsf{cWtn}^{\mathsf{lhs}}_\Sigma(\mathcal{T}_1, \mathcal{T}_2)$ coincide with the corresponding witness sets for the instance difference, which allows us to re-use some results that we have developed for detecting instance differences.

**Lemma 50.** *Let $\mathcal{T}_1$ and $\mathcal{T}_2$ be normalised $\mathcal{ELH}^r$-terminologies and $\Sigma$ a signature. Then the following holds:*

1. $\mathsf{cWtn}^{\mathsf{R}}_\Sigma(\mathcal{T}_1, \mathcal{T}_2) = \mathsf{iWtn}^{\mathsf{R}}_\Sigma(\mathcal{T}_1, \mathcal{T}_2)$,
2. $\mathsf{cWtn}^{\mathsf{lhs}}_\Sigma(\mathcal{T}_1, \mathcal{T}_2) = \mathsf{iWtn}^{\mathsf{lhs}}_\Sigma(\mathcal{T}_1, \mathcal{T}_2)$
3. $\mathsf{cWtn}^{\mathsf{rhs}}_\Sigma(\mathcal{T}_1, \mathcal{T}_2) \subseteq \mathsf{iWtn}^{\mathsf{rhs}}_\Sigma(\mathcal{T}_1, \mathcal{T}_2)$

*Proof.* Point 1 follows directly from the definition. Proving $\mathsf{cWtn}^{\mathsf{lhs}}_\Sigma(\mathcal{T}_1, \mathcal{T}_2) \subseteq \mathsf{iWtn}^{\mathsf{lhs}}_\Sigma(\mathcal{T}_1, \mathcal{T}_2)$ and $\mathsf{cWtn}^{\mathsf{rhs}}_\Sigma(\mathcal{T}_1, \mathcal{T}_2) \subseteq \mathsf{iWtn}^{\mathsf{rhs}}_\Sigma(\mathcal{T}_1, \mathcal{T}_2)$ is similar to Lemma 10. Finally, to prove that $\mathsf{iWtn}^{\mathsf{lhs}}_\Sigma(\mathcal{T}_1, \mathcal{T}_2) \subseteq \mathsf{cWtn}^{\mathsf{lhs}}_\Sigma(\mathcal{T}_1, \mathcal{T}_2)$, assume that $A \in \mathsf{iWtn}^{\mathsf{lhs}}_\Sigma(\mathcal{T}_1, \mathcal{T}_2)$. Then there exists an $\mathcal{EL}$-concept $D(a)$ with $(\{A(a)\}, D(a)) \in \mathsf{iDiff}_\Sigma(\mathcal{T}_1, \mathcal{T}_2)$. But then $\mathcal{T}_1 \models A \sqsubseteq D$ and $\mathcal{T}_2 \not\models A \sqsubseteq D$ and, therefore, $A \in \mathsf{cWtn}^{\mathsf{lhs}}_\Sigma(\mathcal{T}_1, \mathcal{T}_2)$. The argument for $r \in \mathsf{iWtn}^{\mathsf{lhs}}_\Sigma(\mathcal{T}_1, \mathcal{T}_2)$ is similar. □

We have presented polynomial-time algorithms which can compute $\mathsf{iWtn}^{\mathsf{lhs}}_\Sigma(\mathcal{T}_1, \mathcal{T}_2)$ and $\mathsf{iWtn}^{\mathsf{R}}_\Sigma(\mathcal{T}_1, \mathcal{T}_2)$. Thus, it remains to analyse $\mathsf{cWtn}^{\mathsf{rhs}}_\Sigma(\mathcal{T}_1, \mathcal{T}_2)$.

### 6.1 Tractability of $\mathsf{cWtn}^{\mathsf{rhs}}_\Sigma(\mathcal{T}_1, \mathcal{T}_2)$

We prove tractability of $\mathsf{cWtn}^{\mathsf{rhs}}_\Sigma(\mathcal{T}_1, \mathcal{T}_2)$ by modifying the ABox $\mathcal{A}_{\mathcal{T}_2,\Sigma}$ that has been introduced to prove tractability of $\mathsf{iWtn}^{\mathsf{rhs}}_\Sigma(\mathcal{T}_1, \mathcal{T}_2)$. Recall that $A \in \mathsf{iWtn}^{\mathsf{rhs}}_\Sigma(\mathcal{T}_1, \mathcal{T}_2)$ iff $(\mathcal{T}_1, \mathcal{A}_{\mathcal{T}_2,\Sigma}) \models A(\xi_B)$ for some $B \in \mathsf{non\text{-}conj}_{\mathcal{T}_2}(A)$ (cf. Lemma 45). Not all $A$ satisfying this condition are in $\mathsf{cWtn}^{\mathsf{rhs}}_\Sigma(\mathcal{T}_1, \mathcal{T}_2)$ since the ABox $\mathcal{A}_{\mathcal{T}_2,\Sigma}$ cannot always be "captured" by a set of $\mathcal{EL}$-concepts (cf. Example 4). Our modification of $\mathcal{A}_{\mathcal{T}_2,\Sigma}$ is motivated by the observation that if an ABox $\mathcal{A}$ does not contain any individual in the range of two distinct role names, then $\mathcal{EL}$-concepts rather than $\mathcal{C}^{\mathsf{ran}}$-concepts are sufficient to capture the consequences of the ABox. Thus, we are going to modify $\mathcal{A}_{\mathcal{T}_2,\Sigma}$ in a minimal way so that the resulting ABox does *not* contain any individual name in the range of two distinct role names.

**Definition 51.** *An ABox $\mathcal{A}$ is* role-splitting *if there is no pair of assertions $r(a, c), s(b, c) \in \mathcal{A}$, for individual names $a, b, c$ and distinct role names $r$, $s$.*

The following lemma states the main property of role-splitting ABoxes.

**Lemma 52.** *Let $\mathcal{T}_1$ and $\mathcal{T}_2$ be normalised $\mathcal{ELH}^r$-terminologies, $\Sigma$ a signature with $A \in \Sigma$ and let $\mathcal{A}$ be a role-splitting $\Sigma$-ABox such that $(\mathcal{T}_1, \mathcal{A}) \models A(a)$ and $(\mathcal{T}_2, \mathcal{A}) \not\models A(a)$. Then $A \in \mathsf{cWtn}^{\mathsf{rhs}}_\Sigma(\mathcal{T}_1, \mathcal{T}_2)$.*

*Proof.* By Lemma 36, there exists $n \geq 0$ such that $\mathcal{T}_1 \models C^{n,\mathsf{ran}}_{\mathcal{A},a} \sqsubseteq A$ and $\mathcal{T}_2 \not\models C^{n,\mathsf{ran}}_{\mathcal{A},a} \sqsubseteq A$. Assume first that there does not exist $b \in \mathsf{obj}(\mathcal{A})$ and $r \in \Sigma$ such that $r(b, a) \in \mathcal{A}$. Then, by definition and since $\mathcal{A}$ is role-splitting, $\mathsf{ran}(r)$ only occurs in $C^{n,\mathsf{ran}}_{\mathcal{A},a}$ in the direct scope of the existential restriction $\exists r$. Hence $C^{n,\mathsf{ran}}_{\mathcal{A},a}$ is equivalent to an $\mathcal{EL}_\Sigma$-concept, and we are done. Now assume there exists $r(b, a) \in \mathcal{A}$. Then, again since $\mathcal{A}$ is role-splitting, $C^{n,\mathsf{ran}}_{\mathcal{A},a}$ is equivalent to a concept $\mathsf{ran}(r) \sqcap C$, where $C$ is an $\mathcal{EL}_\Sigma$-concept. In this case $\mathcal{T}_1 \models \mathsf{ran}(r) \sqcap C \sqsubseteq A$ and $\mathcal{T}_2 \not\models \mathsf{ran}(r) \sqcap C \sqsubseteq A$, as required. □

For a $\Sigma$-ABox $\mathcal{A}$ such that $\mathsf{sig}(\mathcal{A}) \cap \mathsf{N_R} \neq \emptyset$, we define its *role-splitting unfolding* $\mathcal{A}^{\upuparrows}$ with individuals $\{ a_r \mid a \in \mathsf{obj}(\mathcal{A}), r \in \mathsf{sig}(\mathcal{A}) \cap \mathsf{N_R} \}$ by setting

$$\mathcal{A}^{\upuparrows} = \{ A(a_r) \mid A(a) \in \mathcal{A}, r \in \mathsf{sig}(\mathcal{A}) \cap \mathsf{N_R} \} \cup \{ r(a_s, b_r) \mid r(a,b) \in \mathcal{A}, s \in \mathsf{sig}(\mathcal{A}) \cap \mathsf{N_R} \}.$$

**Example 53.** Consider $\mathcal{T}_1 = \{\mathsf{ran}(r) \sqsubseteq A_1, \mathsf{ran}(s) \sqsubseteq A_2, B \equiv A_1 \sqcap A_2\}$, $\mathcal{T}_2 = \emptyset$, $\Sigma = \{r, s, B\}$ and $\mathcal{A} = \{r(a,c), s(b,c)\}$ from Example 4. We have $(\mathcal{T}_1, \mathcal{A}) \models B(c)$ but $(\mathcal{T}_2, \mathcal{A}) \not\models B(c)$. Notice that the role-splitting unfolding $\mathcal{A}^{\upuparrows} = \{r(a_r, c_r), r(a_s, c_r), s(b_r, c_s), s(b_s, c_s)\}$ does not contain any individual in the range of more than one role and $(\mathcal{T}_1, \mathcal{A}^{\upuparrows}) \not\models B(c_r)$, $(\mathcal{T}_1, \mathcal{A}^{\upuparrows}) \not\models B(c_s)$.

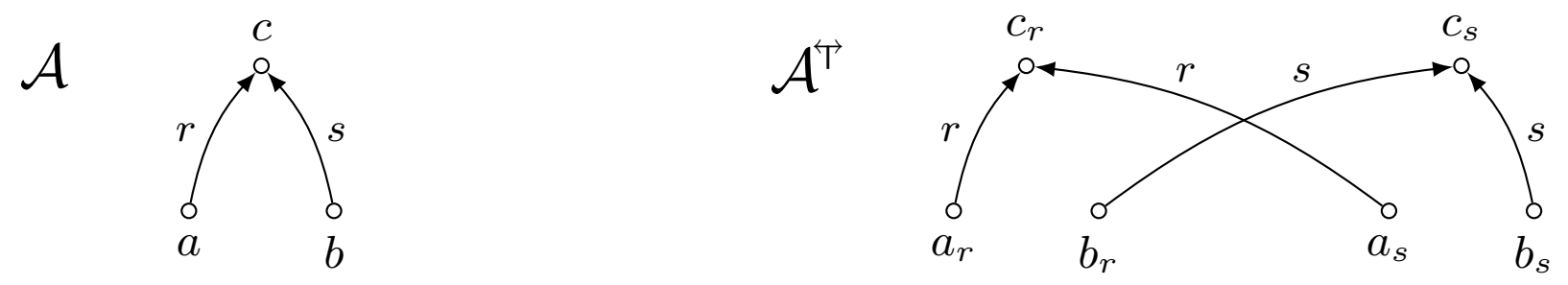

We apply the role-splitting unfolding to the ABox $\mathcal{A}_{\mathcal{T},\Sigma}$ from Figure 6. The following result is the concept version of Lemma 44 and is proved in the appendix by a reduction to Lemma 44. The ABox $\mathcal{A}_C$ corresponding to an $\mathcal{EL}$-concept $C$ has been introduced before Lemma 36. For simplicity, we consider signatures $\Sigma$ containing at least one role name.

**Lemma 54.** *For every normalised $\mathcal{ELH}^r$-terminology $\mathcal{T}$, signature $\Sigma$ such that $\Sigma \cap \mathsf{N_R} \neq \emptyset$, concept name $A$ that is non-conjunctive in $\mathcal{T}$, role name $r \in \Sigma$, and $\mathcal{EL}_\Sigma$-concepts $C$ the following conditions are equivalent for $D = C$ and $D = \mathsf{ran}(r) \sqcap C$:*

- *$\mathcal{T} \not\models D \sqsubseteq A$;*
- *for some $r \in \Sigma \cap \mathsf{N_R}$, $(\xi_A)_r \in \mathsf{obj}(\mathcal{A}^{\upuparrows}_{\mathcal{T},\Sigma})$ and $(\mathcal{A}_D, a_D) \leq^{\mathsf{ran}}_\Sigma (\mathcal{A}^{\upuparrows}_{\mathcal{T},\Sigma}, (\xi_A)_r)$.*

The following lemma can now be proved similarly to Lemma 45, using Lemma 54 instead of Lemma 44.

**Lemma 55.** *Let $\mathcal{T}_1$ and $\mathcal{T}_2$ be normalised $\mathcal{ELH}^r$-terminologies, $\Sigma$ a signature with $A \in \Sigma$ such that $\Sigma \cap \mathsf{N_R} \neq \emptyset$. Then the following conditions are equivalent:*

- *$A \in \mathsf{cWtn}^{\mathsf{rhs}}_\Sigma(\mathcal{T}_1, \mathcal{T}_2)$;*
- *there exists $r \in \Sigma$ such that $(\mathcal{T}_1, \mathcal{A}^{\upuparrows}_{\mathcal{T}_2,\Sigma}) \models A((\xi_B)_r)$ for some $B \in \mathsf{non\text{-}conj}_{\mathcal{T}_2}(A)$.*

*Proof.* Assume that $A \in \mathsf{cWtn}^{\mathsf{rhs}}_\Sigma(\mathcal{T}_1, \mathcal{T}_2)$. Then, there either exists an $\mathcal{EL}_\Sigma$-concept $C$ with $\mathcal{T}_1 \models C \sqsubseteq A$ and $\mathcal{T}_2 \not\models C \sqsubseteq A$, or there additionally exists $r \in \Sigma$ such that $\mathcal{T}_1 \models \mathsf{ran}(r) \sqcap C \sqsubseteq A$ and $\mathcal{T}_2 \not\models \mathsf{ran}(r) \sqcap C \sqsubseteq A$. Hence, for $D = C$ and $D = \mathsf{ran}(r) \sqcap C$, respectively, $\mathcal{T}_2 \not\models D \sqsubseteq B$, for some $B \in \mathsf{non\text{-}conj}_{\mathcal{T}_2}(A)$, and by Lemma 54, there exists $r \in \Sigma$ with $(\xi_A)_r \in \mathsf{obj}(\mathcal{A}^{\upuparrows}_{\mathcal{T},\Sigma})$ and $(\mathcal{A}_D, a_D) \leq^{\mathsf{ran}}_\Sigma (\mathcal{A}^{\upuparrows}_{\mathcal{T}_2,\Sigma}, (\xi_B)_r)$. But then, by Lemma 42 $(\mathcal{T}_1, \mathcal{A}^{\upuparrows}_{\mathcal{T}_2,\Sigma}) \models A((\xi_B)_r)$ as $(\mathcal{T}_1, \mathcal{A}_D) \models A(a_D)$ holds by Lemma 36.

For the converse direction, it is easy to see that $(\xi_B)_r \in \mathsf{obj}(\mathcal{A}^{\upuparrows}_{\mathcal{T}_2,\Sigma})$, $\xi_B \in \mathsf{obj}(\mathcal{A}_{\mathcal{T}_2,\Sigma})$, and $(\mathcal{A}^{\upuparrows}_{\mathcal{T}_2,\Sigma}, (\xi_B)_r) \leq^{\mathsf{ran}}_\Sigma (\mathcal{A}_{\mathcal{T}_2,\Sigma}, \xi_B)$, which implies that $(\mathcal{T}_2, \mathcal{A}^{\upuparrows}_{\mathcal{T}_2,\Sigma}) \not\models A((\xi_B)_r)$ by Lemma 44. Consequently, we obtain $A \in \mathsf{cWtn}^{\mathsf{rhs}}_\Sigma(\mathcal{T}_1, \mathcal{T}_2)$ by applying Lemma 52 and by using the fact that the ABox $\mathcal{A}^{\upuparrows}_{\mathcal{T}_2,\Sigma}$ is role-splitting. $\square$

Finally, we obtain the tractability result.

**Theorem 56.** *Let $\mathcal{T}_1$ and $\mathcal{T}_2$ be $\mathcal{ELH}^r$-terminologies and $\Sigma$ a signature. Then the set $\mathsf{cWtn}^{\mathsf{rhs}}_{\Sigma}(\mathcal{T}_1, \mathcal{T}_2)$ can be computed in polynomial time.*

*Proof.* If $\Sigma \cap \mathsf{N_R} = \emptyset$, then $\mathsf{cWtn}^{\mathsf{rhs}}_{\Sigma}(\mathcal{T}_1, \mathcal{T}_2) = \mathsf{iWtn}^{\mathsf{rhs}}_{\Sigma}(\mathcal{T}_1, \mathcal{T}_2)$, which can be computed in polynomial time by Theorem 46.

Otherwise $\Sigma \cap \mathsf{N_R} \neq \emptyset$ and the result follows from Lemma 55 and the fact that $\mathcal{A}^{\leftrightarrows}_{\mathcal{T}_2,\Sigma}$ can be constructed in polynomial time in the size of $\mathcal{T}_2$. □

## 7. $\mathcal{ELH}^r$-Query Difference

To investigate the query difference between $\mathcal{ELH}^r$-terminologies, we introduce the language $\mathcal{EL}^{\mathsf{ran},\sqcap,u}$ that extends $\mathcal{EL}^{\mathsf{ran}}$ with the universal role and intersections of roles. We show that concept differences in $\mathcal{EL}^{\mathsf{ran},\sqcap,u}$ correspond to query differences in $\mathcal{ELH}^r$. For $\mathcal{EL}^{\mathsf{ran},\sqcap,u}$ we can prove an analogue of Theorem 40, which states that any inclusion in the concept difference "contains" an inclusion in which either the left-hand side or the right-hand side is atomic. Using the correspondence between concept difference in $\mathcal{EL}^{\mathsf{ran},\sqcap,u}$ and query difference in $\mathcal{ELH}^r$ we then obtain a meaningful definition of a succinct representation of the query difference $\mathsf{qDiff}_{\Sigma}(\mathcal{T}_1, \mathcal{T}_2)$. Finally, we provide polynomial-time algorithms deciding $\Sigma$-query inseparability and computing the succinct representation of the query difference.

### 7.1 $\mathcal{EL}^{\mathsf{ran},\sqcap,u}$-Concept Difference

We start this section by defining the language $\mathcal{EL}^{\mathsf{ran},\sqcap,u}$.

**Definition 57** ($\mathcal{EL}^{\mathsf{ran},\sqcap,u}$)**.** Let $u$ (the universal role) be a fresh logical symbol. $\mathcal{C}^{\sqcap,u}$-concepts are constructed using the following syntax rule

$$C := \quad A \quad | \quad C \sqcap D \quad | \quad \exists R.C \quad | \quad \exists u.C,$$

where $A \in \mathsf{N_C}$, $C, D$ range over $\mathcal{C}^{\sqcap,u}$-concepts and $R = r_1 \sqcap \cdots \sqcap r_n$ with $r_1, \ldots, r_n \in \mathsf{N_R}$ for some $n \geq 1$. The set of $\mathcal{EL}^{\mathsf{ran},\sqcap,u}$-inclusions consists of concept inclusions $C \sqsubseteq D$ and role inclusions $r \sqsubseteq s$, where $C$ is a $\mathcal{C}^{\mathsf{ran}}$-concept, $D$ a $\mathcal{C}^{\sqcap,u}$-concept, and $r, s \in \mathsf{N_R}$.

The semantics of the additional constructors is straightforward by setting, for any interpretation $\mathcal{I}$,

- $(r_1 \sqcap \cdots \sqcap r_n)^{\mathcal{I}} = r^{\mathcal{I}_1} \cap \cdots \cap r_n^{\mathcal{I}}$;
- $u^{\mathcal{I}} = \Delta^{\mathcal{I}} \times \Delta^{\mathcal{I}}$.

Note that we regard the universal role $u$ as a logical symbol; i.e., $u \notin \mathsf{N_R}$ and $\mathsf{sig}(\exists u.C) = \mathsf{sig}(C)$ for any concept $C$. Assuming that $u$ is a logical symbol reflects the fact that its first-order translation uses no non-logical symbols. For example, the signature of the first-order translation $\exists x.A(x)$ of $\exists u.A$ does not contain any non-logical symbols with the exception of $A$ itself.

It will be convenient to decompose $\mathcal{C}^{\sqcap,u}$-concepts. The set of $\mathcal{C}^{\sqcap}$-concepts is defined as the set of all $\mathcal{C}^{\sqcap,u}$-concepts without the universal role. Every $\mathcal{C}^{\sqcap,u}$-concept $C$ is equivalent to a concept of the form $D_0 \sqcap \exists u.D_1 \sqcap \cdots \sqcap \exists u.D_k$, where $D_0, \dots, D_k$ are $\mathcal{C}^{\sqcap}$-concepts. To see this, observe that any concept $C$ with a subconcept $\exists u.D$ is equivalent to $\exists u.D \sqcap C'$, where $C'$ is obtained from $C$ by replacing all occurrences of $\exists u.D$ by $\top$. For example, $A \sqcap \exists r.(B \sqcap \exists u.E)$ is equivalent to the concept $\exists u.E \sqcap A \sqcap \exists r.(B \sqcap \top)$.

In the following we denote by $\mathcal{C}^{\sqcap,u}_{\Sigma}$ ($\mathcal{C}^{\sqcap}_{\Sigma}$) the set of all $\mathcal{C}^{\sqcap,u}$ ($\mathcal{C}^{\sqcap}$) concepts whose signature is contained in $\Sigma$.

Clearly, every $\mathcal{EL}^{\mathsf{ran}}$-inclusion is an $\mathcal{EL}^{\mathsf{ran},\sqcap,u}$-inclusion. In addition, role conjunctions and the universal role in $\mathcal{EL}^{\mathsf{ran},\sqcap,u}$-inclusions can be used to capture differences between $\mathcal{ELH}^r$-TBoxes that cannot be captured by $\mathcal{ELH}^r$-inclusions.

**Example 58.** We first reconsider Example 8. Recall that

$$\mathcal{T}_1 = \emptyset, \quad \mathcal{T}_2 = \{A \sqsubseteq \exists r.B\}, \quad \Sigma = \{A, B\}.$$

Then $\mathcal{T}_2 \models A \sqsubseteq \exists u.B$ but $\mathcal{T}_1 \not\models A \sqsubseteq \exists u.B$ and, as the universal role is regarded as a logical symbol, $\mathsf{sig}(A \sqsubseteq \exists u.B) \subseteq \Sigma$. Thus, by employing the universal role $\mathcal{EL}^{\mathsf{ran},\sqcap,u}$ we can simulate the query difference $(\{A(a)\}, \exists x.B(x))$ using the subsumption $A \sqsubseteq \exists u.B$.

Second, we reconsider Example 9. Recall that

$$\mathcal{T}_1 = \{A \sqsubseteq \exists s.\top, s \sqsubseteq r_1, s \sqsubseteq r_2\}, \quad \mathcal{T}_2 = \{A \sqsubseteq \exists r_1.\top \sqcap \exists r_2.\top\}, \quad \Sigma = \{A, r_1, r_2\}.$$

Then $\mathcal{T}_1 \models A \sqsubseteq \exists(r_1 \sqcap r_2).\top$ and $\mathcal{T}_2 \not\models A \sqsubseteq \exists(r_1 \sqcap r_2).\top$. Thus, we can simulate the query difference $(\{A(a)\}, \exists x.(r_1(a,x) \wedge r_2(a,x)))$ using the subsumption $A \sqsubseteq \exists(r_1 \sqcap r_2).\top$.

We introduce the appropriate notion of $\Sigma$-concept difference for $\mathcal{EL}^{\mathsf{ran},\sqcap,u}$.

**Definition 59** ($\mathcal{EL}^{\mathsf{ran},\sqcap,u}_{\Sigma}$-difference)**.** *The $\mathcal{EL}^{\mathsf{ran},\sqcap,u}_{\Sigma}$-difference between $\mathcal{ELH}^r$-TBoxes $\mathcal{T}_1$ and $\mathcal{T}_2$ is the set $\mathsf{cDiff}^{\mathsf{ran},\sqcap,u}_{\Sigma}(\mathcal{T}_1, \mathcal{T}_2)$ of all $\mathcal{EL}^{\mathsf{ran},\sqcap,u}_{\Sigma}$-inclusions $\alpha$ such that $\mathcal{T}_1 \models \alpha$ and $\mathcal{T}_2 \not\models \alpha$.*

We now extend Lemma 39 for concepts that use the universal role or conjunctions of roles.

**Lemma 60.** *Let $\mathcal{T}$ be an $\mathcal{ELH}^r$-terminology and $\exists R.D$ a $\mathcal{C}^{\sqcap}$-concept with $R = t_1 \sqcap \cdots \sqcap t_q$ a conjunction of role names. Assume*

$$\mathcal{T} \models \bigsqcap_{1 \leq i \leq l} \mathsf{ran}(s_i) \sqcap \bigsqcap_{1 \leq j \leq n} A_j \sqcap \bigsqcap_{1 \leq k \leq m} \exists r_k.C_k \sqsubseteq \exists R.D,$$

*where $C_k$, $1 \leq k \leq m$, are $\mathcal{C}^{\mathsf{ran}}$-concepts and $l, m, n \geq 0$. Then at least one of the following conditions holds:*

- **(e1$_{\sqcap}$)** *there exists $r_k$, $1 \leq k \leq m$, such that $r_k \sqsubseteq_{\mathcal{T}} t_1, \dots, r_k \sqsubseteq_{\mathcal{T}} t_q$, and $\mathcal{T} \models C_k \sqcap \mathsf{ran}(r_k) \sqsubseteq D$;*
- **(e2$_{\sqcap}$)** *there exists $A_j$, $1 \leq j \leq n$, such that $\mathcal{T} \models A_j \sqsubseteq \exists R.D$;*
- **(e3$_{\sqcap}$)** *there exists $r_k$, $1 \leq k \leq m$, such that $\mathcal{T} \models \exists r_k.\top \sqsubseteq \exists R.D$;*
- **(e4$_{\sqcap}$)** *there exists $s_i$, $1 \leq i \leq l$, such that $\mathcal{T} \models \mathsf{ran}(s_i) \sqsubseteq \exists R.D$.*

*If $u$ is the universal role and $\mathcal{T} \models C \sqsubseteq \exists u.D$, where $C$ is a $\mathcal{C}^{\mathsf{ran}}$-concept and $D$ is a $\mathcal{C}^{\sqcap}$-concept, then at least one of the following holds:*

**($\mathbf{e1}_u$)** *there exists a subconcept $\exists r.C'$ of $C$ such that $\mathcal{T} \models C' \sqcap \mathsf{ran}(r) \sqsubseteq D$;*

**($\mathbf{e2}_u$)** *there exists a concept name $A$ in $C$ such that $\mathcal{T} \models A \sqsubseteq \exists u.D$;*

**($\mathbf{e3}_u$)** *there exists a role name $r$ in $C$ such that $\mathcal{T} \models \exists r.\top \sqsubseteq \exists u.D$;*

**($\mathbf{e4}_u$)** *there exists a role name $r$ in $C$ such that $\mathcal{T} \models \mathsf{ran}(r) \sqsubseteq \exists u.D$;*

**($\mathbf{e5}_u$)** *$\mathcal{T} \models C \sqsubseteq D$;*

**($\mathbf{e6}_u$)** *there exists a subconcept $(\mathsf{ran}(r) \sqcap C')$ of $C$ such that $\mathcal{T} \models \exists r.C' \sqsubseteq D$.*

**Theorem 61** (Primitive witnesses for $\mathcal{EL}^{\mathsf{ran},\sqcap,u}$). *Let $\mathcal{T}_1$ and $\mathcal{T}_2$ be $\mathcal{ELH}^r$-terminologies and $\Sigma$ a signature. If $\varphi \in \mathsf{cDiff}_\Sigma^{\mathsf{ran},\sqcap,u}(\mathcal{T}_1, \mathcal{T}_2)$, then either there exist $\{r, s\} \subseteq \mathsf{sig}(\varphi)$ with $r \sqsubseteq s \in \mathsf{cDiff}_\Sigma^{\mathsf{ran},\sqcap,u}(\mathcal{T}_1, \mathcal{T}_2)$ or $\varphi$ is of the form $C \sqsubseteq D$, and one of*

1. *$C' \sqsubseteq A$*

2. *$A \sqsubseteq D'$, $\exists r.\top \sqsubseteq D'$ or $\mathsf{ran}(r) \sqsubseteq D'$*

*is a member of $\mathsf{cDiff}_\Sigma^{\mathsf{ran},\sqcap,u}(\mathcal{T}_1, \mathcal{T}_2)$, where $A \in \mathsf{sig}(\varphi)$ is a concept name, $r \in \mathsf{sig}(\varphi)$ is a role name, $C'$ is a $\mathcal{C}^{\mathsf{ran}}$-concept, $D'$ is a $\mathcal{C}^{\sqcap,u}$-concept, and $\mathsf{sig}(C'), \mathsf{sig}(D') \subseteq \mathsf{sig}(\varphi)$.*

*Proof.* Let $C \sqsubseteq D \in \mathsf{cDiff}_\Sigma^{\mathsf{ran},\sqcap,u}(\mathcal{T}_1, \mathcal{T}_2)$, where $C$ is a $\mathcal{C}^{\mathsf{ran}}$-concept and $D$ a $\mathcal{C}^{\sqcap,u}$-concept. We prove the result by induction on the structure of $D$. The proof is very similar to the proof of Theorem 40 and so we consider the case $D = \exists u.D_1$ only. Let $C = \bigsqcap_{1\leq i\leq l} \mathsf{ran}(s_i) \sqcap \bigsqcap_{1\leq j\leq n} A_j \sqcap \bigsqcap_{1\leq k\leq m} \exists r_k.C_k$. Then, by Lemma 60, one of ($\mathbf{e1}_u$)–($\mathbf{e6}_u$) holds.

Cases ($\mathbf{e2}_u$)–($\mathbf{e4}_u$) directly entail the existence of an inclusion from Point 2 of the theorem. In case ($\mathbf{e1}_u$) there exists a subconcept $\exists r.C'$ of $C$ such that $\mathcal{T}_1 \models C' \sqcap \mathsf{ran}(r) \sqsubseteq D_1$. We have that $\mathcal{T}_2 \not\models C' \sqcap \mathsf{ran}(r) \sqsubseteq D_1$ as otherwise we have $\mathcal{T}_2 \models \exists r.C' \sqsubseteq \exists r.D_1$, i.e. $\mathcal{T}_2 \models C \sqsubseteq D$ would hold. Thus, $C' \sqcap \mathsf{ran}(r) \sqsubseteq D_1 \in \mathsf{cDiff}_\Sigma^{\mathsf{ran},\sqcap,u}(\mathcal{T}_1, \mathcal{T}_2)$. We can apply the induction hypothesis to $D_1$ and infer that an inclusion from Point 1 or Point 2 exists.

Similarly, for case ($\mathbf{e5}_u$), we have $C \sqsubseteq D_1 \in \mathsf{cDiff}_\Sigma^{\mathsf{ran},\sqcap,u}(\mathcal{T}_1, \mathcal{T}_2)$ as otherwise $\mathcal{T}_2 \models C \sqsubseteq D_1$, i.e. $\mathcal{T}_2 \models C \sqsubseteq D$ due to $D = \exists u.D_1$. By applying the induction hypothesis to $D_1$, we obtain that an inclusion from Point 1 or Point 2 exists.

Finally, in case ($\mathbf{e6}_u$) there exists a subconcept $\mathsf{ran}(r) \sqcap C'$ of $C$ such that $\mathcal{T}_1 \models \exists r.C' \sqsubseteq D_1$. Observe first that for every model $\mathcal{I}$ of $\mathcal{T}_2$ and for every $d \in C^{\mathcal{I}}$, there exists $d' \in (\mathsf{ran}(r) \sqcap C')^{\mathcal{I}}$, which implies that there exists $d'' \in (\exists r.C')^{\mathcal{I}}$. If we now assume that $\mathcal{T}_2 \models \exists r.C' \sqsubseteq D_1$, it would follow that for every model $\mathcal{I}$ of $\mathcal{T}_2$ and for every $d \in C^{\mathcal{I}}$, there exists $d'' \in D_1^{\mathcal{I}}$, i.e. $\mathcal{T}_2 \models C \sqsubseteq \exists u.D_1$ would hold. We can infer that $\exists r.C' \sqsubseteq D_1 \in \mathsf{cDiff}_\Sigma^{\mathsf{ran},\sqcap,u}(\mathcal{T}_1, \mathcal{T}_2)$ and by applying the induction hypothesis to $D_1$, we conclude that an inclusion from Point 1 or Point 2 exists. □

### 7.2 Query Difference Witnesses

We start by connecting concept differences in $\mathcal{EL}^{\mathsf{ran},\sqcap,u}$ with query differences between $\mathcal{ELH}^r$-terminologies. The direction from query differences in $\mathcal{ELH}^r$ to concept differences in $\mathcal{EL}^{\mathsf{ran},\sqcap,u}$ is straightforward: observe that every assertion $C(a)$ with $C$ a $\mathcal{C}^{\sqcap,u}$-concept can be regarded as a Boolean conjunctive query $q_{C,a}$. For example, the assertion $(\exists u.A \sqcap \exists r.B)(a)$ is equivalent to the conjunctive query $\exists x \exists y.(A(x) \wedge r(a,y) \wedge B(y))$ (details of the translation are provided in the appendix). We obtain (where $\mathcal{A}_C$ is the ABox defined before Lemma 36):

**Lemma 62.** *For any two $\mathcal{ELH}^r$-TBoxes $\mathcal{T}_1$ and $\mathcal{T}_2$ and signature $\Sigma$, we have $C \sqsubseteq D \in \mathsf{cDiff}_\Sigma^{\mathsf{ran},\sqcap,u}(\mathcal{T}_1,\mathcal{T}_2)$ if, and only if, $(\mathcal{A}_C, q_{D,a_C}) \in \mathsf{qDiff}_\Sigma(\mathcal{T}_1,\mathcal{T}_2)$.*

In what follows we will not distinguish between an assertion $C(a)$ with $C$ a $\mathcal{C}^{\sqcap,u}$-concept and the conjunctive query $q_{C,a}$. It follows from Lemma 62 that if $\mathsf{qDiff}_\Sigma(\mathcal{T}_1,\mathcal{T}_2) = \emptyset$, then $\mathsf{cDiff}_\Sigma^{\mathsf{ran},\sqcap,u}(\mathcal{T}_1,\mathcal{T}_2) = \emptyset$.

We come to the (considerably more involved) direction from query differences to concept differences in $\mathcal{EL}^{\mathsf{ran},\sqcap,u}$. The following lemma provides a rather abstract description of how inclusions in $\mathsf{qDiff}_\Sigma(\mathcal{T}_1,\mathcal{T}_2)$ are reflected by members of $\mathsf{cDiff}_\Sigma^{\mathsf{ran},\sqcap,\mathsf{u}}(\mathcal{T}_1,\mathcal{T}_2)$ by stating that they are given in the same signature.

**Lemma 63.** *For any two $\mathcal{ELH}^r$-TBoxes $\mathcal{T}_1$ and $\mathcal{T}_2$ and signature $\Sigma$, if $\varphi \in \mathsf{qDiff}_\Sigma(\mathcal{T}_1,\mathcal{T}_2)$, then there exists $\varphi' \in \mathsf{cDiff}_\Sigma^{\mathsf{ran},\sqcap,\mathsf{u}}(\mathcal{T}_1,\mathcal{T}_2)$ with $\mathsf{sig}(\varphi') \subseteq \mathsf{sig}(\varphi)$.*

The interested reader can extract a more detailed description from the proof given in the appendix. The proof of Lemma 63 given in the appendix is model-theoretic and employs the close relationship between conjunctive query entailment and homomorphisms (Chandra & Merlin, 1977). The intuition behind the result, however, is rather straightforward: if $(\mathcal{T},\mathcal{A}) \models q[\vec{a}]$ for a conjunctive query $q(\vec{x}) = \exists \vec{y} \psi(\vec{x},\vec{y})$ and $\mathcal{ELH}^r$-TBox $\mathcal{T}$, then for every model $\mathcal{I}$ of $(\mathcal{T},\mathcal{A})$ there is a mapping $\pi$ from the variables $\vec{x}$ and $\vec{y}$ into $\Delta^\mathcal{I}$ such that $\vec{a}$ is a $\pi$-match of $q(\vec{x})$ and $\mathcal{I}$. $(\mathcal{T},\mathcal{A})$ has models that are essentially forest-shaped: they consist of tree-shaped models attached to the ABox individuals in $\mathcal{A}$ (cf. Lutz et al., 2009). In such forest-shaped models, the individuals from $\vec{y}$ that are not mapped to individuals in $\mathcal{A}$ are mapped to the trees attached to the ABox individuals. Such a mapping, however, exists already for a conjunctive query $q'$ such that $q$ is a homomorphic image of $q'$ and $q'$ is essentially forest-shaped: the individuals not mapped to ABox individuals form trees that are attached to the core of $q'$ that is mapped to the ABox individuals. In other words, we obtain $q'$ by partitioning $q$ into a core and into subsets that correspond to $\mathcal{C}^{\sqcap,u}$-concepts! Now, if there exists a $\Sigma$-ABox $\mathcal{A}$ and a conjunctive $\Sigma$-query $q(\vec{a})$ such that $(\mathcal{T}_2,\mathcal{A}) \models q[\vec{a}]$ and $(\mathcal{T}_1,\mathcal{A}) \not\models q[\vec{a}]$, then we find such a conjunctive $\Sigma$-query $q'$ with the same behaviour as $q$ that is essentially forest-shaped. From $(\mathcal{A},q')$ one can then obtain the required $\mathcal{EL}^{\mathsf{ran},\sqcap,u}$-inclusion $C \sqsubseteq D$, where $D$ captures some subtree of the query $q'$ (a $\mathcal{C}^{\sqcap,u}$-concept) and $C$ (a $\mathcal{C}^{\mathsf{ran}}$-concept) the ABox $\mathcal{A}$. The intuition for the last step is exactly the same as for Lemma 36.

We note that the result holds for general TBoxes and not only terminologies. From Lemma 63 and Theorem 61, we directly obtain the following description of primitive witnesses for query differences.

**Theorem 64** (Primitive witness for $\mathcal{ELH}^r$-query differences)**.** *Let $\mathcal{T}_1$ and $\mathcal{T}_2$ be $\mathcal{ELH}^r$-terminologies and $\Sigma$ a signature. If $\varphi \in \mathsf{qDiff}_\Sigma(\mathcal{T}_1,\mathcal{T}_2)$, then at least one of the following conditions holds (for some individual names $a,b$):*

1. *$(\{r(a,b)\}, s(a,b)) \in \mathsf{qDiff}_\Sigma(\mathcal{T}_1,\mathcal{T}_2)$, for some $r,s \in \mathsf{sig}(\varphi)$;*

2. $(\mathcal{A}, A(b)) \in \mathsf{qDiff}_\Sigma(\mathcal{T}_1, \mathcal{T}_2)$, *for some concept name* $A \in \mathsf{sig}(\varphi)$ *and ABox* $\mathcal{A}$ *with* $\mathsf{sig}(\mathcal{A}) \subseteq \mathsf{sig}(\varphi)$*;*

3. $(\mathcal{A}, D(b)) \in \mathsf{qDiff}_\Sigma(\mathcal{T}_1, \mathcal{T}_2)$, *for some singleton ABox* $\mathcal{A}$ *and* $\mathcal{C}^{\sqcap,u}$*-concept* $D$ *such that* $\mathsf{sig}(\mathcal{A}), \mathsf{sig}(D) \subseteq \mathsf{sig}(\varphi)$*.*

Observe that Theorem 64 coincides with Theorem 41 with the exception that in Point 3 the concept $D$ can now be a $\mathcal{C}^{\sqcap,u}$-concept. We can, therefore, define the following finite representation of the $\Sigma$-query difference. Assume $\mathcal{T}_1$ and $\mathcal{T}_2$ are given. Define the set

- $\mathsf{qWtn}^{\mathsf{R}}_\Sigma(\mathcal{T}_1, \mathcal{T}_2)$ of *role* $\Sigma$-query difference witnesses as the set of role $\Sigma$-instance difference witnesses; i.e., $\mathsf{qWtn}^{\mathsf{R}}_\Sigma(\mathcal{T}_1, \mathcal{T}_2) = \mathsf{iWtn}^{\mathsf{R}}_\Sigma(\mathcal{T}_1, \mathcal{T}_2)$;
- $\mathsf{qWtn}^{\mathsf{rhs}}_\Sigma(\mathcal{T}_1, \mathcal{T}_2)$ of *right-hand* $\Sigma$-query difference witnesses as the set of right-hand $\Sigma$-instance difference witnesses; i.e., $\mathsf{qWtn}^{\mathsf{rhs}}_\Sigma(\mathcal{T}_1, \mathcal{T}_2) = \mathsf{iWtn}^{\mathsf{rhs}}_\Sigma(\mathcal{T}_1, \mathcal{T}_2)$;
- $\mathsf{qWtn}^{\mathsf{lhs}}_\Sigma(\mathcal{T}_1, \mathcal{T}_2)$ of *left-hand* $\Sigma$-instance difference witnesses as the set of all $A \in \Sigma$ such that there exists a $\mathcal{C}^{\sqcap,u}$-concept $C$ with $(\{A(a)\}, C(a)) \in \mathsf{qDiff}^{\mathsf{lhs}}_\Sigma(\mathcal{T}_1, \mathcal{T}_2)$ and all $r \in \Sigma$ such that there exists a $\mathcal{C}^{\sqcap,u}$-concept $C$ such that $(\{r(a,b)\}, C(c)) \in \mathsf{qDiff}_\Sigma(\mathcal{T}_1, \mathcal{T}_2)$ for $c = a$ or $c = b$.

The set of all $\Sigma$-query difference witnesses is defined as

$$\mathsf{qWtn}_\Sigma(\mathcal{T}_1, \mathcal{T}_2) = (\mathsf{qWtn}^{\mathsf{R}}_\Sigma(\mathcal{T}_1, \mathcal{T}_2), \mathsf{qWtn}^{\mathsf{rhs}}_\Sigma(\mathcal{T}_1, \mathcal{T}_2), \mathsf{qWtn}^{\mathsf{lhs}}_\Sigma(\mathcal{T}_1, \mathcal{T}_2)).$$

By Theorem 64, $\mathsf{qWtn}_\Sigma(\mathcal{T}_1, \mathcal{T}_2) = (\emptyset, \emptyset, \emptyset)$ if, and only if, $\mathsf{qDiff}_\Sigma(\mathcal{T}_1, \mathcal{T}_2) = \emptyset$. Algorithms computing $\mathsf{qWtn}^{\mathsf{R}}_\Sigma(\mathcal{T}_1, \mathcal{T}_2)$ and $\mathsf{qWtn}^{\mathsf{rhs}}_\Sigma(\mathcal{T}_1, \mathcal{T}_2)$ have been presented in the section on instance difference. It thus remains to consider $\mathsf{qWtn}^{\mathsf{lhs}}_\Sigma(\mathcal{T}_1, \mathcal{T}_2)$.

### 7.3 Tractability of $\mathsf{qWtn}^{\mathsf{lhs}}_\Sigma(\mathcal{T}_1, \mathcal{T}_2)$

To prove tractability of $\mathsf{qWtn}^{\mathsf{lhs}}_\Sigma(\mathcal{T}_1, \mathcal{T}_2)$ we first capture the expressive power of $\mathcal{C}^{\sqcap,u}$-concepts using a stronger form of simulation between interpretations. Let $\mathcal{I}_1$ and $\mathcal{I}_2$ be interpretations. A $\Sigma$-simulation $S$ between $\mathcal{I}_1$ and $\mathcal{I}_2$ is called a *global intersection preserving* $\Sigma$*-simulation* if, in addition,

- for every $d \in \Delta^{\mathcal{I}_1}$ there exists a $d' \in \Delta^{\mathcal{I}_2}$ with $(d, d') \in S$;
- if $(d, e) \in S$, $d' \in \Delta^{\mathcal{I}_1}$, and $R = \{r \in \Sigma \mid (d, d') \in r^{\mathcal{I}_1}\} \neq \emptyset$, then there exists $e'$ with $(e, e') \in S$ and $(d', e') \in r^{\mathcal{I}_2}$ for all $r \in R$.

We write $(\mathcal{I}_1, d) \leq^{\sqcap}_\Sigma (\mathcal{I}_2, e)$ if there exists a global intersection preserving $\Sigma$-simulation $S$ between $\mathcal{I}_1$ and $\mathcal{I}_2$ such that $(d, e) \in S$.

**Lemma 65.** *Let* $\mathcal{I}_1$ *and* $\mathcal{I}_2$ *be* finite *interpretations,* $\Sigma$ *a signature,* $d \in \Delta^{\mathcal{I}_1}$*, and* $e \in \Delta^{\mathcal{I}_2}$*. Then*

$$(\mathcal{I}_1, d) \leq^{\sqcap}_\Sigma (\mathcal{I}_2, e) \quad \Leftrightarrow \quad \text{for all } C \in \mathcal{C}^{\sqcap,u}_\Sigma\text{: } d \in C^{\mathcal{I}_1} \Rightarrow e \in C^{\mathcal{I}_2}.$$

*It can be checked in polynomial time whether* $(\mathcal{I}_1, d) \leq^{\sqcap}_\Sigma (\mathcal{I}_2, e)$*.*

The proof is a straightforward extension of the proof of Lemma 29 and the polynomial-time algorithm deciding the existence of Σ-simulations.

We observe that Theorem 2 about the properties of the canonical model $\mathcal{I}_\mathcal{K}$ of a KB $\mathcal{K}$ can be extended to $\mathcal{C}^{\sqcap,u}$-concepts (in the appendix, the proof is given for $\mathcal{C}^{\sqcap,u}$-concepts as well). Namely, we have for all $\mathcal{C}^{\sqcap,u}$-concepts $C_0$:

- $\mathcal{K} \models C_0(a)$ if, and only if, $a^{\mathcal{I}_\mathcal{K}} \in C_0^{\mathcal{I}_\mathcal{K}}$.
- $\mathcal{T} \models C \sqcap D \sqsubseteq C_0$ if, and only if, $x_{C,D} \in C_0^{\mathcal{I}_\mathcal{K}}$.

It follows that for any concept name $A \in \Sigma$, we have

$$A \in \mathsf{qWtn}^{\mathsf{lhs}}_\Sigma(\mathcal{T}_1, \mathcal{T}_2) \quad \Leftrightarrow \quad (\mathcal{I}_{\mathcal{K}_1}, a) \not\leq^{\sqcap}_\Sigma (\mathcal{I}_{\mathcal{K}_2}, a),$$

where $\mathcal{K}_i = (\mathcal{T}_i, \mathcal{A})$ and $\mathcal{A} = \{A(a)\}$, for $i = 1, 2$. We also have for every role name $r \in \Sigma$ that

$$r \in \mathsf{qWtn}^{\mathsf{lhs}}_\Sigma(\mathcal{T}_1, \mathcal{T}_2) \quad \Leftrightarrow \quad (\mathcal{I}_{\mathcal{K}_1}, a) \not\leq^{\sqcap}_\Sigma (\mathcal{I}_{\mathcal{K}_2}, a) \text{ or } (\mathcal{I}_{\mathcal{K}_1}, b) \not\leq^{\sqcap}_\Sigma (\mathcal{I}_{\mathcal{K}_2}, b)$$

where $\mathcal{K}_i = (\mathcal{T}_i, \mathcal{A})$ and $\mathcal{A} = \{r(a, b)\}$, for $i = 1, 2$. Thus, we obtain the following tractability result:

**Theorem 66.** *Let $\mathcal{T}_1$ and $\mathcal{T}_2$ be $\mathcal{ELH}^r$-terminologies and $\Sigma$ a signature. Then the set $\mathsf{qWtn}^{\mathsf{lhs}}_\Sigma(\mathcal{T}_1, \mathcal{T}_2)$ can be computed in polynomial time.*

## 8. Implementation and Experiments

In this section, we describe an experimental evaluation of the theoretical work developed above. Our experiments employ the CEX2 tool.[4] In CEX2, we have implemented polynomial-time algorithms which, given acyclic $\mathcal{ELH}^r$-terminologies $\mathcal{T}_1$ and $\mathcal{T}_2$ and a signature $\Sigma$ as input, compute witnesses for the concept difference $\mathsf{cDiff}_\Sigma(\mathcal{T}_1, \mathcal{T}_2)$ and the instance difference $\mathsf{iDiff}_\Sigma(\mathcal{T}_1, \mathcal{T}_2)$.[5]

CEX2 is written in OCaml and the reasoner CB (Kazakov, 2009) is internally used as classification engine. In the implementation of CEX2, we have employed the algorithms developed in this paper. In more detail, for the instance difference case for acyclic $\mathcal{ELH}^r$-terminologies $\mathcal{T}_1$ and $\mathcal{T}_2$,

- to compute $\mathsf{iWtn}^{\mathsf{R}}_\Sigma(\mathcal{T}_1, \mathcal{T}_2)$, CEX2 performs a straightforward comparison of the role inclusion chains entailed by the terminologies $\mathcal{T}_1$ and $\mathcal{T}_2$;
- to compute $\mathsf{iWtn}^{\mathsf{rhs}}_\Sigma(\mathcal{T}_1, \mathcal{T}_2)$, CEX2 uses the NotWitness algorithm in Figure 7 and then employs Theorem 48;
- to compute $\mathsf{iWtn}^{\mathsf{lhs}}_\Sigma(\mathcal{T}_1, \mathcal{T}_2)$, CEX2 checks for the existence of a Σ-simulation between the canonical models (Theorem 49).

4. Available under an open-source license at `http://www.csc.liv.ac.uk/~michel/software/cex2/`
5. An extended version of CEX2 computing witnesses for the query difference $\mathsf{qDiff}_\Sigma(\mathcal{T}_1, \mathcal{T}_2)$ as well is presented by (Konev, Ludwig, & Wolter, 2012). In addition, Konev et al. describe experiments comparing query difference witnesses with concept and instance difference witnesses that are not presented in this paper.

The output for $\mathsf{iWtn}^{\mathsf{lhs}}_{\Sigma}(\mathcal{T}_1, \mathcal{T}_2)$ is partitioned into three sets:

- the set of *left-hand atomic* $\Sigma$-instance difference witnesses, $\mathsf{iWtn}^{\mathsf{lhs,A}}_{\Sigma}(\mathcal{T}_1, \mathcal{T}_2)$, which is defined as the set of all concept names $A \in \Sigma$ such that there exists an $\mathcal{EL}$-concept $C$ such that $(\{A(a)\}, C(a)\}) \in \mathsf{iDiff}_{\Sigma}(\mathcal{T}_1, \mathcal{T}_2)$ (equivalently $A \sqsubseteq C \in \mathsf{cDiff}_{\Sigma}(\mathcal{T}_1, \mathcal{T}_2)$);
- the set of *left-hand domain* $\Sigma$-instance difference witnesses, $\mathsf{iWtn}^{\mathsf{lhs,dom}}_{\Sigma}(\mathcal{T}_1, \mathcal{T}_2)$, which is defined as the set of all role names $r \in \Sigma$ such that there exists an $\mathcal{EL}$-concept $C$ with $(\{r(a, b)\}, C(a)) \in \mathsf{iDiff}_{\Sigma}(\mathcal{T}_1, \mathcal{T}_2)$ (equivalently, $\exists r.\top \sqsubseteq C \in \mathsf{cDiff}_{\Sigma}(\mathcal{T}_1, \mathcal{T}_2)$); and
- the set of *left-hand range* $\Sigma$-instance difference witnesses, $\mathsf{iWtn}^{\mathsf{lhs,ran}}_{\Sigma}(\mathcal{T}_1, \mathcal{T}_2)$, which is defined as the set of all role names $r \in \Sigma$ such that there exists an $\mathcal{EL}$-concept $C$ with $(\{r(a, b)\}, C(b)) \in \mathsf{iDiff}_{\Sigma}(\mathcal{T}_1, \mathcal{T}_2)$ (equivalently, $\mathsf{ran}(r) \sqsubseteq C \in \mathsf{cDiff}_{\Sigma}(\mathcal{T}_1, \mathcal{T}_2)$).

Obviously, it holds that:

$$\mathsf{iWtn}^{\mathsf{lhs}}_{\Sigma}(\mathcal{T}_1, \mathcal{T}_2) = \mathsf{iWtn}^{\mathsf{lhs,A}}_{\Sigma}(\mathcal{T}_1, \mathcal{T}_2) \cup \mathsf{iWtn}^{\mathsf{lhs,dom}}_{\Sigma}(\mathcal{T}_1, \mathcal{T}_2) \cup \mathsf{iWtn}^{\mathsf{lhs,ran}}_{\Sigma}(\mathcal{T}_1, \mathcal{T}_2).$$

For the concept difference case, recall that

$$\mathsf{cWtn}^{\mathsf{R}}_{\Sigma}(\mathcal{T}_1, \mathcal{T}_2) = \mathsf{iWtn}^{\mathsf{R}}_{\Sigma}(\mathcal{T}_1, \mathcal{T}_2), \quad \mathsf{cWtn}^{\mathsf{lhs}}_{\Sigma}(\mathcal{T}_1, \mathcal{T}_2) = \mathsf{iWtn}^{\mathsf{lhs}}_{\Sigma}(\mathcal{T}_1, \mathcal{T}_2),$$

and so we use the same algorithms as in the instance case. We also set

$$\mathsf{cWtn}^{\mathsf{lhs,X}}_{\Sigma}(\mathcal{T}_1, \mathcal{T}_2) = \mathsf{iWtn}^{\mathsf{lhs,X}}(\mathcal{T}_1, \mathcal{T}_2)$$

for $\mathsf{X} \in \{\mathsf{A}, \mathsf{dom}, \mathsf{ran}\}$. To compute $\mathsf{iWtn}^{\mathsf{rhs}}_{\Sigma}(\mathcal{T}_1, \mathcal{T}_2)$, CEX2 exploits that $\mathsf{cWtn}^{\mathsf{rhs}}_{\Sigma}(\mathcal{T}_1, \mathcal{T}_2) \subseteq \mathsf{iWtn}^{\mathsf{rhs}}_{\Sigma}(\mathcal{T}_1, \mathcal{T}_2)$ (Lemma 50) and first computes $\mathsf{iWtn}^{\mathsf{rhs}}_{\Sigma}(\mathcal{T}_1, \mathcal{T}_2)$ and then checks using a straightforward variant of the NotWitness algorithm for concept differences whether $A \in \mathsf{cWtn}^{\mathsf{rhs}}_{\Sigma}(\mathcal{T}_1, \mathcal{T}_2)$.

In the following three subsections we describe the experiments that we have conducted. The experimental settings were as follows. All programs were run on PCs equipped with an Intel Core 2 Duo E6400 CPU and 3 GiB of main memory. Version 2.0.1 of CEX2 was used.

### 8.1 Comparing Different Versions of Snomed CT

We applied CEX2 to compare a January 2009 (SM09a) and a July 2009 (SM09b) version of SNOMED CT. SM09a and SM09b contain 310013 and 307693 concept names, respectively. Both versions use the same 62 role names, and they contain role inclusions but no domain or range restrictions are present. Consequently, one can infer from Corollary 47 that $\mathsf{iWtn}_{\Sigma}(\mathsf{SM09}b, \mathsf{SM09}a) = \mathsf{cWtn}_{\Sigma}(\mathsf{SM09}b, \mathsf{SM09}a)$. In what follows we consider $\mathsf{cWtn}_{\Sigma}(\mathsf{SM09}b, \mathsf{SM09}a)$ only.

For our experiments we used signatures ranging over so called SNOMED CT *subsets*, which are employed in the UK for the deployment of SNOMED CT in specific areas. We compared SM09a with SM09b on 159 such signatures $\Sigma$ by computing $\mathsf{cWtn}_{\Sigma}(\mathsf{SM09}b, \mathsf{SM09}a)$ for each of these sets $\Sigma$. The considered signatures always contain all of the 62 SNOMED CT role names. The comparisons which resulted in a non-empty difference are reproduced in Table 2. In none of the cases, differences regarding role inclusions have been detected. In Table 2, the second column gives the number of concept names in the respective subset $\Sigma$, and the third and fifth column the number of concept witness differences. Observe that the number of differences does not correlate with the size of the considered signatures $\Sigma$, i.e. there exist signatures that are somewhat comparable in size, but induce a greatly varying number of difference witnesses (see e.g. the subsets "Diagnosis" and "Manumat").

In order to determine how many difference witnesses computed by CEX2 can be obtained from a straightforward comparison of the *class hierarchies* already, we have also computed the sets

$$\mathsf{clsWtn}^{\mathsf{lhs}}_{\Sigma}(\mathsf{SM09}b, \mathsf{SM09}a) = \{ A \in \Sigma \mid \exists B \in \Sigma \colon A \sqsubseteq B \in \mathsf{cDiff}_{\Sigma}(\mathsf{SM09b}, \mathsf{SM09a}) \}$$

and

$$\mathsf{clsWtn}^{\mathsf{rhs}}_{\Sigma}(\mathsf{SM09}b, \mathsf{SM09}a) = \{ B \in \Sigma \mid \exists A \in \Sigma \colon A \sqsubseteq B \in \mathsf{cDiff}_{\Sigma}(\mathsf{SM09b}, \mathsf{SM09a}) \}$$

for each of the considered comparison signatures $\Sigma$. The results that we have obtained are also depicted in Table 2. One can see that often a great number of differences cannot be detected by considering the classification difference only.

In the last three columns of Table 2, we give the CPU times required for computing all concept witnesses:

- first, the times are given when CEX2 is directly applied to the full terminologies SM09a and SM09b;
- second, the times are given when one first extracts $\Sigma$-modules using the module extraction tool MEX (Konev, Lutz, Walther, & Wolter, 2008) from SM09a and, respectively, SM09b and then applies CEX2 to the extracted $\Sigma$-modules. Observe that a $\Sigma$-module extracted by MEX is $\Sigma$-query (and, therefore, $\Sigma$-concept and $\Sigma$-instance) inseparable from the whole terminology. Thus, the computed concept witnesses are the same.
- finally, the times are given if, in addition to computing concept witnesses from the full terminologies SM09a and SM09b, CEX2 also computes examples of concept inclusions in the logical difference that explain the witnesses. We discuss this feature of CEX2 below.

One can observe that extracting MEX modules leads to a significant improvement of the performance of CEX2. Of course, if the signature is very large (e.g., for "Diagnosis" and "Finding"), the resulting modules are almost as large as SNOMED CT itself and the effect is less significant. Secondly, one can observe that the additional computation of example concept inclusions in the logical difference roughly doubles the times needed for the comparison.

Finally, to evaluate the practical feasibility of using the ABox approach to compute the sets $\mathsf{iWtn}^{\mathsf{rhs}}_{\Sigma}(\mathsf{SM09}b, \mathsf{SM09}a)$, we have implemented the computation of ABoxes $\mathcal{A}_{\mathcal{T},\Sigma}$ together with an ABox reasoning algorithm for checking the second condition of Lemma 45. We have then tested our implementation on the subsets $\Sigma$ of SNOMED CT used for evaluating the performance of CEX2. To limit the size of the ABoxes $\mathcal{A}_{\mathcal{T},\Sigma}$ and to speed up computations, we first computed modules using MEX. The results that we obtained are shown in Table 3. The size of the $\Sigma$-modules computed by MEX, i.e. $\mathcal{T}_1$ of $\mathsf{SM09b}$ and $\mathcal{T}_2$ of $\mathsf{SM09a}$, is shown in columns two and three, respectively. As expected from the definition of $\mathcal{A}_{\mathcal{T},\Sigma}$, one can observe that the number of concept and role membership assertions present in the ABoxes $\mathcal{A}_{\mathcal{T}_2,\Sigma}$ can grow very large, even for modules and signatures with only a few thousand concept names.

For 8 of the 41 considered subsets our implementation ran out of available physical memory (indicated by a time value '-') when all possible concept membership consequences of the ABox were to be computed. Overall, we observed the longest execution time of over 5 hours for the set "Specmatyp". In conclusion, one can see that a straightforward implementation of the ABox approach is practically useful "only" for terminologies and signatures of a few thousand concept names.

### 8.2 Comparing Different Versions of the NCI Thesaurus

We have also used the CEX2 tool to compare distinct versions of the NCI Thesaurus. Most distributed releases of the NCI Thesaurus contain language constructs which are not part of $\mathcal{ELH}^r$ (such as disjunction and value restriction). To obtain $\mathcal{ELH}^r$-terminologies, we have removed all inclusions that contain a non-$\mathcal{ELH}^r$ constructor from the original terminologies. Typically, this affected 5%-8% of the inclusions present in each of the distributed NCI versions. Most of the $\mathcal{ELH}^r$-versions generated in this way contain role inclusions as well as domain and range restrictions.

Similarly to the work of Gonçalves et al. (2011), we have compared 71 consecutive $\mathcal{ELH}^r$-versions of the NCI Thesaurus ranging between the versions 03.10J and 10.02d, with the exception of 05.03F and 05.04d, which could not be parsed correctly. Version 10.03h and some later versions of the NCI Thesaurus are not acyclic, and hence, they could not be handled by the CEX2 tool.

For any two consecutive versions $\mathrm{NCI}_n$ and $\mathrm{NCI}_{n+1}$ within the considered range, we computed the sets $\mathsf{cWtn}_\Sigma(\mathrm{NCI}_{n+1}, \mathrm{NCI}_n)$ and $\mathsf{iWtn}_\Sigma(\mathrm{NCI}_{n+1}, \mathrm{NCI}_n)$ on signatures $\Sigma = \mathsf{sig}(\mathrm{NCI}_n) \cap \mathsf{sig}(\mathrm{NCI}_{n+1})$. An overview of the set sizes for $\mathsf{cWtn}^{\mathsf{rhs}}_\Sigma(\mathrm{NCI}_{n+1}, \mathrm{NCI}_n)$ and $\mathsf{cWtn}^{\mathsf{lhs,A}}_\Sigma(\mathrm{NCI}_{n+1}, \mathrm{NCI}_n)$ that we obtained can be found in Figure 8. The comparisons are sorted chronologically along the $x$-axis according to the release dates of the NCI ontology versions, whereas the corresponding number of left-hand atomic difference witnesses or right-hand difference witnesses can be found on the $y$-axis. One can see the number of right-hand difference witnesses remained fairly low throughout the different versions. However, occasional spikes occurred in the number of left-hand atomic difference witnesses with a maximum value of 33487 for comparing the versions 05.01d and 05.03d. Moreover, in none of the comparisons except for those shown in Figure 9 left-hand role domain or left-hand role range difference witnesses were identified. Overall, no witnesses regarding role inclusions were detected and we found that for every two considered consecutive versions $\mathrm{NCI}_n$ and $\mathrm{NCI}_{n+1}$ on $\Sigma = \mathsf{sig}(\mathrm{NCI}_n) \cap \mathsf{sig}(\mathrm{NCI}_{n+1})$,

$$\mathsf{cWtn}_\Sigma(\mathrm{NCI}_{n+1}, \mathrm{NCI}_n) = \mathsf{iWtn}_\Sigma(\mathrm{NCI}_{n+1}, \mathrm{NCI}_n).$$

A running time of 140 seconds and 228 MiB of memory were required on average for computing witnesses and example inclusions for $\mathsf{iDiff}_\Sigma(\mathrm{NCI}_{n+1}, \mathrm{NCI}_n)$. Computing witnesses and example inclusions for $\mathsf{cDiff}_\Sigma(\mathrm{NCI}_{n+1}, \mathrm{NCI}_n)$ on average took 157 seconds and used 228 MiB of memory.

| Subset Name $\Sigma$ | $\lvert\Sigma \cap \mathsf{N_C}\rvert$ | $\lvert\mathsf{cWtn}^{\mathsf{rhs}}_{\Sigma}\rvert$ | $\lvert\mathsf{clsWtn}^{\mathsf{rhs}}_{\Sigma}\rvert$ | $\lvert\mathsf{cWtn}^{\mathsf{lhs,A}}_{\Sigma}\rvert$ | $\lvert\mathsf{clsWtn}^{\mathsf{lhs}}_{\Sigma}\rvert$ | Time (s) $\mathsf{cWtn}_{\Sigma}$ - from full ontologies | Time (s) $\mathsf{cWtn}_{\Sigma}$ - with module extraction | Time (s) $\mathsf{cWtn}_{\Sigma}$ with examples |
|---|---|---|---|---|---|---|---|---|
| Admin | 7684 | 7 | 5 | 29 | 7 | 358.51 | 9.89 | 654.12 |
| Adminproc | 3198 | 0 | 0 | 6 | 0 | 344.60 | 8.24 | 642.23 |
| Cdacarest | 355 | 1 | 1 | 1 | 1 | 337.91 | 6.76 | 556.41 |
| Crcareneur | 1640 | 28 | 8 | 197 | 13 | 399.57 | 15.58 | 704.21 |
| Crcareresp | 1082 | 72 | 18 | 262 | 64 | 377.36 | 12.24 | 680.51 |
| Devicetyp | 6539 | 26 | 26 | 22 | 22 | 369.20 | 8.01 | 589.81 |
| Diagimg | 4162 | 27 | 13 | 13 | 8 | 444.66 | 38.56 | 775.37 |
| Diagnosis | 75879 | 7410 | 881 | 12409 | 5406 | 844.26 | 486.53 | 2699.89 |
| Drgadrcon | 8009 | 131 | 131 | 47 | 47 | 1419.52 | 10.17 | 1708.49 |
| Endosfind | 178 | 0 | 0 | 13 | 0 | 363.23 | 7.86 | 662.67 |
| Endosproc | 73 | 1 | 1 | 5 | 3 | 352.84 | 7.30 | 573.66 |
| Epcream.6a | 403 | 0 | 0 | 3 | 0 | 337.41 | 7.39 | 631.51 |
| Epenema.7a | 25 | 0 | 0 | 3 | 0 | 337.42 | 6.86 | 556.31 |
| Epenema.7b | 6 | 0 | 0 | 2 | 0 | 337.50 | 6.76 | 629.57 |
| Epeye.4 | 223 | 0 | 0 | 6 | 0 | 337.53 | 7.20 | 1236.84 |
| Epiuds16 | 1 | 0 | 0 | 1 | 0 | 337.26 | 6.69 | 1233.89 |
| Famhist | 416 | 8 | 5 | 31 | 4 | 339.36 | 8.84 | 633.94 |
| Finding | 168383 | 11824 | 2497 | 31228 | 20063 | 1559.23 | 1366.08 | 5017.02 |
| Foodadrcon | 2378 | 11 | 11 | 15 | 14 | 481.20 | 7.74 | 1516.47 |
| Ffoodaller | 468 | 1 | 1 | 9 | 9 | 379.42 | 7.03 | 677.97 |
| Invest | 14839 | 1396 | 534 | 5549 | 5441 | 511.12 | 76.90 | 769.93 |
| Labinvest | 3904 | 61 | 45 | 2520 | 133 | 382.32 | 12.47 | 680.94 |
| Labinvmeth | 3794 | 103 | 81 | 3380 | 3374 | 367.20 | 10.70 | 1290.83 |
| Labisolate | 16313 | 150 | 150 | 661 | 661 | 671.36 | 14.14 | 1005.95 |
| Labmorph | 4854 | 32 | 32 | 45 | 45 | 858.11 | 8.28 | 1113.70 |
| Labspec | 1221 | 3 | 3 | 18 | 3 | 360.80 | 13.38 | 1272.51 |
| Labtopog | 27277 | 866 | 220 | 169 | 169 | 1947.19 | 38.05 | 4463.05 |
| Lifestyle | 13090 | 77 | 41 | 826 | 148 | 445.75 | 32.49 | 765.10 |
| Manumat | 90503 | 2 | 0 | 22 | 0 | 349.73 | 15.36 | 1224.92 |
| Nofoodall | 686 | 1 | 1 | 13 | 13 | 421.23 | 7.11 | 721.74 |
| Nonhuman | 1839 | 24 | 11 | 469 | 131 | 678.53 | 12.50 | 1907.70 |
| Pbcl | 5866 | 633 | 116 | 1342 | 402 | 395.27 | 12.18 | 1358.88 |
| Pbhllng | 1113 | 1 | 0 | 27 | 0 | 454.39 | 7.99 | 761.00 |
| Pf | 79 | 0 | 0 | 4 | 0 | 337.68 | 7.12 | 634.26 |
| Provadv | 1052 | 2 | 1 | 158 | 108 | 343.78 | 8.19 | 569.56 |
| Sf | 613 | 0 | 0 | 3 | 0 | 338.13 | 7.44 | 629.50 |
| Socpercir | 6786 | 8 | 8 | 2 | 2 | 366.14 | 8.99 | 1300.47 |
| Specmatyp | 8830 | 10 | 8 | 46 | 10 | 380.19 | 16.10 | 685.35 |
| Treatment | 43660 | 2419 | 1255 | 9251 | 8740 | 793.12 | 330.45 | 1315.23 |
| Vmp | 13667 | 2 | 0 | 22 | 0 | 342.70 | 12.95 | 1247.18 |
| Vtm | 2117 | 0 | 0 | 13 | 0 | 339.14 | 9.45 | 633.50 |

Table 2: Subset Comparisons for $\mathcal{T}_1 = \mathsf{SM09}b$ and $\mathcal{T}_2 = \mathsf{SM09}a$ Resulting in a Non-Empty Difference

| Subset Name $\Sigma$ | $\lvert\Sigma \cap \mathsf{N_C}\rvert$ | $\lvert\mathsf{sig}(\mathcal{T}_1) \cap \mathsf{N_C}\rvert$ | $\lvert\mathsf{sig}(\mathcal{T}_2) \cap \mathsf{N_C}\rvert$ | $\lvert\{ A(a) \mid A(a) \in \mathcal{A}_{\mathcal{T}_2,\Sigma} \}\rvert$ (in thousands) | $\lvert\{ r(a,b) \mid r(a,b) \in \mathcal{A}_{\mathcal{T}_2,\Sigma} \}\rvert$ (in thousands) | Time (s) |
|---|---|---|---|---|---|---|
| Admin | 7684 | 6746 | 6750 | 66942 | 1081 | 9291.75 |
| Adminproc | 3198 | 3071 | 3120 | 12352 | 480 | 1642.41 |
| Cdacarest | 355 | 322 | 323 | 148 | 52 | 3.86 |
| Crcareneur | 1640 | 6484 | 6375 | 8568 | 651 | 3110.07 |
| Crcareresp | 1082 | 5273 | 5206 | 4361 | 503 | 3689.22 |
| Devicetyp | 6539 | 3617 | 3619 | 43743 | 830 | 2381.00 |
| Diagimg | 4162 | 11007 | 11074 | 40817 | 1220 | 12503.63 |
| Diagnosis | 75879 | 156588 | 156441 | 8636801 | 14134 | - |
| Drgadrcon | 8009 | 8323 | 8361 | 70643 | 1095 | 2097.23 |
| Endosfind | 178 | 1487 | 1534 | 210 | 148 | 143.87 |
| Endosproc | 73 | 809 | 826 | 48 | 82 | 17.88 |
| Epcream.6a | 403 | 1425 | 1446 | 641 | 198 | 315.58 |
| Epenema.7a | 25 | 85 | 86 | 4 | 19 | 0.32 |
| Epenema.7b | 6 | 13 | 15 | 0 | 10 | 0.06 |
| Epeye.4 | 223 | 851 | 859 | 214 | 120 | 60.80 |
| Epiuds16 | 1 | 5 | 7 | 0 | 8 | 0.05 |
| Famhist | 416 | 3126 | 3136 | 1003 | 301 | 137.34 |
| Finding | 168383 | 323809 | 324400 | 41381927 | 30521 | - |
| Foodadrcon | 2378 | 2716 | 2723 | 6745 | 353 | 277.80 |
| Foodaller | 468 | 636 | 644 | 326 | 87 | 11.57 |
| Invest | 14839 | 42071 | 42559 | 504618 | 4224 | - |
| Labinvest | 3904 | 9308 | 9302 | 36048 | 1147 | 8632.09 |
| Labinvmeth | 3794 | 10132 | 10147 | 36738 | 1203 | 4131.02 |
| Labisolate | 16313 | 16281 | 16313 | 267268 | 2033 | 7785.59 |
| Labmorph | 4854 | 4575 | 4558 | 24538 | 628 | 1275.48 |
| Labspec | 1221 | 7106 | 7064 | 5566 | 570 | 646.59 |
| Labtopog | 27277 | 27118 | 27142 | 723594 | 3294 | - |
| Lifestyle | 13090 | 26233 | 26473 | 250140 | 2374 | - |
| Manumat | 90503 | 11605 | 11649 | 8851332 | 12127 | - |
| Nofoodall | 686 | 990 | 991 | 727 | 132 | 26.99 |
| Nonhuman | 1839 | 8728 | 8848 | 13698 | 926 | 10110.25 |
| Pbcl | 5866 | 8497 | 8793 | 64174 | 1357 | 16410.12 |
| Pbhllng | 1113 | 2488 | 2487 | 3083 | 344 | 933.00 |
| Pf | 79 | 386 | 389 | 37 | 59 | 3.86 |
| Provadv | 1052 | 3104 | 3014 | 3202 | 378 | 518.47 |
| Sf | 613 | 1856 | 1860 | 1332 | 270 | 249.26 |
| Socpercir | 6786 | 6757 | 6754 | 48627 | 889 | 5819.23 |
| Specmatyp | 8830 | 12928 | 12871 | 112252 | 1580 | 18306.56 |
| Treatment | 43660 | 111178 | 111612 | 3716810 | 10578 | - |
| Vmp | 13667 | 11972 | 12018 | 289683 | 2629 | - |
| Vtm | 2117 | 7655 | 7711 | 16540 | 970 | 2861.37 |

Table 3: Performance of the ABox Approach for Computing $\mathsf{iWtn}^{\mathsf{rhs}}_{\Sigma}(\mathsf{SM09}b, \mathsf{SM09}a)$

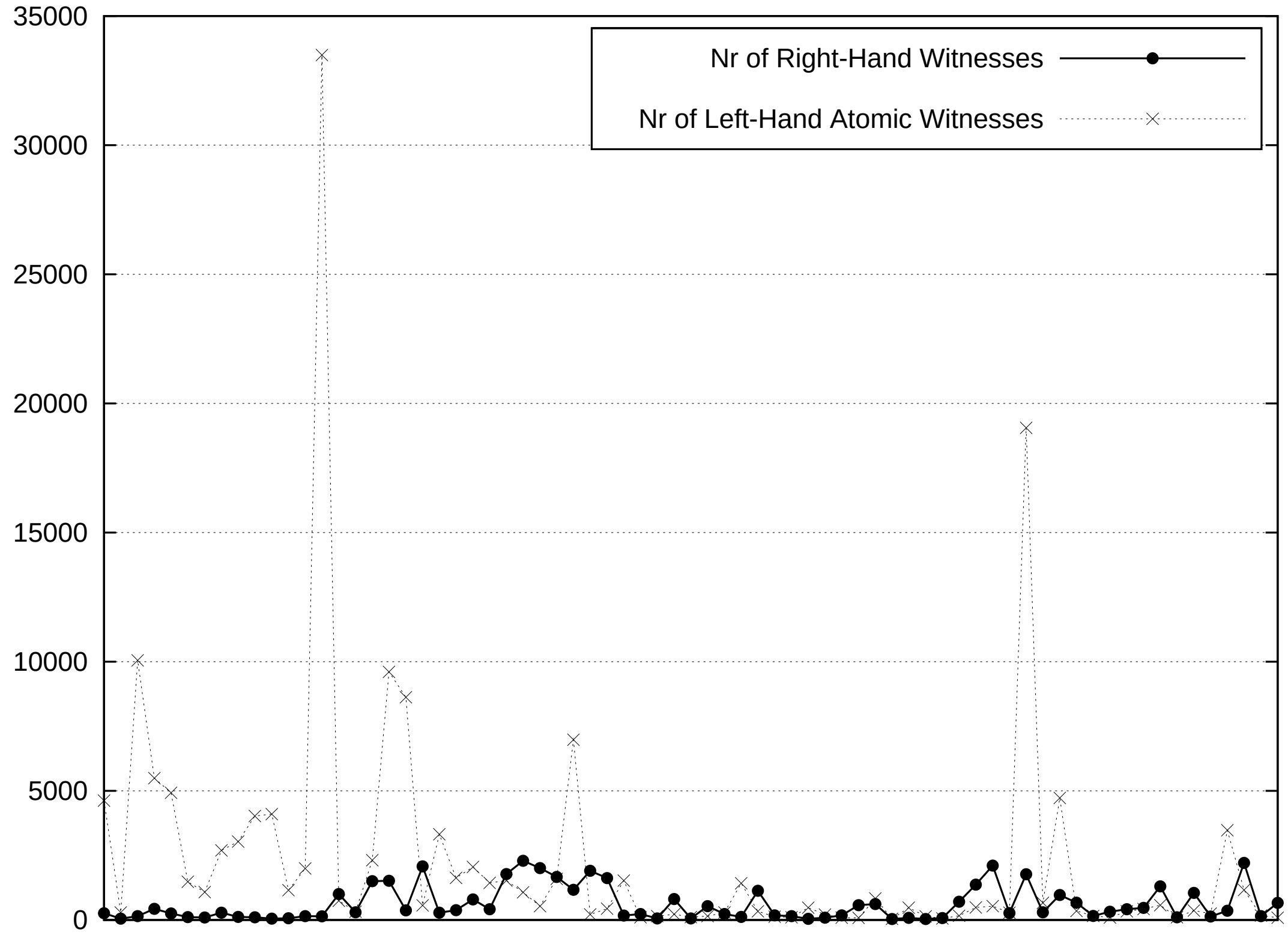

Figure 8: Sizes of $\mathsf{cWtn}^{\mathsf{rhs}}_{\Sigma}(\mathrm{NCI}_{n+1}, \mathrm{NCI}_n)$ and $\mathsf{cWtn}^{\mathsf{lhs,A}}_{\Sigma}(\mathrm{NCI}_{n+1}, \mathrm{NCI}_n)$ between Consecutive $\mathcal{ELH}^r$-versions $\mathrm{NCI}_n$ and $\mathrm{NCI}_{n+1}$ of the NCI Thesaurus

| $\mathcal{T}_1$ | $\mathcal{T}_2$ | $\lvert\Sigma \cap \mathsf{N_C}\rvert$ | $\lvert\Sigma \cap \mathsf{N_R}\rvert$ | $\lvert\mathsf{cWtn}^{\mathsf{rhs}}_{\Sigma}\rvert$ | $\lvert\mathsf{cWtn}^{\mathsf{lhs,A}}_{\Sigma}\rvert$ | $\lvert\mathsf{cWtn}^{\mathsf{lhs,dom}}_{\Sigma}\rvert$ | $\lvert\mathsf{cWtn}^{\mathsf{lhs,ran}}_{\Sigma}\rvert$ |
|---|---|---|---|---|---|---|---|
| 04.04j | 04.03n | 34245 | 76 | 252 | 4926 | 1 | 1 |
| 04.11a | 04.09a | 35976 | 91 | 106 | 4023 | 2 | 2 |
| 05.03d | 05.01d | 38020 | 92 | 138 | 33487 | 92 | 92 |
| 06.02d | 06.01c | 45582 | 113 | 419 | 1438 | 1 | 1 |
| 08.10e | 08.09d | 66052 | 123 | 1774 | 19055 | 113 | 113 |
| 08.12d | 08.11d | 68229 | 123 | 968 | 4726 | 114 | 113 |
| 09.06e | 09.05d | 70493 | 123 | 1305 | 575 | 1 | 1 |

Figure 9: Detailed Results for $\mathsf{cWtn}^{\mathsf{rhs}}_{\Sigma}(\mathcal{T}_1, \mathcal{T}_2)$ and $\mathsf{cWtn}^{\mathsf{lhs}}_{\Sigma}(\mathcal{T}_1, \mathcal{T}_2)$ on Selected Versions of the NCI Thesaurus using Shared Signatures $\Sigma = \mathsf{sig}(\mathcal{T}_1) \cap \mathsf{sig}(\mathcal{T}_2)$

The peaks in atomic left-hand difference witnesses mostly resulted from changes to a few very general concepts. As mentioned above already, Gonçalves et al. (2011) provide an in-depth analysis of NCI versions. A systematic comparison of the methods used by Gonçalves et al. with the logical diff introduced in this paper would be very interesting, but is beyond the scope of this paper. One interesting observation that can be made is, however, that the peak of atomic left-hand witnesses that we observed between the versions 05.01d and 05.03d correlates with the fact that according to Gonçalves et al. a large number of non-redundant axioms were added to version 05.03d. However, a comparable number of non-redundant axioms were also added to version 04.12g, but no peak in atomic left-hand or right-hand witnesses was observed in our analysis.

### 8.3 Scalability Analysis

We demonstrated in the previous sections that CEX2 is capable of finding the logical difference between two unmodified versions of Snomed CT and between distinct versions of the NCI thesaurus restricted to $\mathcal{ELH}^r$. In order to see how CEX2's performance scales, we have also tested it on randomly generated acyclic terminologies of various sizes. Each randomly generated terminology contains a certain number of defined- and primitive concept names and role names. The ratio between concept equations and concept inclusions is fixed, as is the ratio between existential restrictions and conjunctions. The random terminologies were generated for a varying number of defined concept names using the parameters of $\mathsf{SM09a}$: 62 role names; the equality-inclusion ratio is 0.525; and the exists-conjunction ratio is 0.304. For every chosen size, we generated 10 samples consisting of two random terminologies as described above. We then applied CEX2 to find the logical difference of the two terminologies over their joint signature. Figure 10 shows the average memory consumption of CEX2 over 10 randomly generated terminologies of various sizes. In 10(a) the maximum length of conjunctions was fixed as two ($\mathsf{M}$=2), and in 10(b) the number of conjuncts in each conjunction is randomly selected between two and $\mathsf{M}$. It can be seen that the performance of CEX2 crucially depends on the length of conjunctions. In 10(b), the curves break off at the point where CEX2 runs out of physical memory[6]. For instance, in the case of $\mathsf{M}$=22, this happens for terminologies with more than 7 500 defined concept names. Finally, we note that the time required by CEX2 to compare two such random terminologies highly varied across the different samples. The maximum time required by CEX2 was 11 333 seconds.

### 8.4 Additional User Support for Analysing Differences

So far we have discussed experiments with CEX2 in which one computes the set of concept and instance difference witnesses between two terminologies. Clearly, such witnesses do not provide sufficient information for a detailed analysis of the logical difference between two terminologies. For a more thorough analysis, it is required to consider *examples* $\alpha$ from $\mathsf{cDiff}_\Sigma(\mathcal{T}_1, \mathcal{T}_2)$ and $\mathsf{iDiff}_\Sigma(\mathcal{T}_1, \mathcal{T}_2)$ that show why certain concept names are concept/instance difference witnesses. Thus, whenever it searches for concept names $A$ such that there exists a $C$ with $C \sqsubseteq A \in \mathsf{cDiff}_\Sigma(\mathcal{T}_1, \mathcal{T}_2)$, CEX2 can output example concept inclusions $C \sqsubseteq A \in \mathsf{cDiff}_\Sigma(\mathcal{T}_1, \mathcal{T}_2)$. Similarly, if requested, CEX2 can also compute example inclusions

6. In some cases the classification of the terminologies through CB already requires more than 3 GiB of memory.

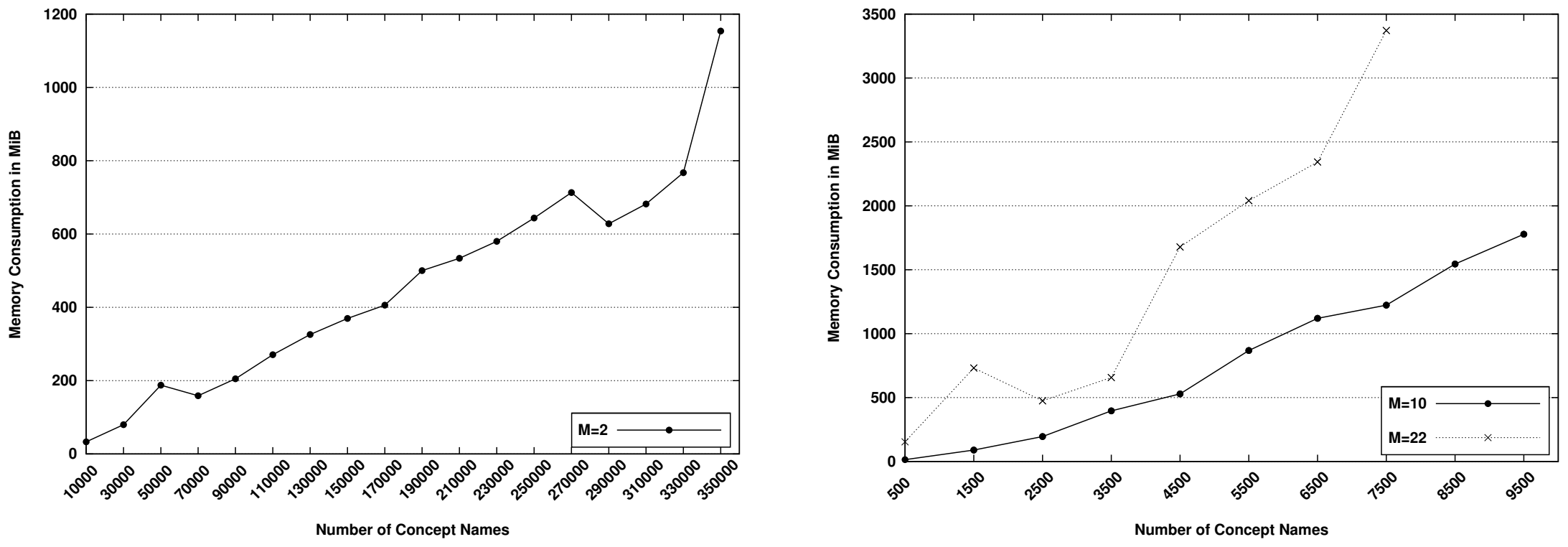

(a) Short Conjunctions (b) Long Conjunctions

Figure 10: Memory Consumption of CEX2 on Randomly Generated Terminologies

illustrating left-hand concept differences $A \sqsubseteq C$, $\exists r.\top \sqsubseteq C$, or $\mathsf{ran}(r) \sqsubseteq C$, and examples for the instance difference case. We know from Example 12 that even minimal such examples can be of exponential size in the input terminologies. In practice, however, for Snomed CT and NCI the additional computation of an example inclusion for every concept/instance difference witness "only" doubles the times required for the computation. As described above already, this can be observed in Table 2, where the computation times with examples are shown in the last column and the computation times without examples are shown in the 7th column. The examples computed by CEX2 are often of reasonable size. For instance, if we consider the subset "Specimen Material Type" (Specmatyp) from Table 2, it holds that

(i) there exist 10 right-hand $\Sigma$-concept witnesses, i.e. $|\mathsf{cWtn}^{\mathsf{rhs}}_{\Sigma}(\mathsf{SM09b}, \mathsf{SM09a})| = 10$;

(ii) the set of left-hand atomic $\Sigma$-concept difference witnesses, $\mathsf{cWtn}^{\mathsf{lhs,A}}_{\Sigma}(\mathsf{SM09b}, \mathsf{SM09a})$, contains 46 concept names.

In Point (i) and (ii), the longest concepts $C$, $D$ for $C \sqsubseteq A \in \mathsf{cDiff}_{\Sigma}(\mathsf{SM09b}, \mathsf{SM09a})$ and $A \sqsubseteq D \in \mathsf{cDiff}_{\Sigma}(\mathsf{SM09b}, \mathsf{SM09a})$ that were computed by CEX2 had twelve concept and role name occurrences (thus were far smaller than the exponential worst case suggests).

Having computed not only difference witnesses but also example concept inclusions for witnesses, it is of interest to explain why an example concept inclusion is entailed by one terminology but not the other. Computing *minimal subsets* of a terminology that entail an example concept inclusion is a promising approach to explaining logical differences that is also known as axiom pinpointing or justification. It is not supported by CEX2, but has been investigated extensively for various description logics including $\mathcal{EL}$ (Schlobach & Cornet, 2003; Baader, Peñaloza, & Suntisrivaraporn, 2007; Kalyanpur, Parsia, Horridge, & Sirin, 2007; Horridge, Parsia, & Sattler, 2010; Peñaloza & Sertkaya, 2010). To illustrate this approach, consider again the subset "Specimen Material Type" (Specmatyp) from Table 2. CEX2 outputs

$$\text{VenipunctureForBloodTest} \in \mathsf{cWtn}^{\mathsf{lhs,A}}_{\Sigma}(\mathsf{SM09b}, \mathsf{SM09a}).$$

$$
\begin{array}{lrl}
(1) & \text{LaboratoryTest} \sqsubseteq & \text{LaboratoryProcedure} \sqcap \text{EvaluationProcedure} \\
(2) & \text{BloodTest} \equiv & \text{LaboratoryTest} \sqcap \exists \text{roleGroup}.\exists \text{hasSpecimen}.\text{BloodSpecimen} \\
(3) & \text{VenipunctureForBloodTest} \equiv & (\exists \text{roleGroup}.\exists \text{hasFocus}.\text{BloodTest}) \\
 & & \sqcap \text{Venipuncture} \\
 & & \sqcap (\exists \text{roleGroup}.((\exists \text{procedureSiteDirect}.\text{VenousStructure}) \\
 & & \qquad \sqcap (\exists \text{method}.\text{PunctureAction})))
\end{array}
$$

Figure 11: Minimal Axiom Set

It also computes the following concept inclusion (slightly simplified by hand) as a member of $\mathsf{cDiff}_\Sigma(\mathsf{SM09b}, \mathsf{SM09a})$:

$$
(*) \quad \begin{array}{l} \text{VenipunctureForBloodTest} \\ \qquad \sqsubseteq \exists \text{roleGroup}.\exists \text{hasFocus}.\text{EvaluationProcedure} \end{array}
$$

Using axiom pinpointing one can then compute a minimal set of inclusions from $\mathsf{SM09b}$ which entails the concept inclusion above; such a set is shown in Figure 11. Axioms 2 and 3 are in both terminologies, but $\mathsf{SM09a}$ contains

$$
\text{LaboratoryTest} \sqsubseteq \text{LaboratoryProcedure}
$$

instead of Axiom 1, which explains this difference between the two terminologies. Note that concept and role names from $\Sigma$ are shaded in grey. It can be seen that the interaction between $\Sigma$-concepts heavily depends on inclusions that are built up mainly from non-$\Sigma$-concepts; actually none of inclusions required to derive $(*)$ is a $\Sigma$-inclusion.

We finally note that CEX2 is a text-based tool. In order to make it more accessible to ontology users, a Protégé plugin, LogDiffViz[7], was created, which calls CEX2 and visualises both ontology versions and the differences as a hierarchical structure. LogDiffViz also provides basic axiom pinpointing. The plugin is distributed as a self-contained Java archive file (JAR) in which CEX2 is bundled.

## 9. Related Work

We describe the relationship between the work presented in this paper and existing work on logical difference and inseparability of ontologies. Related work on versioning and the distinction between syntactical, structural, and logic-based approaches to versioning have been discussed in the introduction already and will not be presented again here. The problem of deciding whether two ontologies are $\Sigma$-inseparable for some signature $\Sigma$ has been investigated for many ontology languages and different notions of inseparability such as concept inseparability, instance inseparability, conjunctive query inseparability, and model-theoretic inseparability (i.e., the $\Sigma$-reducts of models of the first ontology coincide with the $\Sigma$-reducts of models of the second ontology). Inseparability is also closely related to the notion of *conservative extensions* since one ontology is a conservative extension of another ontology if it contains the other ontology as a subset and both are inseparable w.r.t. the signature of

7. Available at `http://protegewiki.stanford.edu/wiki/Logical_Difference_Vizualiser_(LogDiffViz)`

the smaller ontology. Thus, algorithmic results on deciding conservativity are directly relevant for inseparability as well. The tractability results presented in this paper are in sharp contrast to most other known results. We start with general $\mathcal{EL}$-TBoxes: for general $\mathcal{EL}$-TBoxes deciding inseparability and conservative extensions are ExpTime complete problems for concept, instance and conjunctive queries. Both problems are undecidable for model-theoretic inseparability and model-theoretic conservative extensions (Lutz & Wolter, 2010). (We note, however, that in the model-theoretic case unexpected positive algorithmic results have been obtained in Konev, Lutz, et al., 2008, for acyclic $\mathcal{EL}$ and $\mathcal{ALC}$ and their extensions with inverse roles.) For $\mathcal{ALC}$ and its standard extensions without nominals deciding concept inseparability and conservative extensions is 2ExpTime-complete (Ghilardi, Lutz, & Wolter, 2006; Lutz et al., 2007; Lutz & Wolter, 2011) and for $\mathcal{ALCQIO}$ deciding concept inseparability and conservative extensions becomes undecidable (Lutz et al., 2007; Cuenca Grau et al., 2008). Nothing is known for $\mathcal{ALC}$ about the complexity of inseparability for instance and conjunctive queries. For DL-Lite dialects (Calvanese, Giacomo, Lembo, Lenzerini, & Rosati, 2006), the complexity of concept, instance, and query inseparability ranges from PSpace-hard (and in ExpTime) for the description logic underlying the OWL 2 QL standard, NP-complete for DL-Lite$_{horn}$, and $\Pi_2^p$-complete for DL-Lite$_{bool}$ (Konev, Kontchakov, Ludwig, Schneider, Wolter, & Zakharyaschev, 2011; Kontchakov et al., 2010). For DL-Lite$_{bool}$ model-theoretic inseparability is decidable (Kontchakov et al., 2010) and for DL-Lite$_{core}$ concept, instance, and query inseparability are in PTime (Konev et al., 2011). In contrast to the work presented in this paper, however, no attempt is made to present the logical difference to the user if two ontology are not inseparable. As mentioned above, in the work of Konev et al. (2012), CEX2 is extended to the conjunctive query difference case between acyclic $\mathcal{ELH}^r$-terminologies and various experiments based on the NCI thesaurus are discussed.

The work discussed so far is concerned with the logical difference and inseparability between description logic TBoxes. The difference between description logic *concepts* has been investigated, for example, in the work of Teege (1994), and of Brandt, Küsters, and Turhan (2002) but besides of the interest in some kind of difference the problems considered as well as the techniques employed are rather different. Inseparability and conservativity between ontologies given in ontology languages that are more expressive than description logics (including first-order logic) have been considered in the work of Kutz and Mossakowski (2008, 2011). Similar relationships between theories have also been investigated in answer set programming (Pearce & Valverde, 2004; Eiter, Fink, & Woltran, 2007; Pearce & Valverde, 2012).

Finally, we note that Lemma 15 and the ABox constructed in Figure 3 appear to capture and describe fundamental properties of $\mathcal{EL}$ and $\mathcal{ELH}^r$-terminologies. Both have been applied to investigate seemingly unrelated problems such as query containment for ontology based data access using $\mathcal{EL}$-terminologies (Bienvenu, Lutz, & Wolter, 2012b) and first-order rewritability of instance queries (Bienvenu, Lutz, & Wolter, 2012a).

## 10. Conclusion

In this paper, we have presented polytime algorithms that decide concept, instance, and query-inseparability w.r.t. a signature $\Sigma$ for $\mathcal{ELH}^r$-terminologies and compute a represen-tation of the difference if it is non-empty. Experiments using CEX2 based on SNOMED CT and NCI show that the outputs given by our algorithm are mostly of reasonable size and can be analysed by users. Many extensions, applications, and open problems remain to be explored. Here we mention some of them:

(1) We have motivated the study of $\Sigma$-inseparability between terminologies by the problem of comparing different versions of a terminology regarding "what they say" about a certain signature. Other potential and promising applications can be found in the area of decomposing and composing ontologies. For example, when importing an ontology $\mathcal{T}$ into an ontology $\mathcal{T}'$ (i.e., forming $\mathcal{T} \cup \mathcal{T}'$) it is often important to ensure that $\mathcal{T}'$ does not interfere with the signature of $\mathcal{T}$. In other words, $\mathcal{T} \cup \mathcal{T}'$ should be a conservative extension of $\mathcal{T}$ in the sense that the consequences of $\mathcal{T} \cup \mathcal{T}'$ in the signature of $\mathcal{T}$ should coincide with the consequences of $\mathcal{T}$ itself (Cuenca Grau et al., 2008; Ghilardi et al., 2006; Vescovo, Parsia, Sattler, & Schneider, 2011). As observed above already, $\Sigma$-inseparability generalises conservative extensions and, therefore, our algorithms can be used to check whether one terminology is a conservative extension of another terminology. Algorithms checking conservative extensions can also be used to extract modules from ontologies (Cuenca Grau et al., 2008; Kontchakov, Pulina, Sattler, Schneider, Selmer, Wolter, & Zakharyaschev, 2009; Konev et al., 2011). It would be of interest to explore applications of our inseparability testing algorithms to extract modules of terminologies and check conservativity.

(2) Inseparability as defined in this paper does not mean that one terminology can be replaced by another terminology in every context. In various applications of inseparability for modularity it is important to ensure that if $\mathcal{T}_1$ and $\mathcal{T}_2$ are $\Sigma$-inseparable, then $\mathcal{T}_1 \cup \mathcal{T}$ and $\mathcal{T}_2 \cup \mathcal{T}$ are $\Sigma$-inseparable as well, for any ontology $\mathcal{T}$. This is called the replacement property by Konev, Lutz, Walther, and Wolter (2009) and has been exploited and discussed, for example, in the work of Cuenca Grau et al. (2008) and of Kontchakov et al. (2010). The notions of inseparability introduced in this paper do not have the replacement property. To see this, let $\Sigma = \{A, A', B, B'\}$ and

$$\mathcal{T}_1 = \left\{ \begin{array}{c} A \sqsubseteq \exists r.B \\ A' \equiv \exists r.B' \end{array} \right\} \text{ and } \mathcal{T}_2 = \left\{ \begin{array}{c} A \sqsubseteq \exists r.B \\ A' \sqsubseteq \exists r.B' \end{array} \right\}.$$

$\mathcal{T}_1$ and $\mathcal{T}_2$ are $\Sigma$-query inseparable (and, therefore, $\Sigma$-concept and $\Sigma$-instance inseparable), but $\mathcal{T}_1 \cup \mathcal{T}$ is not even $\Sigma$-concept inseparable from $\mathcal{T}_2 \cup \mathcal{T}$, for $\mathcal{T} = \{B \sqsubseteq B'\}$. Indeed, observe that $(\mathcal{T}_1 \cup \mathcal{T}) \models A \sqsubseteq A'$, but $(\mathcal{T}_2 \cup \mathcal{T}) \not\models A \sqsubseteq A'$.

It is an important open research problem to determine the complexity of, and to develop algorithms for strong versions of inseparability with the replacement property for $\mathcal{EL}$ and $\mathcal{ELH}^r$-terminologies.

(3) $\mathcal{ELH}^r$ is a rather weak description logic. It would be of great interest to explore in how far techniques developed for $\mathcal{ELH}^r$ can be applied to ontologies which contain additional constructors, but still consist mainly of $\mathcal{ELH}^r$-inclusions. It is unlikely that tractable sound and complete algorithms for interesting extensions exist, but it seems worth exploring algorithms that are sound and incomplete extensions of the algorithms presented in this paper. Some results in that direction have been presented by Gonçalves, Parsia, and Sattler (2012).

## Acknowledgments

This research was supported by EPSRC grant EP/H043594/1. We would like to thank William Gatens for the development of the LogDiffViz Protégé plugin and three anonymous reviewers for their helpful comments.

## Appendix A. Proofs for Section 2

**Lemma 1** *For every terminology $\mathcal{T}$, one can construct in polynomial time a normalised terminology $\mathcal{T}'$ of polynomial size in $|\mathcal{T}|$ such that $\mathsf{sig}(\mathcal{T}) \subseteq \mathsf{sig}(\mathcal{T}')$, $\mathcal{T}' \models \mathcal{T}$, and for every model $\mathcal{I}$ of $\mathcal{T}$ there exists a model $\mathcal{J}$ of $\mathcal{T}'$ such that $\Delta^{\mathcal{I}} = \Delta^{\mathcal{J}}$ and $X^{\mathcal{I}} = X^{\mathcal{J}}$ for every $X \in \mathsf{sig}(\mathcal{T})$. Moreover, $\mathcal{T}'$ is acyclic if $\mathcal{T}$ is acyclic.*

*Proof.* Given a terminology $\mathcal{T}$, construct a normalised terminology $\mathcal{T}'$ in five steps as follows: First, remove all occurrences of $\top$ in conjunctions, and replace $C$ in each occurrence of $\exists r.C$, where $C$ is not a concept name or $\top$, with a fresh concept name $A$ and add the concept definition $A \equiv C$ to the terminology. Repeat the last step exhaustively.

Second, replace every $\exists r_i.B_i$ in each inclusion with a right-hand side of the form $F \sqcap \exists r_1.B_1 \sqcap \cdots \sqcap \exists r_m.B_m$ ($m \geq 1$), where each $B_i$ is either a concept name or $B_i = \top$, and $F$ is a conjunction of concept names such that $F \neq \top$ or $m \geq 2$, with a fresh concept name $B_i'$ and add the concept definition $B_i' \equiv \exists r_i.B_i$ to the terminology.

Third, replace every inclusion of the form $A \equiv \exists r.\top$ with two inclusions $A \sqsubseteq \exists r.\top$ and $\exists r.\top \sqsubseteq A$ in the terminology.

Fourth, consider any concept name $A$ such that there are sequences $B_0, \ldots, B_{n-1}$ and $F_0, \ldots, F_n$, where the $F_i$ are conjunctions of concept names, such that the terminology contains the concept definitions $A \equiv F_0$ and $B_i \equiv F_{i+1}$, for $i < n$, where $B_i$ is a conjunct of $F_i$ and $A$ a conjunct of $F_n$. Let $F_n'$ be the conjunction of concept names in $F_n$ except $A$. Let, recursively, $F_{i-1}'$ be the result of replacing the conjunct $B_{i-1}$ in $F_{i-1}$ with the conjunction $F_i'$, for $1 \leq i \leq n$. Replace the concept definition $A \equiv F_0$ in the terminology with the primitive concept definition $A \sqsubseteq F_0'$.

Fifth, for each inclusion $A \equiv F$, $A \sqsubseteq F$, $\exists r.\top \sqsubseteq F$, or $\mathsf{ran}(r) \sqsubseteq F$, where $F$ is a conjunction of concept names, replace every conjunct $B$ in $F$ for which there is a $B \equiv F'$ in the terminology, where $F'$ is a conjunction of non-conjunctive concept names, with $F'$.

To see that the construction indeed yields a normalised terminology $\mathcal{T}'$, observe that the steps 1, 2, and 3 ensure that each inclusion has one of the following forms: $A \equiv \exists r.B$, $A \equiv F$, $E \sqsubseteq \exists r.B$, $E \sqsubseteq \exists r.\top$, or $E \sqsubseteq F$, where $B$ is a concept name, $E$ is either a concept name, or is of the form $\exists s.\top$, or $\mathsf{ran}(s)$, and $F$ is a conjunction of (possibly conjunctive) concept names. Step 4 breaks cycles in concept definitions and Step 5 takes care that all conjuncts of the conjunction of concept names $F$ in the right-hand side of each inclusion of the form $A \equiv F$, $A \sqsubseteq F$, $\exists r.\top \sqsubseteq F$, or $\mathsf{ran}(r) \sqsubseteq F$ are non-conjunctive concept names. It is readily verified that $\mathcal{T}'$ is acyclic if $\mathcal{T}$ is acyclic as none of the above steps introduces cycles in concept definitions.

We now show that $\mathcal{T}'$ can be obtained in polynomial time and that $\mathcal{T}'$ is of polynomial size in $|\mathcal{T}|$. Let $n$ be the number of inclusions in $\mathcal{T}$ and $c$ the maximal length of an inclusion's right-hand side in $\mathcal{T}$. Clearly, the steps 1, 2 and 3 each do not increase the number of inclusions by more than $c \cdot n$, raising the total number of inclusions to at most $4nc$. Steps 4

and 5 do not increase the number of inclusions, but the length of their right-hand sides. The length of the right-hand side of an inclusion can increase to at most the sum of the lengths of the right-hand sides of all inclusions, i.e., $4nc^2$ is an upper bound for each right-hand side. The upper bound of the running time for each of the steps in the construction is therefore $16n^2c^3$. Hence, the size of $\mathcal{T}'$ and the running time of the construction are both in $\mathcal{O}(n^2 \cdot c^3)$.

Notice that every new concept name occurs on the left-hand side of a unique concept definition $A \equiv C$ in $\mathcal{T}'$. Thus, every model $\mathcal{I}$ of $\mathcal{T}$ can be expanded to a model $\mathcal{J}$ of $\mathcal{T}'$ by interpreting the fresh concept names in $\mathsf{sig}(\mathcal{T}') \setminus \mathsf{sig}(\mathcal{T})$ by setting $A^{\mathcal{J}} = C^{\mathcal{I}}$.

Moreover, it is readily checked that $\mathcal{T}' \models \mathcal{T}$. □

We prove an extended version of Theorem 2 according to which not only $\mathcal{EL}$-concepts and concepts of the form $\mathsf{ran}(r)$ are evaluated "correctly" in the canonical model $\mathcal{I}_{\mathcal{K}}$, but also $\mathcal{C}^{\sqcap,u}$-concepts (which are introduced in Definition 57).

**Theorem 2**[Extended Version] *Let $\mathcal{K} = (\mathcal{T}, \mathcal{A})$ be an $\mathcal{ELH}^r$-KB. Then*

1. *$\mathcal{I}_{\mathcal{K}}$ is a model of $\mathcal{K}$;*
2. *$\mathcal{I}_{\mathcal{K}}$ can be computed in polynomial time in the size of $\mathcal{K}$;*
3. *for all $x_{C,D} \in \Delta^{\mathcal{I}_{\mathcal{K}}}$ and all $a \in \mathsf{obj}(\mathcal{A})$, if $C_0$ is a $\mathcal{C}^{\sqcap,u}$-concept or of the form $\mathsf{ran}(r)$, then*
   - *$\mathcal{K} \models C_0(a)$ if, and only if, $a^{\mathcal{I}_{\mathcal{K}}} \in C_0^{\mathcal{I}_{\mathcal{K}}}$.*
   - *$\mathcal{T} \models C \sqcap D \sqsubseteq C_0$ if, and only if, $x_{C,D} \in C_0^{\mathcal{I}_{\mathcal{K}}}$.*

*Proof.* Point 2 follows from the fact that instance checking in $\mathcal{ELH}^r$ can be done in polynomial time.

We first prove Point 3 for $\mathcal{EL}$-concepts $C_0 \in \mathsf{sub}(\mathcal{T})$. The proof is by simultaneous induction on the construction of $C_0$. The interesting step is for $C_0 = \exists r.D_0$.

We start with the proof of the direction from left to right. Assume first that $\mathcal{K} \models C_0(a)$. Then $(a, x_{\mathsf{ran}(r),D_0}) \in r^{\mathcal{I}_{\mathcal{K}}}$. We have $\mathcal{T} \models (\mathsf{ran}(r) \sqcap D_0) \sqsubseteq D_0$. Thus, by the induction hypothesis, $x_{\mathsf{ran}(r),D_0} \in D_0^{\mathcal{I}_{\mathcal{K}}}$. But then $a \in C_0^{\mathcal{I}_{\mathcal{K}}}$, as required. Now assume $\mathcal{T} \models C \sqcap D \sqsubseteq C_0$. Then $(x_{C,D}, x_{\mathsf{ran}(r),D_0}) \in r^{\mathcal{I}_{\mathcal{K}}}$. We have $\mathcal{T} \models (\mathsf{ran}(r) \sqcap D_0) \sqsubseteq D_0$. By the induction hypothesis, $x_{\mathsf{ran}(r),D_0} \in D_0^{\mathcal{I}_{\mathcal{K}}}$. But then $x_{C,D} \in C_0^{\mathcal{I}_{\mathcal{K}}}$, as required.

Conversely, assume that $a^{\mathcal{I}_{\mathcal{K}}} \in C_0^{\mathcal{I}_{\mathcal{K}}}$. There exists $d \in \Delta^{\mathcal{I}_{\mathcal{K}}}$ such that $(a^{\mathcal{I}_{\mathcal{K}}}, d) \in r^{\mathcal{I}_{\mathcal{K}}}$ and $d \in D_0^{\mathcal{I}_{\mathcal{K}}}$. Assume first that $d = b \in \mathsf{obj}(\mathcal{A})$. By the induction hypothesis, $\mathcal{K} \models D_0(b)$. There exists $s$ such that $s(a,b) \in \mathcal{A}$ and $s \sqsubseteq_{\mathcal{T}} r$. Thus, $\mathcal{K} \models C_0(a)$, as required. Assume now that $d = x_{\mathsf{ran}(s),F}$. Then $\mathcal{K} \models \exists s.F(a)$, $s \sqsubseteq_{\mathcal{T}} r$ and $x_{\mathsf{ran}(s),F} \in D_0^{\mathcal{I}_{\mathcal{K}}}$. By the induction hypothesis, $\mathcal{T} \models \mathsf{ran}(s) \sqcap F \sqsubseteq D_0$. Thus, $\mathcal{K} \models C_0(a)$, as required.

Now assume $x_{C,D} \in C_0^{\mathcal{I}_{\mathcal{K}}}$. There exists $x_{\mathsf{ran}(s),F}$ with $\mathcal{T} \models C \sqcap D \sqsubseteq \exists s.F$, $s \sqsubseteq_{\mathcal{T}} r$ and $x_{\mathsf{ran}(s),F} \in D_0^{\mathcal{I}_{\mathcal{K}}}$. By the induction hypothesis, $\mathcal{T} \models \mathsf{ran}(s) \sqcap F \sqsubseteq D_0$. Thus $\mathcal{T} \models C \sqcap D \sqsubseteq \exists r.D_0$, as required.

We now prove Point 3 for concepts of the form $C_0 = \mathsf{ran}(r)$. Assume $\mathcal{K} \models (\mathsf{ran}(r))(a)$. Then there exist $b$ and $s$ with $s(b,a) \in \mathcal{A}$ and $s \sqsubseteq_{\mathcal{T}} r$. But then $a \in \mathsf{ran}(r)^{\mathcal{I}_{\mathcal{K}}}$. Conversely, assume that $a \in \mathsf{ran}(r)^{\mathcal{I}_\mathcal{K}}$. Then, by definition of $\mathcal{I}_\mathcal{K}$, there exist $b$ and $s$ with $s(b,a) \in \mathcal{A}$ and $s \sqsubseteq_\mathcal{T} r$. Hence $\mathcal{K} \models (\mathsf{ran}(r))(a)$, as required.

Assume $\mathcal{T} \models C \sqcap D \sqsubseteq \mathsf{ran}(r)$. Then we have, for $C = \mathsf{ran}(s)$, $s \sqsubseteq_\mathcal{T} r$. Then $x_{C,D} \in \mathsf{ran}(r)^{\mathcal{I}_\mathcal{K}}$ since there is a path in $\mathcal{W}_\mathcal{K}$ with tail $x_{C,D}$. The converse direction is similar.

It follows from what has been proved so far that $\mathcal{I}_\mathcal{K}$ is a model of $(\mathcal{T}, \mathcal{A})$. Thus we have proved Point 1, and it remains to prove Point 3.

We prove Point 3 for arbitrary $\mathcal{C}^{\sqcap,u}$-concepts $C_0$. The interesting step is for $C_0 = \exists S.D_0$, where $S = r_1 \sqcap \cdots \sqcap r_n$.

Assume first that $\mathcal{K} \models C_0(a)$. Then $a \in C_0^{\mathcal{I}_\mathcal{K}}$ since $\mathcal{I}_\mathcal{K}$ is a model of $\mathcal{K}$. Similarly, if $\mathcal{T} \models C \sqcap D \sqsubseteq C_0$, then $x_{C,D} \in C_0^{\mathcal{I}_\mathcal{K}}$ since $x_{C,D} \in (C \sqcap D)^{\mathcal{I}_\mathcal{K}}$ and $\mathcal{I}_\mathcal{K}$ is a model of $\mathcal{T}$.

Conversely, assume that $a \in C_0^{\mathcal{I}_\mathcal{K}}$. There exists $d \in \Delta^{\mathcal{I}_\mathcal{K}}$ such that $(a^{\mathcal{I}_\mathcal{K}}, d) \in S^{\mathcal{I}_\mathcal{K}}$ and $d \in D_0^{\mathcal{I}_\mathcal{K}}$. Assume first that $d = b \in \mathsf{obj}(\mathcal{A})$. By the induction hypothesis, $\mathcal{K} \models D_0(b)$. For every $r_i$, $1 \leq i \leq n$, there exists $s_i$ with $s_i(a,b) \in \mathcal{A}$ and $s_i \sqsubseteq_\mathcal{T} r_i$. Thus, $\mathcal{K} \models C_0(a)$, as required.

Assume now that $d = x_{\mathsf{ran}(s),F}$. Then $\mathcal{K} \models \exists s.F(a)$, $s \sqsubseteq_\mathcal{T} r_i$ for $1 \leq i \leq n$ and $x_{\mathsf{ran}(s),F} \in D_0^{\mathcal{I}_\mathcal{K}}$. By the induction hypothesis, $\mathcal{T} \models \mathsf{ran}(s) \sqcap F \sqsubseteq D_0$. Thus, $\mathcal{K} \models C_0(a)$, as required.

Now assume $x_{C,D} \in C_0^{\mathcal{I}_\mathcal{K}}$. There exists $x_{\mathsf{ran}(s),F}$ with $\mathcal{T} \models C \sqcap D \sqsubseteq \exists s.F$, $s \sqsubseteq_\mathcal{T} r_i$, $1 \leq i \leq n$, and $x_{\mathsf{ran}(s),F} \in D_0^{\mathcal{I}_\mathcal{K}}$. By the induction hypothesis, $\mathcal{T} \models \mathsf{ran}(s) \sqcap F \sqsubseteq D_0$. Thus $\mathcal{T} \models C \sqcap D \sqsubseteq \exists S.D_0$, as required. □

## Appendix B. Proofs for Section 5

In some proofs, we require models for infinite sets of concepts. We introduce some notation and a well known result about the existence of "minimal" models. Let $\Gamma$ be a (possibly infinite) set of $\mathcal{C}^{\mathsf{ran}}$-concepts (which are introduced in Definition 32), $\mathcal{T}$ an $\mathcal{ELH}^r$-TBox, and $D$ either a $\mathcal{C}^{\sqcap,u}$-concept (which are introduced in Definition 57) or a $\mathcal{C}^{\mathsf{ran}}$-concept. We write $\mathcal{T} \cup \Gamma \models D$ and say that $\Gamma$ *is included in* $D$ *w.r.t.* $\mathcal{T}$ if, for every model $\mathcal{I}$ of $\mathcal{T}$ and $d \in \Delta^\mathcal{I}$, $d \in D^\mathcal{I}$ follows from $d \in C^\mathcal{I}$ for all $C \in \Gamma$. The following observation follows from the fact that all $\mathcal{C}^{\sqcap,u}$ and $\mathcal{C}^{\mathsf{ran}}$-concepts are equivalent to Horn formulas (in the sense of Chang and Keisler, 1990):

**Lemma 67.** *For all $\mathcal{ELH}^r$-TBoxes $\mathcal{T}$ and sets $\Gamma$ of $\mathcal{C}^{\mathsf{ran}}$-concepts there exists a model $\mathcal{I}$ of $\mathcal{T}$ and $d \in \Delta^\mathcal{I}$ such that the following are equivalent, for all $\mathcal{C}^{\sqcap,u} \cup \mathcal{C}^{\mathsf{ran}}$-concepts $D$:*

- *$\mathcal{T} \cup \Gamma \models D$;*
- *$d \in D^\mathcal{I}$.*

We now come to the proof of Lemma 36. For the convenience of the reader we formulate the result again.

**Lemma 36.** *For every $\mathcal{ELH}^r$-TBox $\mathcal{T}$, ABox $\mathcal{A}$, and all $\mathcal{C}^{\mathsf{ran}}$-concepts $C_0$ and $D_0$, and $a_0 \in \mathsf{obj}(\mathcal{A})$:*

- *$(\mathcal{T}, \mathcal{A}) \models D_0(a_0)$ if, and only if, there exists $n \geq 0$ such that $\mathcal{T} \models C^{n,\mathsf{ran}}_{\mathcal{A},a_0} \sqsubseteq D_0$;*

- $\mathcal{T} \models C_0 \sqsubseteq D_0$ *if, and only if,* $(\mathcal{T}, \mathcal{A}_{C_0}) \models D_0(a_{C_0})$.

*Proof.* We prove Point 1. For the direction from right to left observe that $\mathcal{A} \models C^{n,\mathsf{ran}}_{\mathcal{A},a_0}(a_0)$ for all $n \geq 0$. Thus, $\mathcal{T} \models C^{n,\mathsf{ran}}_{\mathcal{A},a_0} \sqsubseteq D_0$ implies $(\mathcal{T}, \mathcal{A}) \models D_0(a_0)$.

Now assume $(\mathcal{T}, \mathcal{A}) \models D_0(a_0)$. We show that $\mathcal{T} \cup \mathcal{C}^{\mathsf{ran}}_{\mathcal{A},a_0} \models D_0$. Then, using compactness, we find an $n \geq 0$ such that $\mathcal{T} \models C^{n,\mathsf{ran}}_{\mathcal{A},a_0} \sqsubseteq D_0$, as required.

Assume $\mathcal{T} \cup \mathcal{C}^{\mathsf{ran}}_{\mathcal{A},a_0} \not\models D_0$. Take, for every $a \in \mathsf{obj}(\mathcal{A})$, a model $\mathcal{I}_a$ of $\mathcal{T}$ with a point $d_a \in \Delta^{\mathcal{I}_a}$ such that for all $\mathcal{C}^{\mathsf{ran}}$-concepts $C$: $d_a \in C^{\mathcal{I}_a}$ if, and only if, $\mathcal{T} \cup \mathcal{C}^{\mathsf{ran}}_{\mathcal{A},a} \models C$. Such models exist by Lemma 67. We may assume that they are mutually disjoint. Take the following union $\mathcal{I}$ of the models $\mathcal{I}_a$:

- $\Delta^{\mathcal{I}} = \bigcup_{a \in \mathsf{obj}(\mathcal{A})} \Delta^{\mathcal{I}_a}$;
- $A^{\mathcal{I}} = \bigcup_{a \in \mathsf{obj}(\mathcal{A})} A^{\mathcal{I}_a}$, for $A \in \mathsf{N_C}$;
- $r^{\mathcal{I}} = \bigcup_{a \in \mathsf{obj}(\mathcal{A})} r^{\mathcal{I}_a} \cup \{(d_a, d_b) \mid r'(a,b) \in \mathcal{A}, r' \sqsubseteq_{\mathcal{T}} r\}$, for $r \in \mathsf{N_R}$;
- $a^{\mathcal{I}} = d_a$, for $a \in \mathsf{obj}(\mathcal{A})$.

*Claim 1.* For all $\mathcal{C}^{\mathsf{ran}}$-concepts $C$ and all $a \in \mathsf{obj}(\mathcal{A})$ the following holds for all $d \in \Delta^{\mathcal{I}_a}$:

$$d \in C^{\mathcal{I}_a} \text{ iff } d \in C^{\mathcal{I}}.$$

The proof is by induction on the construction of $C$. The interesting cases are $C = \mathsf{ran}(r)$ and $C = \exists r.D$ and the direction from right to left.

Let $d \in C^{\mathcal{I}}$ and assume first that $C = \mathsf{ran}(r)$. Let $d \in C^{\mathcal{I}} \cap \Delta^{\mathcal{I}_a}$ and $(d', d) \in r^{\mathcal{I}}$. For $(d', d) \in \bigcup_{a \in \mathsf{obj}(\mathcal{A})} r^{\mathcal{I}_a}$, the claim follows from the definition. Otherwise, $d = d_a$, $d' = d_b$ for some $b$ with $r'(b,a) \in \mathcal{A}$ and $r' \sqsubseteq_{\mathcal{T}} r$. Thus, $\mathsf{ran}(r') \in \mathcal{C}^{n,\mathsf{ran}}_{\mathcal{A},a}$ for every $n \geq 0$. Hence, $\mathcal{T} \cup \mathcal{C}^{\mathsf{ran}}_{\mathcal{A},a} \models \mathsf{ran}(r)$ and we obtain $d \in C^{\mathcal{I}_a}$.

Assume now that $C = \exists r.D$ and $d \in C^{\mathcal{I}} \cap \Delta^{\mathcal{I}_a}$. Take $d'$ with $(d, d') \in r^{\mathcal{I}}$ and $d' \in D^{\mathcal{I}}$. For $(d, d') \in \bigcup_{a' \in \mathsf{obj}(\mathcal{A})} r^{\mathcal{I}_{a'}}$, $d \in C^{\mathcal{I}_a}$ follows immediately from the induction hypothesis. Otherwise, $d = d_a$ and $d' = d_b$ for some $b$ with $r'(a,b) \in \mathcal{A}$ and $r' \sqsubseteq_{\mathcal{T}} r$. By the induction hypothesis, $d' \in D^{\mathcal{I}_b}$. Hence, $\mathcal{T} \cup \mathcal{C}^{\mathsf{ran}}_{\mathcal{A},b} \models D$. By compactness, there exists a concept $E \in \mathcal{C}^{\mathsf{ran}}_{\mathcal{A},b}$ such that $\mathcal{T} \models E \sqsubseteq D$. From $r'(a,b) \in \mathcal{A}$, we obtain $\exists r'.E \in \mathcal{C}^{n,\mathsf{ran}}_{\mathcal{A},a}$ for every $n > 0$. But then, $\mathcal{T} \cup \mathcal{C}^{\mathsf{ran}}_{\mathcal{A},a} \models \exists r'.D$ and we obtain $d_a \in C^{\mathcal{I}_a}$ using $r' \sqsubseteq_{\mathcal{T}} r$. This finishes the proof of the claim.

Now, for $C \sqsubseteq D \in \mathcal{T}$, let $d \in \Delta^{\mathcal{I}}$ with $d \in C^{\mathcal{I}}$, i.e. $d \in \Delta^{\mathcal{I}_a}$ for some $a \in \mathsf{obj}(\mathcal{A})$. By Claim 1 we have $d \in C^{\mathcal{I}_a}$, which implies that $d \in D^{\mathcal{I}_a}$ as $C^{\mathcal{I}_a} \subseteq D^{\mathcal{I}_a}$. We can conclude that $d \in D^{\mathcal{I}}$ by applying Claim 1 again. Similarly, one can show that $C^{\mathcal{I}} = D^{\mathcal{I}}$ for every $C \equiv D \in \mathcal{T}$ and $r^{\mathcal{I}} \subseteq s^{\mathcal{I}}$ for every $r \sqsubseteq s \in \mathcal{T}$. It follows that $\mathcal{I}$ is a model of $\mathcal{T}$. By construction of $\mathcal{I}$, we have $(a^{\mathcal{I}}, b^{\mathcal{I}}) \in r^{\mathcal{I}}$ for every $r(a,b) \in \mathcal{A}$. Moreover, for $A(a) \in \mathcal{A}$ with $a \in \mathsf{obj}(\mathcal{A})$, it holds that $\mathcal{T} \cup \mathcal{C}^{\mathsf{ran}}_{\mathcal{A},a} \models A$, which implies that $d_a \in A^{\mathcal{I}_a}$ and $a^{\mathcal{I}} \in A^{\mathcal{I}}$ by our claim. We can thus infer that $\mathcal{I}$ is a model of $(\mathcal{T}, \mathcal{A})$ and $\mathcal{I} \not\models D_0(a_0)$ as $\mathcal{T} \cup \mathcal{C}^{\mathsf{ran}}_{\mathcal{A},a_0} \not\models D_0$, which implies that $d_{a_0} \notin D_0^{\mathcal{I}_{a_0}}$ and $a_0^{\mathcal{I}} \notin D_0^{\mathcal{I}}$, by Claim 1. Hence, $(\mathcal{T}, \mathcal{A}) \not\models D_0(a_0)$ and we have derived a contradiction.

The proof of Point 2 is a simple application of the definition. $\square$

Now we prove cut elimination, correctness, and completeness of the calculus for $\mathcal{ELH}^r$ given in Figures 1 and 5. We start with some basic observations, which can be easily proved by induction on the length of derivations.

**Lemma 68.** *For any $\mathcal{ELH}^r$-terminology $\mathcal{T}$, $\mathcal{C}^{\mathsf{ran}}$-concepts $C$, $D$ and any role names $r$, $s$ we have*

1. *if $\mathcal{T} \vdash \top \sqsubseteq D$, then $\mathcal{T} \vdash C \sqsubseteq D$;*
2. *if $\mathcal{T} \vdash C \sqsubseteq A$ and $A \sqsubseteq C_A \in \mathcal{T}$ or $A \equiv C_A \in \mathcal{T}$, then $\mathcal{T} \vdash C \sqsubseteq C_A$;*
3. *if $\mathcal{T} \vdash C \sqsubseteq \exists r.D$ then $\mathcal{T} \vdash C \sqsubseteq \exists r.(D \sqcap \mathsf{ran}(r))$;*
4. *if $\mathcal{T} \vdash C \sqsubseteq \exists r.D$, and $\exists r.\top \sqsubseteq B \in \mathcal{T}$, then $\mathcal{T} \vdash C \sqsubseteq B$;*
5. *if $\mathcal{T} \vdash C \sqsubseteq \mathsf{ran}(r)$ and $\mathsf{ran}(r) \sqsubseteq A \in \mathcal{T}$, then $\mathcal{T} \vdash C \sqsubseteq A$;*
6. *if $\mathcal{T} \vdash C \sqsubseteq \exists r.D$, and $r \sqsubseteq s \in \mathcal{T}$, then $\mathcal{T} \vdash C \sqsubseteq \exists s.D$;*
7. *if $\mathcal{T} \vdash C \sqsubseteq \mathsf{ran}(r)$ and $r \sqsubseteq s \in \mathcal{T}$, then $\mathcal{T} \vdash C \sqsubseteq \mathsf{ran}(s)$.*

**Lemma 69** (Cut elimination)**.** *For any $\mathcal{ELH}^r$-terminology $\mathcal{T}$, $\mathcal{C}^{\mathsf{ran}}$-concepts $C$, $D$, and $E$, if $\mathcal{T} \vdash C \sqsubseteq D$ and $\mathcal{T} \vdash D \sqsubseteq E$ then $\mathcal{T} \vdash C \sqsubseteq E$.*

*Proof.* Let $\mathcal{D}_1$ be the derivation of $C \sqsubseteq D$ and $\mathcal{D}_2$ be the derivation of $D \sqsubseteq E$. Let $L_i$ be the length of $\mathcal{D}_i$, $i = 1, 2$. The proof of the lemma is by induction on the lexicographical ordering on pairs $(L_2, L_1)$.

The case when $L_2 = 0$ or $L_1 = 0$, as well as the cases when $L_2$ ends with one of AndL1, AndL2, AndR, Ex, DefL, DefR or PDefL are virtually the same as in the proof of Hofmann (2005). Assume $\mathcal{D}_2$ ends with Dom, and so its last sequent is of the form $\exists r.D' \sqsubseteq E$, and the sequent above it is $B \sqsubseteq E$. By Lemma 68, Item 4, $\mathcal{T} \vdash C \sqsubseteq \exists r.D'$ implies $\mathcal{T} \vdash C \sqsubseteq B$, so by the induction hypothesis, $\mathcal{T} \vdash C \sqsubseteq E$.

The cases when $\mathcal{D}_2$ ends with ExRan, Ran, Sub, or RanSub can be dealt with in the similar way using Lemma 68, Items 3, 5–7. □

**Theorem 38.** *Let $\mathcal{T}$ be an $\mathcal{ELH}^r$-terminology; $C_0$ and $D_0$ be $\mathcal{C}^{\mathsf{ran}}$-concepts. Then $\mathcal{T} \models C_0 \sqsubseteq D_0$ if, and only if, $\mathcal{T} \vdash C_0 \sqsubseteq D_0$.*

*Proof.* It can be easily checked that the proof system rules are sound and so if $\mathcal{T} \vdash C_0 \sqsubseteq D_0$, then $\mathcal{T} \models C_0 \sqsubseteq D_0$.

Conversely, assume that $\mathcal{T} \models C_0 \sqsubseteq D_0$. To prove $\mathcal{T} \vdash C_0 \sqsubseteq D_0$ we construct an interpretation $\mathcal{I}$ based on the derivability of sequents from $\mathcal{T}$. We show that $\mathcal{I}$ is a model of $\mathcal{T}$. As a consequence we obtain $C_0^{\mathcal{I}} \subseteq D_0^{\mathcal{I}}$ and conclude that $\mathcal{T} \vdash C_0 \sqsubseteq D_0$ based on the properties of $\mathcal{I}$.

The domain $\Delta^{\mathcal{I}}$ is the set of all *well-formed* pairs $x = \langle C, \mathcal{R}_C \rangle$, where $C$ is a $\mathcal{C}^{\mathsf{ran}}$-concept and $\mathcal{R}_C$ is a finite set of role names such that

$$\forall s \in \mathsf{N_R} : \text{if } \mathcal{T} \vdash (C \sqcap \bigsqcap_{r \in \mathcal{R}_C} \mathsf{ran}(r)) \sqsubseteq \mathsf{ran}(s), \text{ then } s \in \mathcal{R}_C.$$

We introduce the following abbreviation. Let

$$\mathsf{Ran}(\mathcal{R}_C) = \bigsqcap_{r\in\mathcal{R}_C} \mathsf{ran}(r).$$

$\mathcal{C}^{\mathsf{ran}}$-concepts $C$ are interpreted as

$$\mathcal{I}(C) = \{\langle D, \mathcal{R}_D\rangle \in \Delta^{\mathcal{I}} \mid \mathcal{T} \vdash (D \sqcap \mathsf{Ran}(\mathcal{R}_D)) \sqsubseteq C\},$$

and $r \in \mathsf{N_R}$ are interpreted as

$$\mathcal{I}(r) = \{(\langle C, \mathcal{R}_C\rangle, \langle D, \mathcal{R}_D\rangle) \in \Delta^{\mathcal{I}} \times \Delta^{\mathcal{I}} \mid r \in \mathcal{R}_D \text{ and } \mathcal{T} \vdash (C \sqcap \mathsf{Ran}(\mathcal{R}_C)) \sqsubseteq \exists r.(D \sqcap \mathsf{Ran}(\mathcal{R}_D))\}.$$

Note that $\mathcal{I}(C)$ is nonempty for every $C$: consider $\mathcal{R}^0_C = \{s \in \mathsf{N_R} \mid \mathcal{T} \vdash C \sqsubseteq \mathsf{ran}(s)\}$. As $\mathcal{T}$ is finite, $\mathcal{R}^0_C$ is finite. Notice that, by Ax and AndR, $\mathcal{T} \vdash C \sqsubseteq C \sqcap \mathsf{Ran}(\mathcal{R}^0_C)$ so, by Lemma 69, if $\mathcal{T} \vdash (C \sqcap \bigsqcap_{r\in\mathcal{R}^0_C} \mathsf{ran}(r)) \sqsubseteq \mathsf{ran}(s)$, for some $s$, then $\mathcal{T} \vdash C \sqsubseteq \mathsf{ran}(s)$, so $s \in \mathcal{R}^0_C$. That is, $\langle C, \mathcal{R}^0_C\rangle$ is a well-formed pair and, obviously, $\langle C, \mathcal{R}^0_C\rangle \in \mathcal{I}(C)$.

We now show that $\mathcal{I}(C) = C^{\mathcal{I}}$ for all $\mathcal{C}^{\mathsf{ran}}$-concepts $C$. The proof is by induction on the construction of $C$.

1. $\mathcal{I}(\top) = \Delta^{\mathcal{I}}$.
For any well-formed pair $\langle C, \mathcal{R}_C\rangle$, $\mathcal{T} \vdash C \sqcap \mathsf{Ran}(\mathcal{R}_C) \sqsubseteq \top$ is an axiom.

2. $\mathcal{I}(C \sqcap D) = \mathcal{I}(C) \cap \mathcal{I}(D)$.
Let $\langle C, \mathcal{R}_C\rangle \in \mathcal{I}(D_1 \sqcap D_2)$, that is $\mathcal{T} \vdash (C \sqcap \mathsf{Ran}(\mathcal{R}_C)) \sqsubseteq (D_1 \sqcap D_2)$. Since $\mathcal{T} \vdash (D_1 \sqcap D_2) \sqsubseteq D_1$, by Lemma 69, we have $\mathcal{T} \vdash (C \sqcap \mathsf{Ran}(\mathcal{R}_C)) \sqsubseteq D_1$, that is, $\langle C, \mathcal{R}_C\rangle \in \mathcal{I}(D_1)$. Similarly, $\langle C, \mathcal{R}_C\rangle \in \mathcal{I}(D_2)$.

Conversely, suppose $\langle C, \mathcal{R}_C\rangle \in \mathcal{I}(D_1)$ and $\langle C, \mathcal{R}_C\rangle \in \mathcal{I}(D_2)$ holds, that is, $\mathcal{T} \vdash (C \sqcap \mathsf{Ran}(\mathcal{R}_C)) \sqsubseteq D_1$ and $\mathcal{T} \vdash (C \sqcap \mathsf{Ran}(\mathcal{R}_C)) \sqsubseteq D_2$. By AndR, $\mathcal{T} \vdash (C \sqcap \mathsf{Ran}(\mathcal{R}_C)) \sqsubseteq (D_1 \sqcap D_2)$, that is, $\langle C, \mathcal{R}_C\rangle \in \mathcal{I}(D_1 \sqcap D_2)$.

3. $\mathcal{I}(\exists r.C) = \{x \in \Delta^{\mathcal{I}} \mid \exists y \in \mathcal{I}(C) : (x, y) \in \mathcal{I}(r)\}$.
Suppose for some well-formed pair $\langle D, \mathcal{R}_D\rangle$ we have $\langle D, \mathcal{R}_D\rangle \in \mathcal{I}(\exists r.C)$, that is $\mathcal{T} \vdash (D \sqcap \mathsf{Ran}(\mathcal{R}_D)) \sqsubseteq \exists r.C$. Then, by Lemma 68, Item 3, $\mathcal{T} \vdash (D \sqcap \mathsf{Ran}(\mathcal{R}_D)) \sqsubseteq \exists r.(C \sqcap \mathsf{ran}(r))$. Consider $\mathcal{R}^r_C = \{s \in \mathsf{N_R} \mid \mathcal{T} \vdash (C \sqcap \mathsf{ran}(r)) \sqsubseteq \mathsf{ran}(s)\}$. Clearly, $r \in \mathcal{R}^r_C$ and, similarly to the argument for $\mathcal{R}^0_C$ above, $\langle C, \mathcal{R}^r_C\rangle$ is a well-formed pair. By Ax and AndR, $\mathcal{T} \vdash C \sqcap \mathsf{ran}(r) \sqsubseteq C \sqcap \mathsf{Ran}(\mathcal{R}^r_C)$, by Ex, $\mathcal{T} \vdash \exists r.(C \sqcap \mathsf{ran}(r)) \sqsubseteq \exists r.(C \sqcap \mathsf{Ran}(\mathcal{R}^r_C))$ and by Lemma 69, $\mathcal{T} \vdash (D \sqcap \mathsf{Ran}(\mathcal{R}_D)) \sqsubseteq \exists r.(C \sqcap \mathsf{Ran}(\mathcal{R}^r_C))$. Then, by definition, $(\langle D, \mathcal{R}_D\rangle, \langle C, \mathcal{R}^r_C\rangle) \in \mathcal{I}(r)$ and, since $\mathcal{T} \vdash (C \sqcap \mathsf{Ran}(\mathcal{R}^r_C)) \sqsubseteq C$, we have $\langle C, \mathcal{R}^r_C\rangle \in \mathcal{I}(C)$.

Conversely, let $(\langle D_1, \mathcal{R}_{D_1}\rangle, \langle D_2, \mathcal{R}_{D_2}\rangle) \in \mathcal{I}(r)$ and $\langle D_2, \mathcal{R}_{D_2}\rangle \in \mathcal{I}(C)$, that is, $\mathcal{T} \vdash (D_1 \sqcap \mathsf{Ran}(\mathcal{R}_{D_1})) \sqsubseteq \exists r.(D_2 \sqcap \mathsf{Ran}(\mathcal{R}_{D_2}))$, $r \in \mathcal{R}_{D_2}$, and $\mathcal{T} \vdash (D_2 \sqcap \mathsf{Ran}(\mathcal{R}_{D_2})) \sqsubseteq C$. By Ex we have $\mathcal{T} \vdash \exists r.(D_2 \sqcap \mathsf{Ran}(\mathcal{R}_{D_2})) \sqsubseteq \exists r.C$, and, by Lemma 69, we have $\mathcal{T} \vdash (D_1 \sqcap \mathsf{Ran}(\mathcal{R}_{D_1})) \sqsubseteq \exists r.C$, that is, $\langle D_1, \mathcal{R}_{D_1}\rangle \in \mathcal{I}(\exists r.C)$.

4. $\mathcal{I}(\mathsf{ran}(r)) = \{y \in \Delta^{\mathcal{I}} \mid \exists x : (x, y) \in \mathcal{I}(r)\}$.
First we show that $\mathcal{I}(\mathsf{ran}(r)) = \{\langle C, \mathcal{R}_C\rangle \in \Delta^{\mathcal{I}} \mid r \in \mathcal{R}_C\}$. If $r \in \mathcal{R}_C$, we have $\mathcal{T} \vdash C \sqcap \mathsf{Ran}(\mathcal{R}_C) \sqsubseteq \mathsf{ran}(r)$, that is, $\mathcal{I}(\mathsf{ran}(r)) \supseteq \{\langle C, \mathcal{R}_C\rangle \in \Delta^{\mathcal{I}} \mid r \in \mathcal{R}_C\}$. Suppose $\langle C, \mathcal{R}_C\rangle \in \mathcal{I}(\mathsf{ran}(r))$, that is, $\mathcal{T} \vdash (C \sqcap \mathsf{Ran}(\mathcal{R}_C)) \sqsubseteq \mathsf{ran}(r)$. Then, since $\langle C, \mathcal{R}_C\rangle$ is a well-formed pair, $r \in \mathcal{R}_C$, that is, $\mathcal{I}(\mathsf{ran}(r)) \subseteq \{\langle C, \mathcal{R}_C\rangle \in \Delta^{\mathcal{I}} \mid r \in \mathcal{R}_C\}$.

Suppose now that $\langle C, \mathcal{R}_C \rangle \in \mathcal{I}(\mathsf{ran}(r))$, that is, $\langle C, \mathcal{R}_C \rangle$ is such that $r \in \mathcal{R}_C$. Let $D$ denote $(C \sqcap \mathsf{Ran}(\mathcal{R}_C))$. By induction on the length of derivations one can see that a sequent of the form $\exists r.D \sqsubseteq \mathsf{ran}(s)$ is not derivable for any $s \in \mathsf{N_R}$. Therefore, $\langle \exists r.D, \emptyset \rangle$ is a well-formed pair and $(\langle \exists r.D, \emptyset \rangle, \langle C, \mathcal{R}_C \rangle) \in \mathcal{I}(r)$. Conversely, let $(\langle D_1, \mathcal{R}_{D_1} \rangle, \langle D_2, \mathcal{R}_{D_2} \rangle) \in \mathcal{I}(r)$ then, in particular, $r \in \mathcal{R}_{D_2}$. That is, $\langle D_2, \mathcal{R}_{D_2} \rangle \in \mathcal{I}(\mathsf{ran}(r))$.

Now we show that $\mathcal{I}$ is a model of $\mathcal{T}$. We need to show that all axioms of $\mathcal{T}$ are true in $\mathcal{I}$.

1. $\mathcal{I}(X) \subseteq \mathcal{I}(C_X)$, whenever $X \equiv C_X \in \mathcal{T}$ or $X \sqsubseteq C_X \in \mathcal{T}$.
Let $\langle C, \mathcal{R}_C \rangle \in \mathcal{I}(X)$, that is, $\mathcal{T} \vdash (C \sqcap \mathsf{Ran}(\mathcal{R}_C)) \sqsubseteq X$. By Lemma 68, Item 2, $\mathcal{T} \vdash (C \sqcap \mathsf{Ran}(\mathcal{R}_C)) \sqsubseteq C_X$, that is, $\langle C, \mathcal{R}_C \rangle \in \mathcal{I}(C_X)$.

2. $\mathcal{I}(C_X) \subseteq \mathcal{I}(X)$, whenever $X \equiv C_X \in \mathcal{T}$.
Let $\langle C, \mathcal{R}_C \rangle \in \mathcal{I}(C_X)$, that is, $\mathcal{T} \vdash (C \sqcap \mathsf{Ran}(\mathcal{R}_C)) \sqsubseteq C_X$. Since by Ax and DefR $\mathcal{T} \vdash C_X \sqsubseteq X$, by Lemma 69, $\mathcal{T} \vdash (C \sqcap \mathsf{Ran}(\mathcal{R}_C)) \sqsubseteq X$, that is $\langle C, \mathcal{R}_C \rangle \in \mathcal{I}(X)$.

3. $(x, y) \in \mathcal{I}(r) \Rightarrow y \in \mathcal{I}(A)$, whenever $\mathsf{ran}(r) \sqsubseteq A \in \mathcal{T}$.
Let $(\langle C, \mathcal{R}_C \rangle, \langle D, \mathcal{R}_D \rangle) \in \mathcal{I}(r)$, that is, $\mathcal{T} \vdash (C \sqcap \mathsf{Ran}(\mathcal{R}_C)) \sqsubseteq \exists r.(D \sqcap \mathsf{Ran}(\mathcal{R}_D))$ and $r \in \mathcal{R}_D$. Since $r \in \mathcal{R}_D$ and, as, by Ax and Ran, $\mathcal{T} \vdash \mathsf{ran}(r) \sqsubseteq A$, by AndL1, AndL2 we have $\mathcal{T} \vdash (D \sqcap \mathsf{Ran}(\mathcal{R}_D)) \sqsubseteq A$, that is, $\langle D, \mathcal{R}_D \rangle \in \mathcal{I}(A)$.

4. $(x, y) \in \mathcal{I}(r) \Rightarrow x \in \mathcal{I}(B)$, whenever $\exists r.\top \sqsubseteq B \in \mathcal{T}$.
Let $(\langle C, \mathcal{R}_C \rangle, \langle D, \mathcal{R}_D \rangle) \in \mathcal{I}(r)$, that is, $\mathcal{T} \vdash (C \sqcap \mathsf{Ran}(\mathcal{R}_C)) \sqsubseteq \exists r.(D \sqcap \mathsf{Ran}(\mathcal{R}_D))$ and $r \in \mathcal{R}_D$. Notice that, by Lemma 68, Item 4, we have $\mathcal{T} \vdash (C \sqcap \mathsf{Ran}(\mathcal{R}_C)) \sqsubseteq B$, that is, $\langle C, \mathcal{R}_C \rangle \in \mathcal{I}(B)$.

5. $\mathcal{I}(s) \subseteq \mathcal{I}(r)$, whenever $s \sqsubseteq r \in \mathcal{T}$.
Let $(\langle C, \mathcal{R}_C \rangle, \langle D, \mathcal{R}_D \rangle \in \mathcal{I}(r))$, that is $\mathcal{T} \vdash (C \sqcap \mathsf{Ran}(\mathcal{R}_C)) \sqsubseteq \exists r.(D \sqcap \mathsf{Ran}(\mathcal{R}_D))$ and $r \in \mathcal{R}_D$. By Lemma 68, Item 6, $\mathcal{T} \vdash (C \sqcap \mathsf{Ran}(\mathcal{R}_C)) \sqsubseteq \exists s.(D \sqcap \mathsf{Ran}(\mathcal{R}_D))$. Since $r \sqsubseteq s \in \mathcal{T}$, by Ax and RanSub, $\mathcal{T} \vdash \mathsf{ran}(r) \sqsubseteq \mathsf{ran}(s)$ and $\mathcal{T} \vdash (D \sqcap \mathsf{Ran}(\mathcal{R}_D)) \sqsubseteq \mathsf{ran}(s)$ by AndL1 and AndL2. Since $\langle D, \mathcal{R}_D \rangle$ is well-formed, $s \in \mathcal{R}_D$. Thus, $(\langle C, \mathcal{R}_C \rangle, \langle D, \mathcal{R}_D \rangle) \in \mathcal{I}(s)$

As $\mathcal{T} \models C_0 \sqsubseteq D_0$, we have $\mathcal{I}(C_0) \subseteq \mathcal{I}(D_0)$. Since $\langle C_0, \mathcal{R}^0_{C_0} \rangle \in \mathcal{I}(C_0)$, we have $\langle C_0, \mathcal{R}^0_{C_0} \rangle \in \mathcal{I}(D_0)$, that is $\mathcal{T} \vdash (C_0 \sqcap \mathsf{Ran}(\mathcal{R}^0_{C_0})) \sqsubseteq D_0$. As $\mathcal{T} \vdash C_0 \sqsubseteq C_0 \sqcap \mathsf{Ran}(\mathcal{R}^0_{C_0})$, we have $\mathcal{T} \vdash C_0 \sqsubseteq D_0$ by Lemma 69. □

**Proof of Lemma 44.** Let $\mathcal{T}$ be a normalised $\mathcal{ELH}^r$-terminology and $\Sigma$ a signature. Additionally, let $\mathcal{A}$ be a $\Sigma$-ABox, $A \in \mathsf{sig}(\mathcal{T}) \cup \Sigma$ non-conjunctive in $\mathcal{T}$ and $a \in \mathsf{obj}(\mathcal{A})$.

For the direction "(1.) $\Rightarrow$ (2.)", it is a direct consequence of the construction of $\mathcal{A}_{\mathcal{T},\Sigma}$ that for all $b \in \mathsf{obj}(\mathcal{A})$ and $B \in \mathsf{sig}(\mathcal{T}) \cup \Sigma$ non-conjunctive in $\mathcal{T}$ if $(\mathcal{T}, \mathcal{A}) \not\models B(b)$ then $\xi_B \in \mathsf{obj}(\mathcal{A}_{\mathcal{T},\Sigma})$.

Assume that $(\mathcal{T}, \mathcal{A}) \not\models A(a)$. Then $\xi_A \in \mathsf{obj}(\mathcal{A}_{\mathcal{T},\Sigma})$. We now define a $\Sigma$-range simulation $S$ by setting,

- for $b \in \mathsf{obj}(\mathcal{A})$ and for $B \in \mathsf{sig}(\mathcal{T}) \cup \Sigma$ non-conjunctive in $\mathcal{T}$ with $\xi_B \in \mathsf{obj}(\mathcal{A}_{\mathcal{T},\Sigma})$: $(b, \xi_B) \in S$ if, and only if, $(\mathcal{T}, \mathcal{A}) \not\models B(b)$,
- $(b, \xi_\Sigma) \in S$ for all $b \in \mathsf{obj}(\mathcal{A})$.

We show that $S$ is indeed a $\Sigma$-range simulation with $(a, \xi_A) \in S$ by verifying that the conditions (S1)–(S3) and (RS) introduced on page 663 hold.

**(S1)** As $(\mathcal{T},\mathcal{A}) \not\models A(a)$ and $\xi_A \in \mathsf{obj}(\mathcal{A}_{\mathcal{T},\Sigma})$, it immediately follows that $(a,\xi_A) \in S$.

**(S2)** Let now $(b,\xi) \in S$ and $\tilde{B}(b) \in \mathcal{A}$ with $\tilde{B} \in \Sigma$. We have to prove that $\tilde{B}(\xi) \in \mathcal{A}_{\mathcal{T},\Sigma}$. For $\xi = \xi_B$ with $B \in \mathsf{sig}(\mathcal{T}) \cup \Sigma$ non-conjunctive in $\mathcal{T}$, we obtain from the definition of $S$ that $(\mathcal{T},\mathcal{A}) \not\models B(b)$. Moreover, it holds that $\tilde{B} \notin \mathsf{preC}^{\Sigma}_{\mathcal{T}}(B)$ as otherwise $(\mathcal{T},\mathcal{A}) \models B(b)$. Thus, by the definition of $\mathcal{A}_{\mathcal{T},\Sigma}(B)$ we have $\tilde{B}(\xi_B) \in \mathcal{A}_{\mathcal{T},\Sigma}$. For $\xi = \xi_\Sigma$, it immediately follows that $\tilde{B}(\xi_\Sigma) \in \mathcal{A}_{\mathcal{T},\Sigma}$ by the definition of $\mathcal{A}_{\mathcal{T},\Sigma}$.

**(S3)** Now, let $(b,\xi) \in S$ and $r(b,b') \in \mathcal{A}$ with $r \in \Sigma$. We have to prove that there exists $\xi' \in \mathsf{obj}(\mathcal{A}_{\mathcal{T},\Sigma})$ with $(b',\xi') \in S$ and $r(\xi,\xi') \in \mathcal{A}_{\mathcal{T},\Sigma}$. For $\xi = \xi_\Sigma$, it immediately follows from the definition of $\mathcal{A}_{\mathcal{T},\Sigma}$ that $r(\xi_\Sigma,\xi_\Sigma) \in \mathcal{A}_{\mathcal{T},\Sigma}$ and $(b',\xi_\Sigma) \in S$ holds by the definition of $S$.

For $\xi = \xi_B$ with $B \in \mathsf{sig}(\mathcal{T}) \cup \Sigma$ non-conjunctive in $\mathcal{T}$ it follows from the definition of $S$ that $(\mathcal{T},\mathcal{A}) \not\models B(b)$. Additionally, we can infer that $r \notin \mathsf{preDom}^{\Sigma}_{\mathcal{T}}(B)$ as otherwise $(\mathcal{T},\mathcal{A}) \models (\exists r.\top)(b)$ would imply that $(\mathcal{T},\mathcal{A}) \models B(b)$.

Consider cases how $B$ is defined in $\mathcal{T}$. If $B$ is pseudo-primitive in $\mathcal{T}$, we obtain from the definition of $\mathcal{A}_{\mathcal{T},\Sigma}(B)$ that $r(\xi_B,\xi_\Sigma) \in \mathcal{A}_{\mathcal{T},\Sigma}$ and it holds that $(b',\xi_\Sigma) \in S$ by the definition of $S$.

For $B \equiv \exists r'.B' \in \mathcal{T}$, we have to distinguish between the following two cases. If $r \notin \mathsf{preRole}^{\Sigma}_{\mathcal{T}}(r')$, we obtain $r \in \Sigma \setminus (\mathsf{preRole}^{\Sigma}_{\mathcal{T}}(r') \cup \mathsf{preDom}^{\Sigma}_{\mathcal{T}}(B))$ and thus $r(\xi_B,\xi_\Sigma) \in \mathcal{A}_{\mathcal{T},\Sigma}$ by the definition of $\mathcal{A}_{\mathcal{T},\Sigma}$ and it holds again that $(b',\xi_\Sigma) \in S$ by the definition of $S$. In the case where $r \in \mathsf{preRole}^{\Sigma}_{\mathcal{T}}(r')$, we have $r \in \mathsf{preRole}^{\Sigma}_{\mathcal{T}}(r') \setminus \mathsf{preDom}^{\Sigma}_{\mathcal{T}}(B)$. Furthermore, as $(\mathcal{T},\mathcal{A}) \not\models B(b)$ and so $(\mathcal{T},\mathcal{A}) \not\models (\exists r'.B')(b)$, it is easy to see that there must exist $B''_i \in \mathsf{non\text{-}conj}_{\mathcal{T}}(B')$ with $r \notin \mathsf{preRan}^{\Sigma}_{\mathcal{T}}(B''_i)$ and $(\mathcal{T},\mathcal{A}) \not\models B''_i(b')$. Then we have $r(\xi_B,\xi_{B''_i}) \in \mathcal{A}_{\mathcal{T},\Sigma}$ by the definition of $\mathcal{A}_{\mathcal{T},\Sigma}(B)$ and $(b',\xi_{B''_i}) \in S$ by the definition of $S$.

**(RS)** Let now $(b,\xi) \in S$ such that $r(c,b) \in \mathcal{A}$ for $r \in \Sigma$. We have to show that there exists $\xi'$ with $r(\xi',\xi) \in \mathcal{A}_{\mathcal{T},\Sigma}$. For $\xi = \xi_B$ with $B \in \mathsf{sig}(\mathcal{T}) \cup \Sigma$ non-conjunctive in $\mathcal{T}$, we obtain again from the definition of $S$ that $(\mathcal{T},\mathcal{A}) \not\models B(b)$. Furthermore, we have $r \notin \mathsf{preRan}^{\Sigma}_{\mathcal{T}}(B)$ as otherwise $(\mathcal{T},\mathcal{A}) \models B(b)$. Thus, by the definition of $\mathcal{A}_{\mathcal{T},\Sigma}(B)$ we have $r(\xi_\Sigma,\xi_B) \in \mathcal{A}_{\mathcal{T},\Sigma}$. For $\xi = \xi_\Sigma$, it follows by the definition of $\mathcal{A}_{\mathcal{T},\Sigma}$ that $r(\xi_\Sigma,\xi_\Sigma) \in \mathcal{A}_{\mathcal{T},\Sigma}$.

For the converse direction "(2.) $\Rightarrow$ (1.)", we assume that $\xi_A \in \mathsf{obj}(\mathcal{A}_{\mathcal{T},\Sigma})$ and $(\mathcal{A},a) \leq^{\mathsf{ran}}_{\Sigma} (\mathcal{A}_{\mathcal{T},\Sigma},\xi_A)$. It is then sufficient to show for all $n$ that

$$\mathcal{T} \not\models C^{n,\mathsf{ran}}_{\mathcal{A}_{\mathcal{T},\Sigma},\xi_A} \sqsubseteq A$$

as this implies that $(\mathcal{T},\mathcal{A}_{\mathcal{T},\Sigma}) \not\models A(\xi_A)$ by Lemma 36. We then obtain from Lemma 42 that $(\mathcal{T},\mathcal{A}) \not\models A(a)$ holds.

Thus, we now prove by induction on $n$ that for every concept name $B \in \mathsf{sig}(\mathcal{T}) \cup \Sigma$ non-conjunctive in $\mathcal{T}$ with $\xi_B \in \mathsf{obj}(\mathcal{A}_{\mathcal{T},\Sigma})$, we have $\mathcal{T} \not\models C^{n,\mathsf{ran}}_{\mathcal{A}_{\mathcal{T},\Sigma},\xi_B} \sqsubseteq B$.

Let $n = 0$ and $B \in \mathsf{sig}(\mathcal{T}) \cup \Sigma$ non-conjunctive in $\mathcal{T}$ with $\xi_B \in \mathsf{obj}(\mathcal{A}_{\mathcal{T},\Sigma})$. It then follows that

$$C^{0,\mathsf{ran}}_{\mathcal{A}_{\mathcal{T},\Sigma},\xi_B} = \bigsqcap_{B' \in \Sigma \setminus \mathsf{preC}^{\Sigma}_{\mathcal{T}}(B)} B' \sqcap \bigsqcap_{s \in \Sigma \setminus \mathsf{preRan}^{\Sigma}_{\mathcal{T}}(B)} \mathsf{ran}(s) \sqcap \bigsqcap_{\substack{\tilde{A} \equiv \exists \tilde{r}.\tilde{B} \in \mathcal{T} \\ B \in \mathsf{non\text{-}conj}_{\mathcal{T}}(\tilde{B}) \\ s \in \mathsf{preRole}^{\Sigma}_{\mathcal{T}}(\tilde{r}) \setminus (\mathsf{preDom}^{\Sigma}_{\mathcal{T}}(\tilde{A}) \cup \mathsf{preRan}^{\Sigma}_{\mathcal{T}}(B))}} \mathsf{ran}(s)$$

Hence, one can see that for every subconcept of the form $\mathsf{ran}(s)$ that occurs in $C^{0,\mathsf{ran}}_{\mathcal{A}_{\mathcal{T},\Sigma},\xi_B}$, we obtain that $s \notin \mathsf{preRan}^{\Sigma}_{\mathcal{T}}(B)$. As $B$ it non-conjunctive in $\mathcal{T}$, it holds that either $B$ is pseudo-primitive in $\mathcal{T}$ or that $B \equiv \exists r'.B' \in \mathcal{T}$. Hence, by Lemma 39 we can conclude that $\mathcal{T} \not\models C^{0,\mathsf{ran}}_{\mathcal{A}_{\mathcal{T},\Sigma},\xi_B} \sqsubseteq B$.

For $n > 0$, let again $B \in \mathsf{sig}(\mathcal{T}) \cup \Sigma$ non-conjunctive in $\mathcal{T}$ with $\xi_B \in \mathsf{obj}(\mathcal{A}_{\mathcal{T},\Sigma})$. We then distinguish between the following two cases. If $B$ is pseudo-primitive in $\mathcal{T}$, we obtain

$$C^{n,\mathsf{ran}}_{\mathcal{A}_{\mathcal{T},\Sigma},\xi_B} = \bigsqcap_{B' \in \Sigma \setminus \mathsf{preC}^{\Sigma}_{\mathcal{T}}(B)} B' \sqcap \bigsqcap_{s \in \Sigma \setminus \mathsf{preRan}^{\Sigma}_{\mathcal{T}}(B)} \mathsf{ran}(s) \sqcap \bigsqcap_{\substack{\tilde{A} \equiv \exists \tilde{r}.\tilde{B} \in \mathcal{T} \\ B \in \mathsf{non\text{-}conj}_{\mathcal{T}}(\tilde{B}) \\ s \in \mathsf{preRole}^{\Sigma}_{\mathcal{T}}(\tilde{r}) \setminus (\mathsf{preDom}^{\Sigma}_{\mathcal{T}}(\tilde{A}) \cup \mathsf{preRan}^{\Sigma}_{\mathcal{T}}(B))}} \mathsf{ran}(s) \sqcap \bigsqcap_{s \in \Sigma \setminus \mathsf{preDom}^{\Sigma}_{\mathcal{T}}(B)} \exists s.C_s$$

for $\mathcal{C}^{\mathsf{ran}}$-concepts $C_s$. It follows again from Lemma 39 that $\mathcal{T} \not\models C^{n,\mathsf{ran}}_{\mathcal{A}_{\mathcal{T},\Sigma},\xi_B} \sqsubseteq B$.

For $B \equiv \exists r'.B'$, we obtain

$$C^{n,\mathsf{ran}}_{\mathcal{A}_{\mathcal{T},\Sigma},\xi_B} = \bigsqcap_{B' \in \Sigma \setminus \mathsf{preC}^{\Sigma}_{\mathcal{T}}(B)} B' \sqcap \bigsqcap_{s \in \Sigma \setminus \mathsf{preRan}^{\Sigma}_{\mathcal{T}}(B)} \mathsf{ran}(s) \sqcap \bigsqcap_{\substack{\tilde{A} \equiv \exists \tilde{r}.\tilde{B} \in \mathcal{T} \\ B \in \mathsf{non\text{-}conj}_{\mathcal{T}}(\tilde{B}) \\ s \in \mathsf{preRole}^{\Sigma}_{\mathcal{T}}(\tilde{r}) \setminus (\mathsf{preDom}^{\Sigma}_{\mathcal{T}}(\tilde{A}) \cup \mathsf{preRan}^{\Sigma}_{\mathcal{T}}(B))}} \mathsf{ran}(s)$$
$$\sqcap \bigsqcap_{s \in \Sigma \setminus (\mathsf{preRole}^{\Sigma}_{\mathcal{T}}(r') \cup \mathsf{preDom}^{\Sigma}_{\mathcal{T}}(B))} \exists s.C_s \sqcap \bigsqcap_{\substack{B'' \in \mathsf{non\text{-}conj}_{\mathcal{T}}(B') \\ s \in \mathsf{preRole}^{\Sigma}_{\mathcal{T}}(r') \setminus (\mathsf{preDom}^{\Sigma}_{\mathcal{T}}(B) \cup \mathsf{preRan}^{\Sigma}_{\mathcal{T}}(B''))}} \exists s.C^{n-1,\mathsf{ran}}_{\mathcal{A}_{\mathcal{T},\Sigma},\xi_{B''}}$$

for $\mathcal{C}^{\mathsf{ran}}$-concepts $C_s$. It is easy to see that the conditions (e2), (e3) and (e4) of Lemma 39 do not hold. Thus, for $\mathcal{T} \models C^{n,\mathsf{ran}}_{\mathcal{A}_{\mathcal{T},\Sigma},\xi_B} \sqsubseteq B$ to hold, condition (e1) would have to be fulfilled. We observe that for every subconcept $\exists s.C^{n-1,\mathsf{ran}}_{\mathcal{A}_{\mathcal{T},\Sigma},\xi_{B''}}$ of $C^{n,\mathsf{ran}}_{\mathcal{A}_{\mathcal{T},\Sigma},\xi_B}$ with $B'' \in \mathsf{non\text{-}conj}_{\mathcal{T}}(B')$ and $s \in \mathsf{preRole}^{\Sigma}_{\mathcal{T}}(r') \setminus (\mathsf{preDom}^{\Sigma}_{\mathcal{T}}(B) \cup \mathsf{preRan}^{\Sigma}_{\mathcal{T}}(B''))$, we obtain $\mathcal{T} \not\models C^{n-1,\mathsf{ran}}_{\mathcal{A}_{\mathcal{T},\Sigma},\xi_{B''}} \sqsubseteq B''$ from the induction hypothesis. Thus, we have $\mathcal{T} \not\models C^{n-1,\mathsf{ran}}_{\mathcal{A}_{\mathcal{T},\Sigma},\xi_{B''}} \sqcap \mathsf{ran}(s) \sqsubseteq B'$ by Lemma 39 for every such $B''$ and $s$. We can infer that condition (e1) does not hold and, therefore, $\mathcal{T} \not\models C^{n,\mathsf{ran}}_{\mathcal{A}_{\mathcal{T},\Sigma},\xi_B} \sqsubseteq B$. □

## Appendix C. Proofs for Section 6

**Proof of Lemma 54.** Let $\mathcal{T}$ be a normalised $\mathcal{ELH}^r$-terminology and $\Sigma$ a signature such that $\Sigma \cap \mathsf{N_R} \neq \emptyset$. Additionally, let $A \in \mathsf{N_C}$ be a concept name that is non-conjunctive in $\mathcal{T}$, let $r \in \Sigma$ be a role name, and let $C$ be an $\mathcal{EL}_\Sigma$-concept. Finally, let $D = C$ or $D = \mathsf{ran}(r) \sqcap C$.

First observe that we obtain from Lemma 36 that $\mathcal{T} \not\models D \sqsubseteq A$ holds if, and only if, $(\mathcal{T}, \mathcal{A}_D) \not\models A(a_D)$. Additionally, by Lemma 44, we have $(\mathcal{T}, \mathcal{A}_D) \not\models A(a_D)$ if, and only if, $\xi_A \in \mathsf{obj}(\mathcal{A}_{\mathcal{T},\Sigma})$ and $(\mathcal{A}_D, a_D) \leq^{\mathsf{ran}}_{\Sigma} (\mathcal{A}_{\mathcal{T},\Sigma}, \xi_A)$. Thus, it is sufficient to show the following equivalence:

$$(\mathcal{A}_D, a_D) \leq^{\mathsf{ran}}_{\Sigma} (\mathcal{A}_{\mathcal{T},\Sigma}, \xi_A) \;\Leftrightarrow\; \exists\, r \in \Sigma\colon (\xi_A)_r \in \mathsf{obj}(\mathcal{A}^{\top}_{\mathcal{T},\Sigma}) \text{ and } (\mathcal{A}_D, a_D) \leq^{\mathsf{ran}}_{\Sigma} (\mathcal{A}^{\top}_{\mathcal{T},\Sigma}, (\xi_A)_r)$$

Next note that the ABox $\mathcal{A}_D$ is role-splitting as $C$ is an $\mathcal{EL}$-concept and if $D = \mathsf{ran}(r) \sqcap C$, then $\{ s(b, a_D) \in \mathcal{A}_D \mid b \in \mathsf{obj}(\mathcal{A}_D), s \in \mathsf{sig}(\mathcal{A}_D) \} = \{r(a_{\mathsf{ran}}, a_D)\}$.

Assume first $\xi_A \in \mathsf{obj}(\mathcal{A}_{\mathcal{T},\Sigma})$, $(\mathcal{A}_D, a_D) \leq^{\mathsf{ran}}_{\Sigma} (\mathcal{A}_{\mathcal{T},\Sigma}, \xi_A)$ and let $S \subseteq \mathsf{obj}(\mathcal{A}_D) \times \mathsf{obj}(\mathcal{A}_{\mathcal{T},\Sigma})$ be the corresponding $\Sigma$-range simulation. We define a relation $S^* \subseteq \mathsf{obj}(\mathcal{A}_D) \times \mathsf{obj}(\mathcal{A}^{\dagger}_{\mathcal{T},\Sigma})$ by setting for every $a \in \mathsf{obj}(\mathcal{A}_D)$, every $\xi \in \mathsf{obj}(\mathcal{A}_{\mathcal{T},\Sigma})$ and every role name $r \in \Sigma$ such that $\xi_r \in \mathsf{obj}(\mathcal{A}^{\dagger}_{\mathcal{T},\Sigma})$:

$$(a, \xi_r) \in S^* \quad \Leftrightarrow \quad (a, \xi) \in S \text{ and if } s(c, a) \in \mathcal{A}_D \text{ for some } s \in \mathsf{sig}(\mathcal{A}_D) \text{ and } c \in \mathsf{obj}(\mathcal{A}_D), \text{ then } s = r$$

Note that $S^*$ is well-defined as $\mathcal{A}_D$ is role-splitting.

To show that $S^*$ is a $\Sigma$-range simulation such that there exists $r \in \mathsf{sig}(\mathcal{A}_{\Sigma,\mathcal{T}})$ with $(\xi_A)_r \in \mathsf{obj}(\mathcal{A}^{\dagger}_{\mathcal{T},\Sigma})$ and $(a_D, (\xi_A)_r) \in S^*$, we prove that the conditions (S1)–(S3) and condition (RS) from page 663 hold.

**(S1)** If there exists $s(c, a_D) \in \mathcal{A}_D$ for some $s \in \mathsf{sig}(\mathcal{A}_D) \subseteq \Sigma$ and $c \in \mathsf{obj}(\mathcal{A}_D)$, then there exists $\xi' \in \mathsf{obj}(\mathcal{A}_{\mathcal{T},\Sigma})$ with $s(\xi', \xi_A) \in \mathcal{A}_{\mathcal{T},\Sigma}$ as $(a_D, \xi_A) \in S$ and $S$ is a $\Sigma$-range simulation, i.e. $s((\xi')_s, (\xi_A)_s) \in \mathcal{A}^{\dagger}_{\mathcal{T},\Sigma}$ and $(\xi_A)_s \in \mathsf{obj}(\mathcal{A}^{\dagger}_{\mathcal{T},\Sigma})$. Hence, $(a_D, (\xi_A)_s) \in S^*$.

Otherwise, it is easy to see that there exists $r \in \Sigma$ with $(\xi_A)_r \in \mathsf{obj}(\mathcal{A}^{\dagger}_{\mathcal{T},\Sigma})$ as $\xi_A \in \mathsf{obj}(\mathcal{A}_{\mathcal{T},\Sigma})$ and $\mathsf{sig}(\mathcal{A}_{\mathcal{T},\Sigma}) \subseteq \Sigma$. Thus, as $(a_D, \xi_A) \in S$, we have $(a_D, (\xi_A)_r) \in S^*$.

**(S2)** Let $(a, \xi_r) \in S^*$ and $A(a) \in \mathcal{A}_D$ for $a \in \mathsf{obj}(\mathcal{A}_D)$, $\xi \in \mathsf{obj}(\mathcal{A}_{\mathcal{T},\Sigma})$, $A \in \Sigma$ and $r \in \mathsf{sig}(\mathcal{A}_{\mathcal{T},\Sigma})$. It follows from the definition of $S^*$ that $(a, \xi) \in S$. Hence, as $S$ is a $\Sigma$-range simulation, we have $A(\xi) \in \mathcal{A}_{\mathcal{T},\Sigma}$, which implies that $A(\xi_r) \in \mathcal{A}^{\dagger}_{\mathcal{T},\Sigma}$ by the definition of $\mathcal{A}^{\dagger}_{\mathcal{T},\Sigma}$.

**(S3)** Let $(a, \xi_r) \in S^*$ and $s(a, a') \in \mathcal{A}_D$ for $a, a' \in \mathsf{obj}(\mathcal{A}_D)$, $\xi \in \mathsf{obj}(\mathcal{A}_{\mathcal{T},\Sigma})$, $r \in \mathsf{sig}(\mathcal{A}_{\mathcal{T},\Sigma})$ and $s \in \Sigma$. From the definition of $S^*$ we obtain $(a, \xi) \in S$. Additionally, as $S$ is a $\Sigma$-range simulation, there exists $\xi' \in \mathsf{obj}(\mathcal{A}_{\mathcal{T},\Sigma})$ such that $(a', \xi') \in S$ and $s(\xi, \xi') \in \mathcal{A}_{\mathcal{T},\Sigma}$. Thus, we have $s(\xi_r, \xi'_s) \in \mathcal{A}^{\dagger}_{\mathcal{T},\Sigma}$ by the definition of $\mathcal{A}^{\dagger}_{\mathcal{T},\Sigma}$ and $(a', \xi'_s) \in S^*$ by the definition of $S^*$ as $\mathcal{A}_D$ is role-splitting.

**(RS)** Let $(a, \xi_r) \in S^*$ and $s(c, a) \in \mathcal{A}_D$ for $a, c \in \mathsf{obj}(\mathcal{A}_D)$, $\xi \in \mathsf{obj}(\mathcal{A}_{\mathcal{T},\Sigma})$, $r \in \mathsf{sig}(\mathcal{A}_{\mathcal{T},\Sigma})$ and $s \in \Sigma$. By the definition of $S^*$, $(a, \xi) \in S$ holds and $r = s$. As $S$ is a $\Sigma$-range simulation, there exists $\xi' \in \mathsf{obj}(\mathcal{A}_{\mathcal{T},\Sigma})$ with $s(\xi', \xi) = r(\xi', \xi) \in \mathcal{A}_{\mathcal{T},\Sigma}$. Hence, $r(\xi'_r, \xi_r) \in \mathcal{A}^{\dagger}_{\mathcal{T},\Sigma}$ holds by the definition of $\mathcal{A}^{\dagger}_{\mathcal{T},\Sigma}$.

For the converse direction, we assume that there exists $\tilde{r} \in \Sigma$ such that $(\xi_A)_{\tilde{r}} \in \mathsf{obj}(\mathcal{A}^{\dagger}_{\mathcal{T},\Sigma})$ and $(\mathcal{A}_D, a_D) \leq^{\mathsf{ran}}_{\Sigma} (\mathcal{A}^{\dagger}_{\mathcal{T},\Sigma}, (\xi_A)_{\tilde{r}})$ holds. Let $S^* \subseteq \mathsf{obj}(\mathcal{A}_D) \times \mathsf{obj}(\mathcal{A}^{\dagger}_{\mathcal{T},\Sigma})$ be the corresponding $\Sigma$-range simulation. We define a relation $S \subseteq \mathsf{obj}(\mathcal{A}_D) \times \mathsf{obj}(\mathcal{A}_{\mathcal{T},\Sigma})$ by setting for every $a \in \mathsf{obj}(\mathcal{A}_D)$ and every $\xi \in \mathsf{obj}(\mathcal{A}_{\mathcal{T},\Sigma})$:

$$(a, \xi) \in S \quad \Leftrightarrow \quad \exists\, r \in \mathsf{sig}(\mathcal{A}_{\mathcal{T},\Sigma})\colon (a, \xi_r) \in S^*.$$

It is straightforward to verify that $\xi_A \in \mathsf{obj}(\mathcal{A}_{\mathcal{T},\Sigma})$ and that $S$ is a $\Sigma$-range simulation with $(a_D, \xi_A) \in S$. □

## Appendix D. Proofs for Section 7

**Proof of Lemma 60.** We require some preliminary observations. Let $\mathcal{A}_C$ be the ABox associated with a $\mathcal{C}^{\mathsf{ran}}$-concept $C$ (Lemma 36). Then, for any $\mathcal{ELH}^r$-terminology $\mathcal{T}$, $\mathcal{C}^{\mathsf{ran}}$-concept $C$ and $\mathcal{C}^{\sqcap,u}$ concept $D$, we have $\mathcal{T} \models C \sqsubseteq D$ if, and only if $\mathcal{K} \models D(a_C)$, where $\mathcal{K} = (\mathcal{T}, \mathcal{A}_C)$. By Theorem 2 (extended version),

- $\mathcal{T} \models C \sqsubseteq D$ if, and only if, $\mathcal{I}_\mathcal{K} \models D(a_C)$, where $\mathcal{I}_\mathcal{K}$ is the canonical model for $\mathcal{K}$.

Note that $\mathcal{T} \models C \sqsubseteq \exists u.D$ if, and only if, $D^{\mathcal{I}_\mathcal{K}} \neq \emptyset$ and that for any $d, d' \in \Delta^{\mathcal{I}_\mathcal{K}}$ and $R = t_1 \sqcap \cdots \sqcap t_n$, we have $(d, d') \in R^{\mathcal{I}_\mathcal{K}}$ if, and only if, there exists a role name $s$ such that $(d, d') \in s^{\mathcal{I}_\mathcal{K}}$ and $s \sqsubseteq_\mathcal{T} t_i$, for $i = 1, \ldots, n$. We summarise the consequences we require in the proof below:

(i) if $D$ is a $\mathcal{C}^{\sqcap}$-concept with occurrences $S_i = r_{i,1} \sqcap \ldots \sqcap r_{i,m_i}$ of intersections of roles, $1 \leq i \leq k$, then $\mathcal{T} \models C \sqsubseteq D$ if, and only if, there exist role names $s_i$, $1 \leq i \leq k$, such that $s_i \sqsubseteq_\mathcal{T} r_{i,j}$ for $1 \leq i \leq k$, $1 \leq j \leq m_i$ and $\mathcal{T} \models C \sqsubseteq D'$, where $D'$ is obtained from $D$ by replacing $S_i$ with $s_i$.

(ii) If $D$ is a $\mathcal{C}^{\sqcap}$-concept, then $\mathcal{T} \models C \sqsubseteq \exists u.D$ if, and only if, there exists a sequence $r'_1, \ldots, r'_n$ such that $\mathcal{I}_\mathcal{K} \models (\exists r'_1. \cdots \exists r'_n.D)(a_{\mathsf{ran}})$ or $\mathcal{I}_\mathcal{K} \models (\exists r'_1. \cdots \exists r'_n.D)(a_C)$. In the first case, there exists a subconcept $(\mathsf{ran}(r) \sqcap C')$ of $C$ (up to commutativity and associativity of $\sqcap$) such that $\mathcal{T} \models \exists r.C' \sqsubseteq \exists r'_1. \cdots \exists r'_n.D$. In the second case $\mathcal{T} \models C \sqsubseteq \exists r'_1. \cdots \exists r'_n.D$.

Now assume that $C = \bigsqcap_{1 \leq i \leq l} \mathsf{ran}(s_i) \sqcap \bigsqcap_{1 \leq j \leq n} A_j \sqcap \bigsqcap_{1 \leq k \leq m} \exists r_k.C_k$ and $\mathcal{T} \models C \sqsubseteq \exists R_1.D$. Let $R_1, \ldots, R_k$ be all the occurrences of role intersections in $\exists R_1.D$, where $R_i = r_{i,1} \sqcap \ldots \sqcap r_{i,m_i}$, for $1 \leq i \leq k$. By (i), we find role names $s_i$, $1 \leq i \leq k$, such that $s_i \sqsubseteq_\mathcal{T} r_{i,j}$ for $1 \leq i \leq k$, $1 \leq j \leq m_i$ and $\mathcal{T} \models C \sqsubseteq D'$, where $D'$ is obtained from $D$ by replacing $R_i$ with $s_i$. By applying Lemma 39 to $\mathcal{T} \models C \sqsubseteq \exists s_1.D'$ and by using that $t_1 \sqsubseteq_\mathcal{T} r_{1,j}$, for $1 \leq j \leq m_1$ and $\mathcal{T} \models D' \sqsubseteq D$, we obtain that one of the conditions (e1$_\sqcap$), (e2$_\sqcap$), (e3$_\sqcap$), or (e4$_\sqcap$) must hold.

For the second part of the lemma, we first prove by induction on $n \geq 1$ for every $\mathcal{C}^{\mathsf{ran}}$-concept $C$ and for every $\mathcal{C}^{\sqcap}$-concept $D$ with $\mathcal{T} \models C \sqsubseteq \exists r_1. \cdots \exists r_n.D$ that at least one of the following conditions holds

**(e1$_n$)** there exists a subconcept $\exists r.C'$ of $C$ such that $\mathcal{T} \models C' \sqcap \mathsf{ran}(r) \sqsubseteq D$;

**(e2$_n$)** there exists a concept name $A$ in $C$ such that $\mathcal{T} \models A \sqsubseteq \exists u.D$;

**(e3$_n$)** there exists a role name $r$ in $C$ such that $\mathcal{T} \models \exists r.\top \sqsubseteq \exists u.D$;

**(e4$_n$)** there exists a role name $r$ in $C$ such that $\mathcal{T} \models \mathsf{ran}(r) \sqsubseteq \exists u.D$.

For $n = 1$, let $C$ be a $\mathcal{C}^{\mathsf{ran}}$ concept and $D$ be a $\mathcal{C}^{\sqcap}$-concept with $\mathcal{T} \models C \sqsubseteq \exists r_1.D$. We then obtain that at least one of the conditions (e1$_\sqcap$), (e2$_\sqcap$), (e3$_\sqcap$), or (e4$_\sqcap$) must hold from the first part of the lemma, and hence, one of (e1$_n$), (e2$_n$), (e3$_n$), or (e4$_n$) is satisfied. For $n > 1$, let $C$ now be a $\mathcal{C}^{\mathsf{ran}}$ concept and $D$ be a $\mathcal{C}^{\sqcap}$-concept such that $\mathcal{T} \models C \sqsubseteq \exists r_1. \cdots \exists r_n.D$. We can apply the first part of the lemma again, and if conditions (e2$_\sqcap$), (e3$_\sqcap$), or (e4$_\sqcap$) are fulfilled, then we can conclude that conditions (e2$_n$), (e3$_n$), or (e4$_n$) are also satisfied. In the case where ($e1_\sqcap$) holds, there exists a subconcept $\exists r.C'$ of $C$ such that $\mathcal{T} \models C' \sqcap \mathsf{ran}(r) \sqsubseteq \exists r_2.\cdots\exists r_n.D$. From the induction hypothesis we obtain that at least one of the conditions ($e1_n$), ($e2_n$), ($e3_n$), or ($e4_n$) is fulfilled for $\mathcal{T} \models C' \sqcap \mathsf{ran}(r) \sqsubseteq \exists r_2.\cdots\exists r_n.D$, and thus also for $\mathcal{T} \models C \sqsubseteq \exists r_1.\cdots\exists r_n.D$ as $r \in \mathsf{sig}(C)$ and as every subconcept of $C'$ is also a subconcept of $C$.

Now, if $\mathcal{T} \models C \sqsubseteq \exists u.D$ for a $\mathcal{C}^{\mathsf{ran}}$-concept $C$ and a $\mathcal{C}^{\sqcap}$-concept $D$, then by (ii) we have to distinguish between the following two cases:

- There exists a subconcept $\mathsf{ran}(r) \sqcap C'$ of $C$ and a sequence $r'_1, \ldots, r'_{n'}$ such that $\mathcal{T} \models \exists r.C' \sqsubseteq \exists r'_1.\cdots\exists r'_{n'}.D$. For $n' = 0$, we have $\mathcal{T} \models \exists r.C' \sqsubseteq D$ and condition ($e6_u$) holds. For $n' \geq 1$ we obtain that at least one of the conditions ($e1_n$), ($e2_n$), ($e3_n$), or ($e4_n$) is satisfied. If ($e1_n$) holds, then there exists a subconcept $\exists r'.C''$ of $\exists r.C'$ such that $\mathcal{T} \models C'' \sqcap \mathsf{ran}(r') \sqsubseteq D$. If $\exists r.C' = \exists r'.C''$, we have $\mathcal{T} \models C' \sqcap \mathsf{ran}(r) \sqsubseteq D$. If $(C' \sqcap \mathsf{ran}(r))$ occurs at the top-level of the concept $C$, then $\mathcal{T} \models C \sqsubseteq D$ holds, and thus, condition ($e5_u$). Otherwise, there exists a subconcept $\exists s.((C' \sqcap \mathsf{ran}(r)) \sqcap E)$ in $C$ and ($e1_u$) is satisfied as $\mathcal{T} \models C' \sqcap \mathsf{ran}(r) \sqcap E \sqcap \mathsf{ran}(s) \sqsubseteq D$. If $\exists r.C' \neq \exists r'.C''$, $\exists r'.C''$ is a subconcept of $C'$ (thus, of $C$) and so condition ($e1_u$) holds. Finally, if one of the conditions ($e2_n$), ($e3_n$), or ($e4_n$) is satisfied, then one of ($e2_u$), ($e3_u$), or ($e4_u$) holds by (ii).
- There exists a sequence $r'_1, \ldots, r'_{n'}$ with $\mathcal{T} \models C \sqsubseteq \exists r'_1.\cdots\exists r'_{n'}.D$. For $n' = 0$ condition ($e5_u$) holds. If $n' \geq 1$, then at least one of the conditions ($e1_n$), ($e2_n$), ($e3_n$), or ($e4_n$) holds. Then, by (ii), we can conclude that one of the conditions ($e1_u$), ($e2_u$), ($e3_u$), or ($e4_u$) is satisfied as well.

□

We give the translation of $\mathcal{C}^{\sqcap,u}$-assertions to conjunctive queries. It is similar to the construction of an ABox from a $\mathcal{C}^{\mathsf{ran}}$-concept given in Section 5.1. First, given a $\mathcal{C}^{\sqcap}$-concept $C$, we define a *path in* $C$ as a finite sequence $C_0 \cdot R_1 \cdot C_1 \ldots R_n \cdot C_n$, where $C_0 = C$, $n \geq 0$, and $\exists R_{i+1}.C_{i+1}$ is a conjunct of $C_i$, for $1 \leq i < n$ ($R_i$ are conjunctions of role names). Let $x_p$ for $p \in \mathsf{paths}(C)$ be pairwise distinct variable names and set

$$
\begin{aligned}
X_C = \quad & \{\, s(x_p, x_q) \mid p, q \in \mathsf{paths}(C); q = p \cdot R \cdot C',\ s \text{ conjunct of } R \,\} \\
& \cup \{\, A(x_p) \mid A \text{ is a conjunct of } \mathsf{tail}(p), p \in \mathsf{paths}(C) \,\}
\end{aligned}
$$

Let $\vec{x}$ be the sequence of all variables in $X_C$ except $x_C$. Then the conjunctive query $q_{C,a}$ is obtained from $\exists \vec{x}. \bigwedge_{\varphi \in X_C} \varphi$ by replacing $x_C$ with $a$. Finally, for $D = D_0 \sqcap \exists u.D_1 \sqcap \cdots \sqcap \exists u.D_k$ we obtain the conjunctive query $q_{D,a}$ from $\exists \vec{x}.(\bigwedge_{0 \leq i \leq k} \bigwedge_{\varphi \in X_{D_i}} \varphi)$, (we assume that distinct variables are used in every $X_{D_i}$, $0 \leq i \leq k$, and that $\vec{x}$ is a sequence of all variables except $x_{D_0}$) by replacing $x_{D_0}$ with $a$.

To prove Lemma 63 we require some preparation. Query answering is closely related to the existence of certain homomorphisms between interpretations. Let $\Sigma$ be a signature, $O$ a set of individual names, and $\mathcal{I}_1, \mathcal{I}_2$ interpretations. A function $f : \Delta^{\mathcal{I}_1} \to \Delta^{\mathcal{I}_2}$ is called a $(O, \Sigma)$-homomorphism if

- $f(a^{\mathcal{I}_1}) = f(a^{\mathcal{I}_2})$ for all $a \in O$;

- $d \in A^{\mathcal{I}_1}$ implies $f(d) \in A^{\mathcal{I}_2}$ for all $A \in \Sigma$;
- $(d_1, d_2) \in r^{\mathcal{I}_1}$ implies $(f(d_1), f(d_2)) \in r^{\mathcal{I}_2}$ for all $r \in \Sigma$.

It is known (Chandra & Merlin, 1977) that if there exists a $(O, \Sigma)$-homomorphism from $\mathcal{I}_1$ to $\mathcal{I}_2$ and $\mathcal{I}_1 \models q[\vec{a}]$ for a conjunctive $\Sigma$-query $q$ using only individual names from $O$ and $\vec{a} = a_1, \ldots, a_k$ from $O$, then $\mathcal{I}_2 \models q[\vec{a}]$.

For the proof below we slightly refine the notion of an $(O, \Sigma)$-homomorphism by considering *partial* $(O, \Sigma)$-homomorphisms with domains that satisfy certain conditions. Namely, for every $n \geq 0$, we will call a partial $(O, \Sigma)$-homomorphism a level $n$ homomorphism if its domain contains all elements reachable by a $\Sigma$-role chain of length at most $n$ from either a named individual or from an element without a $\Sigma$-predecessor. We then prove that if for every $\mathcal{EL}^{\mathsf{ran},\sqcap,u}$-inclusion $C \sqsubseteq D$ with $\mathsf{depth}(C), \mathsf{depth}(D) \leq n$, $\mathcal{T}_1 \models C \sqsubseteq D$ implies $\mathcal{T}_2 \models C \sqsubseteq D$, then there exists a such a partial level $n$ homomorphism from a certain model of $(\mathcal{T}_1, \mathcal{A})$ to a certain model of $(\mathcal{T}_2, \mathcal{A})$.

We consider such partial homomorphisms on certain interpretations only, which we introduce first. Let $O$ be a finite set of individual names and $\mathcal{I}$ an interpretation. $d \in \Delta^{\mathcal{I}}$ is called $O$-named if there exists $a \in O$ with $d = a^{\mathcal{I}}$. A model $\mathcal{I}$ is called an $O$-forest if

**(F1)** for every $d \in \Delta^{\mathcal{I}}$ which is not $O$-named, there exists at most one $d' \in \Delta^{\mathcal{I}}$ such that $(d', d) \in \bigcup_{r \in \mathsf{N_R}} r^{\mathcal{I}}$;

**(F2)** there are no infinite sequences $d_0, d_1, \ldots$ with $(d_{i+1}, d_i) \in \bigcup_{r \in \mathsf{N_R}} r^{\mathcal{I}}$ for all $i \geq 0$ such that no $d_i$ is $O$-named.

**(F3)** if $(d, d') \in \bigcup_{r \in \mathsf{N_R}} r^{\mathcal{I}}$ and $d'$ is $O$-named, then $d$ is $O$-named.

Let $O$ be a finite set of individual names, $n \geq 0$, and $\Sigma$ a signature. A partial function $f$ from an $O$-forest $\mathcal{I}$ to a model $\mathcal{I}'$ is called an *$(O, n, \Sigma)$-homomorphism* if

**(H1)** for all $a \in O$: $a^{\mathcal{I}}$ is in the domain of $f$ and $f(a^{\mathcal{I}}) = a^{\mathcal{I}'}$;

**(H2)** for all $d, d'$ in the domain of $f$ and $r \in \Sigma$: $(d, d') \in r^{\mathcal{I}}$ implies $(f(d), f(d')) \in r^{\mathcal{I}'}$;

**(H3)** for all $d$ in the domain of $f$ and $A \in \Sigma$: $d \in A^{\mathcal{I}}$ implies $f(d) \in A^{\mathcal{I}'}$;

**(H4)** for all $d$ if there does not exist a chain $d_1, \ldots, d_m = d$ with $(d_i, d_{i+1}) \in \bigcup_{r \in \Sigma} r^{\mathcal{I}}$ of length $m > n$ of not $O$-named $d_i$, then $d$ is in the domain of $f$.

Now one can prove the following

**Lemma 70.** *Suppose $\mathcal{I}$ is an $O$-forest, $\mathcal{I}'$ an interpretation and for every $m > 0$ there exists a $(O, m, \Sigma)$-homomorphism from $\mathcal{I}$ to $\mathcal{I}'$. Assume as well that $\mathcal{I} \models q[\vec{a}]$ with $q$ a conjunctive $\Sigma$-query using only individual names from $O$ and $\vec{a} = a_1, \ldots, a_k$ from $O$. Then $\mathcal{I}' \models q[\vec{a}]$.*

*Proof.* Assume that $\vec{a}$ is a $\pi$-match of $\mathcal{I}$ and $q(\vec{x}) = \exists \vec{y}.q'(\vec{x}, \vec{y})$ such that $\vec{a}$ consists of elements of $O$. By (F2) and (F3) in the definition of $O$-forests and (H1) and (H4) in the definition of partial homomorphisms, there exists $m > 0$ such that all $\pi(v)$, $v$ from $\vec{x} \cup \vec{y}$, are in the domain of any $(O, m, \Sigma)$-homomorphism $f$. Take a $(O, m, \Sigma)$-homomorphism $f$. Then $\vec{a}$ is a $\pi'$-match of $q(\vec{x})$ and $\mathcal{I}'$, where $\pi'(v) = f(\pi(v))$, for all $v \in \vec{x} \cup \vec{y}$. $\square$

Finally, we also need a technique for constructing $(O, m, \Sigma)$-homomorphisms. Let $\mathcal{I}$ be an interpretation. For each $d \in \Delta^{\mathcal{I}}$ and $m > 0$, let

$$t_{\mathcal{I}}^{m,\Sigma,\sqcap}(d) = \{C \in \mathcal{C}_{\Sigma}^{\sqcap} \mid \mathsf{depth}(C) \leq m, d \in C^{\mathcal{I}}\},$$

where, as above, $\mathsf{depth}(C)$ is the role-depth of $C$; i.e., the number of nestings of existential restrictions in $C$.

**Lemma 71.** *Let $\Sigma$ be a finite signature and let $m > 0$ Suppose $\mathcal{I}$ is an $O$-forest and $\mathcal{I}'$ an interpretation such that*

*(in0)* $(a^{\mathcal{I}}, b^{\mathcal{I}}) \in r^{\mathcal{I}}$ *implies* $(a^{\mathcal{I}'}, b^{\mathcal{I}'}) \in r^{\mathcal{I}'}$*, for all* $a, b \in O$ *and* $r \in \Sigma$*;*

*(in1)* $t_{\mathcal{I}}^{m,\Sigma,\sqcap}(a^{\mathcal{I}}) \subseteq t_{\mathcal{I}'}^{m,\Sigma,\sqcap}(a^{\mathcal{I}'})$*, for all* $a \in O$*;*

*(in2) for all* $d \in \Delta^{\mathcal{I}}$ *there exists* $d' \in \Delta^{\mathcal{I}'}$ *such that* $t_{\mathcal{I}}^{m,\Sigma,\sqcap}(d) \subseteq t_{\mathcal{I}'}^{m,\Sigma,\sqcap}(d')$*;*

*Then there exists a* $(O, m, \Sigma)$*-homomorphism* $g$ *from* $\mathcal{I}$ *to* $\mathcal{I}'$*.*

*Proof.* We construct $g$ by constructing a sequence of functions $f_0, \ldots, f_m$, where $f_i : \mathcal{I} \to \mathcal{I}'$, as follows: the domain $\mathsf{dom}(f_0)$ of $f_0$ consists of all $a^{\mathcal{I}}$ with $a \in O$ and all $d \in \Delta^{\mathcal{I}}$ such that there does not exist a $d'$ with $(d', d) \in \bigcup_{r \in \Sigma} r^{\mathcal{I}}$. For $a^{\mathcal{I}}$ with $a \in O$ we set $f_0(a^{\mathcal{I}}) = a^{\mathcal{I}'}$. For every remaining $d \in \mathsf{dom}(f_0)$ choose a $d'$ according to (in2) and set $f_0(d) = d'$. Observe that $t_{\mathcal{I}}^{m,\Sigma,\sqcap}(d) \subseteq t_{\mathcal{I}'}^{m,\Sigma,\sqcap}(f_0(d))$ for all $d \in \mathsf{dom}(f_0)$.

Now suppose that $f_n$ has been constructed and

(in3) $t_{\mathcal{I}}^{m-n,\Sigma,\sqcap}(d) \subseteq t_{\mathcal{I}'}^{m-n,\Sigma,\sqcap}(f_n(d))$ for all $d \in \mathsf{dom}(f_n)$;

(in4) for $n > 0$: $d \in \mathsf{dom}(f_n)$ if, and only if, $d$ is not $O$-named and there exists a sequence $d_0 r_1^{\mathcal{I}} d_1 r_2^{\mathcal{I}} \cdots r_n^{\mathcal{I}} d_n = d$ of which at most $d_0$ is $O$-named such that $r_i \in \Sigma$ and $d_0 \in \mathsf{dom}(f_0)$.

To construct $f_{n+1}$ consider a $d \in \mathsf{dom}(f_n)$ and a not $O$-named $d'$ such that $(d, d') \in \bigcup_{r \in \Sigma} r^{\mathcal{I}}$. The domain of $f_{n+1}$ consists of all such $d'$. Let $R_{d,d'} = \{r \in \Sigma \mid (d, d') \in r^{\mathcal{I}}\}$ and $R_{d,d'}^{\sqcap} = (\bigsqcap_{r \in R_{d,d'}} r)$. Then

$$\exists R_{d,d'}^{\sqcap}. \bigsqcap_{D \in t_{\mathcal{I}}^{m-n-1,\Sigma,\sqcap}(d')} D \in t_{\mathcal{I}}^{m-n,\Sigma,\sqcap}(d)$$

By (in3),

$$\exists R_{d,d'}^{\sqcap}. \bigsqcap_{D \in t_{\mathcal{I}}^{m-n-1,\Sigma,\sqcap}(d')} D \in t_{\mathcal{I}'}^{m-n,\Sigma,\sqcap}(f_n(d))$$

Thus, we can choose an $e$ with $(f_n(d), e) \in r^{\mathcal{I}'}$ for all $r \in R_{d,d'}$ and $t_{\mathcal{I}}^{m-n-1,\Sigma,\sqcap}(d') \subseteq t_{\mathcal{I}'}^{m-n-1,\Sigma,\sqcap}(e)$ and set $f_{n+1}(d') = e$. This defines $f_{n+1}$. Observe further that $f_{n+1}$ is well-defined by (F1). Observe that $f_{n+1}$ has the properties (in3) and (in4), by (F3).

Now we set $g = \bigcup_{0 \leq n \leq m} f_m$. It is readily checked that $g$ is as required. ☐

We are now in the position to prove Lemma 63.

**Lemma 63** If $\varphi \in \mathsf{qDiff}_\Sigma(\mathcal{T}_1, \mathcal{T}_2)$, then there exists $\varphi' \in \mathsf{cDiff}^{\mathsf{ran},\sqcap,\mathsf{u}}_\Sigma(\mathcal{T}_1, \mathcal{T}_2)$ with $\mathsf{sig}(\varphi') \subseteq \mathsf{sig}(\varphi)$.

*Proof.* Assume $\mathcal{T}_1$ and $\mathcal{T}_2$ are given and let $(\mathcal{A}, q(\vec{a})) \in \mathsf{qDiff}_\Sigma(\mathcal{T}_1, \mathcal{T}_2)$. Let $\Sigma' = \mathsf{sig}(\mathcal{A}) \cup \mathsf{sig}(q)$. Assume that, in contrast to what is to be shown,

$$(*) \quad \mathcal{T}_1 \models \alpha \quad \Rightarrow \quad \mathcal{T}_2 \models \alpha$$

for all $\mathcal{EL}^{\mathsf{ran},\sqcap,u}$-inclusions $\alpha$ with $\mathsf{sig}(\alpha) \subseteq \Sigma'$.

Consider a model $\mathcal{I}'$ of $(\mathcal{T}_2, \mathcal{A})$ with $\mathcal{I}' \not\models q[\vec{a}]$. By Lemma 70, we obtain a contradiction if there exists an $\mathsf{obj}(\mathcal{A})$-forest $\mathcal{I}$ which is a model of $(\mathcal{T}_1, \mathcal{A})$ and such that for every $n > 0$ there exists an $(\mathsf{obj}(\mathcal{A}), n, \Sigma')$-homomorphism $f_n$ from $\mathcal{I}$ to $\mathcal{I}'$.

Take, for every $a \in \mathsf{obj}(\mathcal{A})$ a model $\mathcal{I}'_a$ of $\mathcal{T}_1$ with $d_a \in \Delta^{\mathcal{I}'_a}$ such that for all $\mathcal{C}^{\mathsf{ran}} \cup \mathcal{C}^{\sqcap,u}$-concepts $C$:

$$d_a \in C^{\mathcal{I}'_a} \quad \Leftrightarrow \quad \mathcal{T}_1 \cup t_{\mathcal{I}'}(a) \models C$$

where

$$t_{\mathcal{I}'}(a) = \{C \in \mathcal{C}^{\mathsf{ran}}_{\Sigma'} \mid a^{\mathcal{I}'} \in C^{\mathcal{I}'}\}.$$

Such interpretations $\mathcal{I}'_a$ exist by Lemma 67. We now define the unfolding $\mathcal{I}_a$ of $\mathcal{I}'_a$. A path in $\mathcal{I}'_a$ is a finite sequence $d_0 R_1 d_1 \ldots R_n d_n$, $n \geq 0$, such that $R_{i+1} = \bigsqcap \mathcal{R}_{i+1}$ for a set $\mathcal{R}_{i+1}$ of role names with $r \in \mathcal{R}_{i+1}$ iff $(d_i, d_{i+1}) \in r^{\mathcal{I}'_a}$, for all $i < n$. For a path $p$, $\mathsf{tail}(p)$ denotes the last element of $p$. Now let $\Delta^{\mathcal{I}_a}$ consist of all paths in $\mathcal{I}'_a$ and set

- $A^{\mathcal{I}_a} = \{p \in \Delta^{\mathcal{I}_a} \mid \mathsf{tail}(p) \in A^{\mathcal{I}'_a}\}$;
- $r^{\mathcal{I}_a} = \{(d, dRd') \in \Delta^{\mathcal{I}_a} \times \Delta^{\mathcal{I}_a} \mid r \in R\}$.

Then $\mathcal{I}_a$ is an $O$-forest for $O = \emptyset$. Moreover, for all $\mathcal{C}^{\sqcap,u}$-concepts $C$ and all $p \in \mathcal{I}_a$:

$$(**) \quad p \in C^{\mathcal{I}_a} \quad \Leftrightarrow \quad \mathsf{tail}(p) \in C^{\mathcal{I}'_a}.$$

In particular, $\mathcal{I}_a$ is still a model of $\mathcal{T}_1$.

Take the following (disjoint) union $\mathcal{I}$ of the interpretations $\mathcal{I}_a$:

- $\Delta^{\mathcal{I}} = \bigcup_{a \in \mathsf{obj}(\mathcal{A})} \Delta^{\mathcal{I}_a}$;
- $A^{\mathcal{I}} = \bigcup_{a \in \mathsf{obj}(\mathcal{A})} A^{\mathcal{I}_a}$, for $A \in \mathsf{N_C}$;
- $r^{\mathcal{I}} = \bigcup_{a \in \mathsf{obj}(\mathcal{A})} r^{\mathcal{I}_a} \cup \{(d_a, d_b) \mid r'(a, b) \in \mathcal{A}, r' \sqsubseteq_{\mathcal{T}_1} r\}$, for $r \in \mathsf{N_R}$;
- $a^{\mathcal{I}} = d_a$, for $a \in \mathsf{obj}(\mathcal{A})$.

We show that $\mathcal{I}$ is an $\mathsf{obj}(\mathcal{A})$-forest, a model of $(\mathcal{T}_1, \mathcal{A})$ and that there exist $(\mathsf{obj}(\mathcal{A}), n, \Sigma)$-homomorphisms from $\mathcal{I}$ to $\mathcal{I}'$ for all $n > 0$. First observe the following:

Claim 1. For all $\mathcal{EL}$ concepts $C$ and $d \in \Delta^{\mathcal{I}_a}$:

$$d \in C^{\mathcal{I}} \Leftrightarrow d \in C^{\mathcal{I}_a}$$

The proof is by induction on the construction of $C$. The interesting case is $C = \exists r.D$ and the direction from left to right. Assume that $d \in C^{\mathcal{I}} \cap \Delta^{\mathcal{I}_a}$. Take $d'$ with $(d, d') \in r^{\mathcal{I}}$ and $d' \in D^{\mathcal{I}}$. For $(d, d') \in \bigcup_{a' \in \mathsf{obj}(\mathcal{A})} r^{\mathcal{I}_{a'}}$, $d \in C^{\mathcal{I}_a}$ follows immediately from the induction hypothesis. Otherwise, $d = d_a$, $d' = d_b$ for some $b$ with $r'(a, b) \in \mathcal{A}$ and $r' \sqsubseteq_{\mathcal{T}_1} r$. By the induction hypothesis, $d' \in D^{\mathcal{I}_b}$. Hence, by $(**)$, $\mathcal{T}_1 \cup t_{\mathcal{I}'}(b) \models D$. By compactness, there exists a concept $E \in t_{\mathcal{I}'}(b)$ such that $\mathcal{T}_1 \models E \sqsubseteq D$. We obtain $\exists r'.E \in t_{\mathcal{I}'}(a)$. But then $\mathcal{T}_1 \models \exists r'.E \sqsubseteq \exists r'.D$ and we obtain $d_a \in C^{\mathcal{I}_a}$ using $r' \sqsubseteq_{\mathcal{T}_1} r$ and $(**)$.

Claim 2. $\mathcal{I}$ is an $\mathsf{obj}(\mathcal{A})$-forest and a model of $(\mathcal{T}_1, \mathcal{A})$.

That $\mathcal{I}$ is an $\mathsf{obj}(\mathcal{A})$-forest and a model of $\mathcal{A}$ follows from the construction. It remains to show that $\mathcal{I}$ is a model of $\mathcal{T}_1$. For role inclusions $r \sqsubseteq s \in \mathcal{T}_1$ it follows from the construction that $r^{\mathcal{I}} \subseteq s^{\mathcal{I}}$. Suppose $C_1 \sqsubseteq C_2 \in \mathcal{T}_1$. If $C_1$ is an $\mathcal{EL}$-concept, then $\mathcal{I} \models C_1 \sqsubseteq C_2$ follows from Claim 1 and the condition that the $\mathcal{I}_a$ are models of $\mathcal{T}_1$. Now assume that $C_1 = \mathsf{ran}(r)$ and let $d \in \mathsf{ran}(r)^{\mathcal{I}}$. If $d \neq d_a$ for any $a$, then $d \in C_2^{\mathcal{I}}$ since the $\mathcal{I}_a$ are models of $\mathcal{T}_1$. If $d = d_a$, there exists $r'(b, a) \in \mathcal{A}$ with $r' \sqsubseteq_{\mathcal{T}_1} r$. We have $\mathsf{ran}(r') \in t_{\mathcal{I}'}(a)$, and so $\mathcal{T}_1 \cup t_{\mathcal{I}'}(a) \models C_2$. Hence, by $(**)$, $d_a \in C_2^{\mathcal{I}_a}$, i.e. $d_a \in C_2^{\mathcal{I}}$ by Claim 1.

Claim 3. For every $n > 0$ there exists an $(\mathsf{obj}(\mathcal{A}), n, \Sigma')$-homomorphism from $\mathcal{I}$ to $\mathcal{I}'$.

By Lemma 71, it is sufficient to show conditions (in0), (in1), and (in2). Condition (in0) follows directly from $(*)$. Condition (in1) is proved by induction on the construction of $C$. The interesting step is for $C = \exists S.D$ with $S = r_1 \sqcap \cdots \sqcap r_m$. Let $a \in \mathsf{obj}(\mathcal{A})$ and $C \in t_{\mathcal{I}}^{n,\Sigma',\sqcap}(a^{\mathcal{I}})$. Take $d'$ with $(a^{\mathcal{I}}, d') \in S^{\mathcal{I}}$ and $d' \in D^{\mathcal{I}}$. If $d' \in \Delta^{\mathcal{I}_a}$, then, by $(**)$, $\mathcal{T}_1 \cup t_{\mathcal{I}'}(a) \models \exists S.D$. By $(*)$ and compactness, $\mathcal{T}_2 \cup t_{\mathcal{I}'}(a) \models \exists S.D$. Hence $C \in t_{\mathcal{I}'}^{n,\Sigma',\sqcap}(a^{\mathcal{I}'})$. Now assume $d' \notin \Delta^{\mathcal{I}_a}$. Then there are $r'_1, \ldots, r'_k$ and $b$ with $d' = b^{\mathcal{I}}$ such that $r'_i(a, b) \in \mathcal{A}$ for $1 \leq i \leq k$ and for every $1 \leq i \leq m$ there exists an $1 \leq j \leq m$ with $r'_j \sqsubseteq_{\mathcal{T}_1} r_i$. We have $D \in t_{\mathcal{I}}^{n,\Sigma',\sqcap}(b^{\mathcal{I}})$. By the induction hypothesis $D \in t_{\mathcal{I}'}^{n,\Sigma',\sqcap}(b^{\mathcal{I}'})$. By $(*)$, for every $1 \leq j \leq m$ there exists an $1 \leq j \leq k$ with $r'_j \sqsubseteq_{\mathcal{T}_2} r_i$. But then $C \in t_{\mathcal{I}'}^{n,\Sigma',\sqcap}(a^{\mathcal{I}'})$, as required.

For (in2), let $d \in \Delta^{\mathcal{I}}$ and $C = \bigsqcap_{D \in t_{\mathcal{I}}^{n,\Sigma',\sqcap}(d)} D$. If $d \neq a^{\mathcal{I}}$ for any $a \in \mathsf{obj}(\mathcal{A})$, then by $(**)$ there exists $b \in \mathsf{obj}(\mathcal{A})$ such that $\mathcal{T}_1 \cup t_{\mathcal{I}'}(b) \models \exists u.C$. By compactness and $(*)$, $\mathcal{T}_2 \cup t_{\mathcal{I}'}(b) \models \exists u.C$. Hence, there exists $d' \in \Delta^{\mathcal{I}'}$ with $t_{\mathcal{I}}^{n,\Sigma',\sqcap}(d) \subseteq t_{\mathcal{I}'}^{n,\Sigma',\sqcap}(d')$, as required. If $d = a^{\mathcal{I}}$ for some $a \in \mathsf{obj}(\mathcal{A})$, then, by (in1) shown above, $d' = a^{\mathcal{I}'}$ is as required.

This finishes the proof of Lemma 63. □